\documentclass[11pt]{article}
\usepackage{style}

\title{
Constant-Query PCPs of length
\texorpdfstring{$O(n\log n)$}{O(n log n)}\\
for Circuit-SAT via Code-to-PCP Holography
}
\author{
William Gay\thanks{
University of Illinois Urbana-Champaign. Email: \texttt{\{whgay2,granha\}@illinois.edu}} 
\and 
Fernando Granha Jeronimo\footnotemark[1]
}
\date{\today}

\begin{document}

\maketitle

\begin{flushright}
\footnotesize
``We may say most aptly, that the Analytical Engine\\
\emph{weaves algebraical patterns}\ \ldots''\\[0.3em]
--- Ada Lovelace (1843)
\end{flushright}

\begin{abstract}
We construct constant-query PCPs for Circuit-SAT of length
\(O(n\log n)\) bits, where \(n\) is the number of gates and wires of the
circuit, reducing the previously known polylogarithmic overhead to a
single logarithmic factor. Our PCPs have perfect completeness, constant
soundness error, and logarithmic randomness (and linear number of bits in the binary encoding of the input circuit).

At the heart of our construction is a new family of codes that reconciles
two fundamental paradigms in coding theory: the rich algebraic structure
of polynomial codes and the strong combinatorial guarantees of
asymptotically good locally testable codes. These codes simultaneously
possess constant rate, constant relative distance, constant-query local
testability, a multiplication property, and a large nonabelian group of
symmetries. Remarkably, they admit systematic encoding on an entire
group orbit. We uncover this structure in the geometry of arithmetic
cubical complexes used in recent constructions of quantum locally testable codes.

We introduce a general framework, which we call \emph{code-to-PCP
holography}, that transforms codes with these properties into short PCPs.
A key insight is that the group algebra underlying the codes enables
constant-query certification of identities on an entire message orbit,
even when an identity fails at a single coordinate. By combining this
phenomenon with a space--time encoding of circuit wiring, we obtain a
locally verifiable representation of an arbitrary computation with
constant encoding overhead. The sole logarithmic factor in our proof
length arises from the depth of the routing network.

Our results reveal a new connection between locally testable codes, probabilistically checkable proofs, and high-dimensional expansion via 
the Code-to-PCP local-to-global holography.
\end{abstract}

\newpage

\section{Introduction}\label{sec:intro}

The PCP theorem \cite{AS98,ALMSS98} shows that an NP witness can be
transformed into a proof that can be checked with a constant soundness
gap by reading only a constant number of bits. How much redundancy does this transformation
require? For more than three decades, the problem of constructing short
PCPs has brought together two central themes of theoretical computer
science: the local verification of computations and the construction of
codes that support such verification.
Polishchuk and Spielman \cite{PS94}, in their work on nearly-linear size
holographic proofs, obtained length \(n^{1+\varepsilon}\). Subsequent
advances \cite{GS06,BGHSV06,BS08,Din07} reduced the overhead to a
polylogarithmic factor: the construction of Ben-Sasson and Sudan
\cite{BS08}, together with Dinur's gap amplification \cite{Din07}, gives
constant-query PCPs of length \(n\cdot\polylog(n)\).
Obtaining length \(O(n)\) with a constant number of queries remains a
fundamental open problem.

In this work we show that a single logarithmic factor suffices. The
advance comes from a structural discovery in coding theory: the
algebraic operations that make polynomial codes so powerful in proof
systems can coexist with the combinatorial guarantees of asymptotically
good locally testable codes. We find this structure in arithmetic
cubical complexes developed for quantum coding theory. It allows us to
construct short classical PCPs by encoding and certifying the identities
of a computation directly.

\begin{theorem*}[Restated as \Cref{thm:explicit-pcp}]
Circuit-SAT has a polynomial-time PCP verifier with perfect
completeness and soundness error at most \(1/2\).
For a circuit \(C\) of size \(n\), measured by its number of
gates and wires, the proof has length \(O(n\log n)\) bits,
and the verifier makes \(O(1)\) nonadaptive bit queries and
uses \(O(\log n)\) random bits.
An accepting proof can be constructed deterministically from
\(\cenc(C)\) and a satisfying assignment in time polynomial
in \(n\).
\end{theorem*}

For every fixed \(0<\varepsilon<1\), independent repetitions
on the same proof reduce the soundness error to at most
\(\varepsilon\), with \(O(\log(1/\varepsilon))\) queries and
\(O(\log(1/\varepsilon)\log n)\) random bits.

A distinctive feature of our construction is that the logarithm has a
single, identifiable source. We represent the circuit's wiring by a
logarithmic-depth routing network and encode the resulting space--time
history at constant overhead. Thus the cost of encoding is linear in
the size of the routed computation. There is no further polylogarithmic
loss from composing or amplifying proof systems.

\subsection{The algebraic structure of good locally testable codes}

Locally testable codes are often regarded as the combinatorial core of
PCPs \cite{GS06}. With the constructions of \cite{DELLM22,PK22}, it is
now possible to check membership in a code of constant rate and constant
relative distance with a constant number of queries. Yet testing code
membership and testing a computation are different tasks. The latter
requires us to relate several encoded objects, not merely to establish
that each of them is well encoded.

Two properties of algebraic codes make these relations accessible to a
verifier. A \emph{multiplication property} allows a nonlinear gate
constraint to be represented by an identity in a code of controlled
distance. A rich group of \emph{symmetries} permits values to be moved
between positions by operations that preserve the code, making it
possible to certify the wiring of a circuit. Low-degree polynomial
codes provide both properties, but not simultaneously with constant
rate, constant distance and constant-query local testability. The
transitive algebraic-geometry codes used by Ben-Sasson, Kaplan,
Kopparty and Meir \cite{BKKM16} have both properties and constant rate,
but are not known to admit constant-query local tests. Conversely, no
such multiplication structure is known for the good locally testable
codes of \cite{DELLM22,PK22}.

This suggests a structural question underlying the problem of short PCPs:

\begin{center}
\begin{minipage}{0.91\textwidth}\centering\itshape
Can asymptotically good locally testable codes possess the algebraic
structure that makes polynomial codes so useful in proof systems?
\end{minipage}
\end{center}

\noindent We answer this question in the affirmative. In fact, we obtain
an additional property that is particularly important for computation:
systematic encoding on an entire free orbit of a large nonabelian group.
One can prescribe an arbitrary function on this orbit and extend it to
a codeword, in a way that commutes with the action of the group.
The orbit carries the data of a computation. The other coordinates
provide redundancy for local verification.

More precisely, for a finite group \(G\) and a constant integer \(b\),
consider codes on \([b]\times G\), where \(G\) acts by left translation
on each of the \(b\) copies of itself. We call the first copy the
\emph{data fibre} and write
\(\res\colon\bbK^{[b]\times G}\to\bbK^G\) for restriction to it.
The essential object is a pair of codes at two grades: a code in which
we encode the data and a second code that contains pointwise products of
words of the first. Both grades retain the same desirable combinatorial
properties.

\begin{theorem*}[Restated as \Cref{thm:explicit-multiplication-codes}]
There are a finite field \(\bbK\) of characteristic two, integers
\(\ell,b,w,h_{\mathrm{chk}},k_0\ge1\) and constants
\(\Delta,\sigma\in(0,1]\) such that the following holds for every
sufficiently large \(r\in 2k_0\N\). Put \(R=2^r\) and
\(G_R=\operatorname{SL}_2(\bbF_R)\). There are linear codes
\(C_\ell,C_{2\ell}\le\bbK^{[b]\times G_R}\) such that:
\begin{enumerate}[label=\textnormal{(\roman*)},ref=\textnormal{(\roman*)}]
\item\label{it:codes-inv} \emph{Symmetry and distance.} Both codes are
\(G_R\)-invariant and have relative distance at least \(\Delta\).
\item\label{it:codes-mult} \emph{Multiplication.} The pointwise product
of any two words of \(C_\ell\) lies in \(C_{2\ell}\), i.e.,
\(C_\ell*C_\ell\le C_{2\ell}\).
\item\label{it:codes-sys} \emph{Systematic data fibre.} There are words
\(\eta_\ell\in C_\ell\) and
\(\eta_{2\ell}=\eta_\ell\odot\eta_\ell\in C_{2\ell}\) with
\(\res(\eta_\ell)=\res(\eta_{2\ell})=\delta_{1_{G_R}}\).
For \(a\in\{\ell,2\ell\}\), the map
\[
  \Enc_a(f):=\sum_{g\in G_R} f(g)(g\cdot\eta_a)
\]
is a \(G_R\)-equivariant linear map from \(\bbK^{G_R}\) to \(C_a\)
satisfying \(\res\circ\Enc_a=\id\). In particular, both codes have
rate at least \(1/b\).
\item\label{it:codes-ltc} \emph{Local testability.}
For each \(a\in\{\ell,2\ell\}\), the code \(C_a\) has a list of \(h_{\mathrm{chk}}\) linear checks of width at most \(w\).
The tester samples a uniformly random check and a uniformly random
translate by \(G_R\), accepts every \(y\in C_a\) with probability one,
and rejects every \(y\notin C_a\) with probability at least \(\sigma\dist(y,C_a)\).
\end{enumerate}
The multiplication table of \(G_R\), the check lists and the words
\(\eta_\ell,\eta_{2\ell}\) can be constructed deterministically in time
polynomial in \(R\). The encoders can be evaluated within the same time
bound.
\end{theorem*}

The precise construction is given in
\Cref{thm:explicit-multiplication-codes}. Note that the message space in
the theorem is the whole space \(\bbK^{G_R}\), not a low-degree
subspace of functions on the group. The existence of a single codeword
restricting to the delta function at the identity, together with the
symmetry of the code, gives an entire systematic encoder: translating
this word gives a lift of every point indicator. This is a particularly
strong way to obtain constant rate, and is crucial for the
certificates we develop below. Multiplication preserves membership in
the designated higher-grade code, it need not preserve the particular
systematic lift.

The distinctive feature of \Cref{thm:explicit-multiplication-codes} is systematic encoding on
an entire free group orbit, together with constant rate, constant
relative distance, constant-query local testability and multiplication.
The construction also reveals an unexpected
connection between quantum and classical coding theory. 
We use the arithmetic cubical complexes of \cite{RSV19},
related to those used for good quantum locally testable codes
in \cite{GJ26}, and the multiplicative sheaves and
explicit global sections developed by Li, Li, Han and Liu
\cite{LLHL26} for transversal non-Clifford gates. Their construction
already supplies linearly many global zero-cocycles on an intermediate
quotient. The new ingredients here
are systematic interpolation on a principal chamber fibre and local
testability of the classical code of global sections on the principal
cover. These are the properties that allow the quantum-code geometry
to serve as a foundation for short classical PCPs. We expect the codes
of \Cref{thm:explicit-multiplication-codes} to be of independent interest.

\subsection{Code-to-PCP Holography}

How can a constant-query verifier certify an identity that is required
to hold at every message coordinate? A violation at just one coordinate
may be invisible to random sampling, even when all the objects involved
are genuine codewords. Local testability alone does not overcome this
obstacle, and the distance of the ambient code says little about the
weight of a codeword restricted to the data fibre.

Our central observation is that the group algebra of a systematic
equivariant code provides a way to certify such an identity. An
auxiliary algebraic witness forces the restrictions of several encoded
words to constitute a codeword of another code of constant relative
distance. A single nonzero entry in the original restriction then
forces a constant fraction of entries of this auxiliary tuple to be
nonzero, unless a locally verifiable codeword identity or membership
condition already fails. In this sense, the certificate makes a
violation confined to one data coordinate visible throughout a
redundant representation of the data. This is the mechanism behind
what we call \emph{code-to-PCP holography}.

The principle applies both to gate relations and to the local rules
of a space--time representation of circuit wiring. It yields a
black-box transformation whose input consists of equivariant
multiplication codes and a routing property of their symmetry groups.
In particular, none of its hypotheses refer to the arithmetic geometry
used to construct our codes. We state the theorem informally here, its
precise formulation is \Cref{thm:code-to-pcp}.

\begin{theorem*}[Restated as \Cref{thm:code-to-pcp}]
Suppose that, over a fixed finite field \(\bbK\), there is a family of
equivariant multiplication pairs \((C_1,C_2)\) with common parameters
\((b,\Delta,\sigma,w,h)\), as in \Cref{def:multiplication-pair}.
Assume that the family contains a group of size between \(u\) and
\(cu\) for every \(u\ge1\), for some absolute constant \(c\).
Suppose also that each \(G\) has a symmetric generating set \(T\) of
bounded size such that every permutation of \(G\) can be realized by
\(O(\log|G|)\) rounds of disjoint swaps between elements \(g\) and \(tg\),
with \(t\in T\). Assume that the levels, code data, generating sets and
routing schedules are deterministically computable in polynomial time,
as specified in \Cref{def:admissible-family}. Then Circuit-SAT has PCPs
with the parameters of \Cref{thm:explicit-pcp}.
\end{theorem*}

The separation identifies a concrete coding-theoretic target for
other constructions of short PCPs: any family meeting these hypotheses
can replace our arithmetic codes.

\subsection{Overview of the proof}\label{sec:overview}

\paragraph{Gate relations as algebraic identities.}
Store the left inputs, right inputs and outputs of NAND gates in
separate tables on the spatial data fibre, and encode the tables as
\(x_{\mathrm A},x_{\mathrm B},x_{\mathrm Z}\in C_{\mathrm s,1}\).
Write \(f_\kappa=\res_{\mathrm s}(x_\kappa)\) and let
\(u_{\mathrm s}=\Enc_{\mathrm s}(\Ind_{G_{\mathrm s}})\).
The multiplication property gives the codeword
\[
  p=x_{\mathrm A}\odot x_{\mathrm B}
    +u_{\mathrm s}\odot x_{\mathrm Z}
    -u_{\mathrm s}\odot u_{\mathrm s}
  \in C_{\mathrm s,2}.
\]
On the data fibre this is simply the residual of the NAND equation,
\(f_{\mathrm A}\odot f_{\mathrm B}+f_{\mathrm Z}-\Ind_{G_{\mathrm s}}\).
By filling unused positions with a satisfying NAND triple, the gate
constraints become the vanishing of an encoded algebraic expression
on the entire data fibre. The full arithmetization also imposes
Booleanity, the designated output and equality between occurrences of
the same wire
(\Cref{lem:circuit-arithmetization}). A query to an algebraic residual
can be simulated with a constant number of queries to the input
codewords. The remaining difficulty is to certify vanishing on the
fibre itself, where the residual may have very small support.

\paragraph{From a single violation to a constant gap.}
In polynomial-based PCPs, divisibility by a vanishing polynomial or a
sum-check can be used to certify global identities. Our alternative
uses the group algebra and does not require a polynomial description of
the data orbit.

Identify functions on \(G\) with elements of \(\bbK[G]\). A
\(G\)-invariant code \(C\) on \(b\) fibres is a left submodule of
\(\bbK[G]^b\). Its translated parity checks determine a sparse matrix
\(H\in\bbK[G]^{h\times b}\). Inverting each group element in each fibre
gives an \emph{opposite code} \(C^{\mathrm{op}}=\iota(C)\), a right
submodule of the same distance, with
\[
  C^{\mathrm{op}}=\ker H.
\]
Let \(D\) be an invariant locally testable code of positive relative
distance (in our application, \(C=C_1\) and \(D=C_2\)). To certify that
\(p\in D\) vanishes on the data fibre, the prover supplies \(b-1\)
auxiliary words \(w_2,\ldots,w_b\), with \(w_1=p\). The verifier tests
membership of the words in \(D\), the group-algebra identities
\(Hw=0\), and vanishing on the data fibres of the words.

The key is the interaction of two distance properties. For genuine
words of \(D\), a failed group-algebra identity is a nonzero codeword
of \(D\), so a random coordinate detects it with constant probability.
If the identities hold, however, the tuple of restrictions
\[
  (\res(w_1),\ldots,\res(w_b))\in C^{\mathrm{op}}.
\]
If even one coordinate of \(\res(p)\) is nonzero, this tuple is a
nonzero codeword of the opposite code, so a constant fraction of its
coordinates are nonzero. The restriction checks now detect the
violation with constant probability.

Completeness is equally algebraic. Let \(\eta\in C\) be the word
restricting to \(\delta_{1_G}\). Its image
\(\eta^{\mathrm{op}}=\iota(\eta)\) is in \(\ker H\) and has first
component \(1_{\bbK[G]}\). When \(\res(p)=0\), the honest prover uses
\[
  w_j=\eta_j^{\mathrm{op}}\cdot p.
\]
Equivariance ensures that these words are in \(D\), and their
restrictions vanish. The equation \(H\eta^{\mathrm{op}}=0\) gives the
required identities among them. In particular, no circuit-specific
algebraic machinery is needed for the zero test. The formal fibre
certificate of \Cref{thm:fibre-certificate} includes robustness to
arbitrary, possibly non-codeword, inputs: its rejection probability is
at least \(\gamma\dist(p,D_0)\) for a constant \(\gamma>0\), where
\(D_0=\{p\in D:\res(p)=0\}\). The certificate has constant query
complexity and constant length overhead.

\paragraph{From wiring to local rules in space--time.}
A computation is more than a collection of valid gates: their inputs
and outputs must be connected according to the circuit. Reorganizing
the occurrence tables reduces this requirement to a constant number of
masked permutation constraints. For public \(\pi\) and \(J\), such a
constraint has the form
\[
  \res_{\mathrm s}(y)(g)=\res_{\mathrm s}(x)(\pi^{-1}(g))
  \qquad (g\in J).
\]
An arbitrary permutation need not be a symmetry of the code. Rather
than trying to verify it in one step, we express it as a sequence of
local swaps along a constant-degree Cayley graph. The routing model is
due to Alon, Chung and Graham \cite{ACG94}, Nenadov's theorem
\cite{Nen23} gives logarithmic-depth routing on sufficiently good
spectral expanders. We apply a deterministic version of the construction
to Morgenstern's Ramanujan Cayley graphs \cite{Mor94} of
\(G_R=\operatorname{SL}_2(\bbF_R)\)
(\Cref{lem:arithmetic-routing}).

The prover records the routing history, one table per round. Each
successive table is obtained from its predecessor by choosing, at each
position, one of a bounded number of group translations (including the
identity). Which translation applies is specified by public masks.
Testing the steps separately would either require logarithmically many
tests or dilute the effect of a single bad transition. We avoid this
by treating time as another encoded direction.
The spatial group has size \(O(n)\), the temporal group has size
\(O(\log n)\), and the history is encoded in their tensor-product code
\[
  C_1(I_\otimes)=C_{\mathrm s,1}\otimes C_{\mathrm t,1},
  \qquad I_\otimes=I_{\mathrm s}\times I_{\mathrm t},
  \quad I_v=[b]\times G_v.
\]
A spanning-tree traversal of the temporal Cayley graph lets us organize
the history into a constant number of words with transitions along a
bounded set of translations. The entire history is then described by a
constant number of quadratic identities on the tensor-code data fibre,
together with the endpoint conditions. Each identity is certified by
the same group-algebra mechanism
(\Cref{thm:quadratic-fibre-identities,thm:permutation-certificate}).

The tensor products retain multiplication and equivariance, and we
prove that they retain local testability using an argument in the
spirit of Dinur, Sudan and Wigderson \cite{DSW06}, exploiting the
sparsity of the translated checks. The number of spatial and
space--time words, and the number of identities, are all constant. Since
the alphabet is fixed, the resulting length in bits is
\[
  O(n)+O(n\log n)=O(n\log n).
\]
The entire routing history is thus certified at constant-factor
encoding overhead.

\paragraph{An arithmetic source of the codes.}
It remains to construct codes with the properties used by holography.
Our starting point is the six-dimensional arithmetic cubical complex
\(\widehat X_R\), obtained from a product of six regular trees by a
quaternionic arithmetic lattice over \(\bbF_2(t)\). The complex is a
principal cover of a fixed finite complex \(X_0\), with deck group
\(G_R=\operatorname{SL}_2(\bbF_R)\). Its chambers, the coordinates of
our code, are therefore \(b=|X_0(6)|\) copies of \(G_R\). This supplies
the required symmetry. Around every vertex the incident chambers are
indexed by a product of six projective lines. A codeword is a global
section whose restriction to every such neighborhood lies in a tensor
product of projective Reed--Solomon codes. The local polynomial degrees
add under multiplication, giving the multiplication property.

We use the quantum-code geometry of \cite{GJ26} (see also
\cite{BLN26,CHLT26}) and the six-direction multiplicative sheaves of
\cite{LLHL26}, including their explicit global sections constructed
from Moore determinants. The remaining challenge is to prove
systematic interpolation, distance and local testability.

\paragraph{A delta function on a free orbit.}
The rate of the local Reed--Solomon factors is too small for a naive
constraint count to prove positive global rate. Instead, we seek one
global section whose restriction to the fibre of a single chamber is
the indicator of one point. Equivariance will then supply a full copy of
the regular representation as message data.

The determinant sections of \cite{LLHL26} initially live on an
intermediate cover \(X_R=U_R\backslash\widehat X_R\), where the chamber
fibre is \(\Omega=\bbF_R^2\setminus\{0\}\). Specializing their rank
calculation to one physical chamber, we obtain a restriction space
\(W\le\bbL^\Omega\) of dimension proportional to \(R^2\), where
\(\bbL=\bbK\bbF_R\), using products at a bounded grade
(\Cref{thm:large-fibre-restriction}). This space contains the constants
and separates points. A bounded-product lemma of \cite{LLHL26},
followed by linear descent to the fixed alphabet, yields global
sections whose restrictions are point indicators on a positive
fraction of \(\Omega\).

Here we exploit the passage to the principal cover. A matrix in
\(\operatorname{SL}_2(\bbF_R)\) is determined by the two columns of its
inverse. The set of points for which we have indicators is dense, we
show it contains a pair of vectors with determinant one. Taking the
product of two suitably translated indicator sections consequently
isolates a single group element on the principal chamber fibre.
Translating it to the identity and padding the grade produces
\(\eta_\ell\) with restriction \(\delta_{1_{G_R}}\), squaring it gives
\(\eta_{2\ell}\). The translates of these two words provide the
systematic encoders of \Cref{thm:explicit-multiplication-codes}
(\Cref{thm:fibre-interpolation}). In this way a geometric construction
of global sections yields arbitrary message data on a large nonabelian
group, while preserving the local polynomial constraints.

\paragraph{Local agreement and directional mixing.}
Finally we prove distance and local testability of the global sections.
The natural tester reads the chambers around a random vertex and
checks membership in the appropriate tensor code. With at least three
directions, local views that agree on most of their overlaps are close
to a single tensor codeword
(\Cref{lem:tensor-patch-agreement}). This is the phenomenon behind the
robustness of three-fold tensor products \cite{BS06,Vid15}.
A local-minimality argument turns agreement into syndrome expansion
when the complex has sufficient directional mixing
(\Cref{lem:locally-minimal-support}). The argument does not require
product expansion of the local codes. The mixing follows from the
Ramanujan property of congruence quotients of quaternionic lattices
\cite{LSV05,BR17}, transported from one-place lattices to the relevant
components of the complex using nonnegative quadratic-form bounds
under finite covers
(\Cref{lem:nonnegative-mixing-covers,lem:mixing-components}). The
translated bounded-width linear tester required by holography is
obtained from these vertex checks
(\Cref{cor:chamber-translated-tester}). This completes the connection
between the arithmetic construction and the abstract PCP theorem.

\subsection{Related work}

\paragraph{Short PCPs.}
Polishchuk and Spielman \cite{PS94} gave the first constant-query
PCPs of nearly linear length \(n^{1+\varepsilon}\), and
arithmetized arbitrary wiring by routing on de Bruijn graphs,
an idea used in most later short PCPs.
Goldreich and Sudan \cite{GS06} and Ben-Sasson, Goldreich, Harsha, Sudan
and Vadhan \cite{BGHSV06} obtained constant-query PCPs of length
\(n\cdot\exp(\tilde O(\sqrt{\log n}))\) and
\(n\cdot\exp((\log n)^{\varepsilon})\), respectively. The
constructions of \cite{BS08,Din07} discussed above give
\(n\cdot\polylog(n)\). The work closest in spirit to ours is
\cite{BKKM16}, which separates the cost of routing from the cost of
encoding, replaces low-degree polynomials by transitive
algebraic-geometry codes, and uses the automorphisms of the code to check
wiring. It achieves length \(O(n)\) with \(n^{\varepsilon}\) queries and a
non-uniform verifier. We share the viewpoint of separating routing from encoding. Our new
locally testable multiplication codes and group-algebra certificates
make it possible to verify the routed array with constant queries and
constant encoding overhead, with one logarithm for the routing itself.

Linear proof length is known under two relaxations of the target
parameters: \cite{BKKM16} allows \(n^\varepsilon\) queries, and linear
length with constant query complexity is attainable with interaction in
the interactive oracle proof model \cite{BCGRS17,RR20}. Neither
relaxation is used in \Cref{thm:explicit-pcp}.

\paragraph{PCPs from high-dimensional expanders.}
A different line of work obtains two-query PCPs with small soundness
error and quasilinear size from agreement tests on high-dimensional
expanders. Bafna, Minzer, Vyas and Yun \cite{BMVY25} achieve every
constant soundness \(\varepsilon>0\), with the swap-expansion bound of
Dikstein, Oppenheim and Valentiner-Branth \cite{DOV26}, their length is
\(n(\log n)^{\exp(\poly(1/\varepsilon))}\). Dikstein, Dinur, Hopkins and
Singer \cite{DDHS26} obtain length \(n(\log n)^{C}\) for an absolute
constant \(C\), with alphabet size \(\exp(\poly(1/\varepsilon))\) and
soundness \(\varepsilon\) as small as inverse polylogarithmic in \(n\).
Both amplify the soundness of a base PCP such as that of \cite{BS08} and
add polylogarithmic factors to its length. They optimize soundness at two
queries, whereas \Cref{thm:explicit-pcp} optimizes length at constant soundness.

\paragraph{Codes with multiplication and local tests.}
Multiplication codes were isolated as the algebraic input to proof systems
by Meir \cite{Mei13}. Dinur, Liu and Zhang \cite{DLZ23} constructed codes
on high-dimensional expanders with a multiplication property, these are
locally testable in a low-degree regime, in which constant rate is not
known. Bafna and Vyas \cite{BV26} construct codes supporting
multiplication from tensor products of Reed--Solomon codes and use them
for private PCPs of length \(O(n\log n)\) over a polynomial-size alphabet
with \(O(\sqrt n)\) queries. The good quantum locally testable codes of
\cite{GJ26} and the related constructions \cite{BLN26,CHLT26}, together
with the multiplicative sheaves and determinant sections of
\cite{LLHL26}, provide the geometric and algebraic framework for our
codes. The combinatorial framework of high-dimensional cubical complexes
with local codes is due to Dinur, Lin and Vidick \cite{DLV24}, building
on \cite{DELLM22,PK22}. What we add is the classical statement of
\Cref{thm:explicit-multiplication-codes}: surjectivity onto a free orbit of coordinates, and
local testability of the code of global sections on the principal cover.

\subsection{Discussion and open problems}

\paragraph{Toward linear-length PCPs.}
The construction isolates a particular obstacle to linear length. The
algebraic encoding of the computation and the fibre certificates have
constant overhead, the routing history is the only super-linear
component. A direct implementation of arbitrary permutations through a
constant-degree routing network needs logarithmic depth in the worst
case. Can one certify an arbitrary permutation without recording the
whole history, or encode the history with linear rather than
space--time overhead? The separation of these issues from the
construction of the codes gives a concrete direction for pursuing the
linear-length PCP conjecture.

\paragraph{Improving the codes.}
The constants in \Cref{thm:explicit-pcp,thm:explicit-multiplication-codes} are absolute but astronomical.
For example, the local tester reads all chambers around a vertex, and
the rate guarantee depends on the size of a fixed arithmetic base
complex. We have not attempted to optimize these parameters. The
present result is qualitative: it shows that the single logarithm can
be achieved with constant bit-query complexity. It would be interesting
to obtain the same algebraic and combinatorial structure with reasonable
parameters, perhaps using complexes of lower dimension or smaller local
codes. By \Cref{thm:code-to-pcp}, such a family with suitable density,
constructibility and routing properties would immediately give another
construction of \(O(n\log n)\)-length PCPs.

\paragraph{Efficient verification and proximity.}
Our uniform verifier preprocesses the circuit in polynomial time, in
particular to compute a routing schedule and public mask codewords. We
do not know whether the construction admits a verifier running in
polylogarithmic time given suitable oracle access to the input, as in
\cite{BGHSV05}, nor do we address PCPs of proximity. For nondeterministic
time \(t\), composing with the standard reduction to circuits of size
\(O(t\log t)\) gives length \(O(t\log^2t)\).

\subsection{Organization}

\Cref{sec:preliminaries} fixes notation and computational conventions.
\Cref{sec:fibre-certificates} proves the vanishing certificate,
\Cref{sec:algebraic-identities} establishes tensor-product local
testability and certification of algebraic identities,
\Cref{sec:permutation-certificates} certifies permutation constraints
using routing histories. These ingredients establish code-to-PCP holography
in \Cref{sec:code-to-pcp}.
\Cref{sec:chamber-codes} defines the multiplicative chamber codes, and
\Cref{sec:chamber-testability} proves distance and local testability from
local agreement and directional mixing.
\Cref{sec:arithmetic-covers} constructs arithmetic covers satisfying the
required mixing bounds and proves deterministic logarithmic-depth routing.
\Cref{sec:determinant-restrictions} constructs a large restriction space
from determinant sections, and \Cref{sec:fibre-interpolation} proves
surjectivity onto a principal chamber fibre and obtains the equivariant
encoders. Finally, \Cref{sec:explicit-codes-pcps} verifies uniformity and
completes the proofs of \Cref{thm:explicit-pcp,thm:explicit-multiplication-codes}.
Appendix~\ref{app:preliminaries} proves auxiliary preliminary lemmas.

\section{Preliminaries}
\label{sec:preliminaries}

We introduce the notation and basic facts used throughout the paper.
Proofs of the auxiliary lemmas are deferred to Appendix~\ref{app:preliminaries}.

\subsection{Notation, Linear Codes, and Tensor Codes}
\label{subsec:codes-and-tensors}

For an integer $n\ge 0$, write $[n]:=\set{1,\ldots,n}$, with
$[0]=\varnothing$.
For a finite set $I$ and a field $\bbK$, the vector space $\bbK^I$
consists of all functions $I\to\bbK$, which we call \emph{words}.
For $A\subseteq I$, let $\Ind_A\in\bbK^I$ be its indicator.
In particular, $\Ind_I$ is the all-ones word.
We also write $\delta_i$ for the indicator of the singleton $\set{i}$.
For nonempty $I$ and $x,y\in\bbK^I$, write
\[
  \supp(x):=\set*{i\in I:x(i)\ne 0},\qquad
  \weight(x):=\frac{\abs{\supp(x)}}{|I|},\qquad
  \dist(x,y):=\weight(x-y).
\]
For a nonempty set $S\subseteq\bbK^I$, put
$\dist(x,S):=\min_{c\in S}\dist(x,c)$.
A subspace $C\le\bbK^I$ is a \emph{linear code}, of block-length $|I|$.
Its \emph{rate} and \emph{relative distance} are
\[
  R(C):=\frac{\dim_{\bbK}C}{|I|},\qquad
  \Delta(C):=\min_{c\in C\setminus\set{0}}\weight(c),
\]
with the convention $\Delta(\set{0})=1$.
A map $\phi:J\to I$ induces the \emph{pullback}
$\phi^*:\bbK^I\to\bbK^J$, defined by $\phi^*f=f\circ\phi$.
If $A \subseteq I$, the pullback of the inclusion $A\hookrightarrow I$ is
called \emph{restriction}, and denoted $x|_A$.
An \emph{information set} for a code $C$ of dimension $k$ is a set
$A\subseteq I$ of size $k$ such that restriction is an isomorphism
$C\to\bbK^A$.

\begin{lemma}\label[lemma]{lem:information-sets}
Let $C\le\bbK^I$ be a linear code and $S\subseteq I$.
If $|I\setminus S|<\Delta(C)|I|$, then restriction
$C\to\bbK^S$ is injective and $S$ contains an information set for $C$.
\end{lemma}

For $x,y\in\bbK^I$, their pointwise (or Hadamard) product is
$(x\odot y)(i)=x(i)y(i)$.
The \emph{Schur product} of linear codes $C,D\le\bbK^I$ is
\[
  C*D:=\linspan_{\bbK}\set*{x\odot y:x\in C,\ y\in D}.
\]
The \emph{tensor product} of $C\le\bbK^I$ and $D\le\bbK^J$ is
identified with the subspace of $\bbK^{I\times J}$ spanned by
$(c\otimes d)(i,j):=c(i)d(j)$, for $c\in C$ and $d\in D$.
More generally, $\bigotimes_{i=1}^d C_i$, with
$C_i\le\bbK^{I_i}$, is a code indexed by $I=\prod_{i=1}^d I_i$.
An $i$-line of $x\in\bbK^I$ is a word indexed by $I_i$ of the form
\[
  t \mapsto x(a_1,\ldots,a_{i-1},t,a_{i+1},\ldots,a_d),
\]
where all $a_j\in I_j$ for $j\ne i$ are fixed.

\begin{lemma}\label[lemma]{lem:tensor-codes}
Let $d\ge 1$ and let $C_i\le\bbK^{I_i}$ be linear codes for
$i\in[d]$. A word belongs to $\bigotimes_{i=1}^d C_i$ if and only if
every $i$-line lies in $C_i$, for every $i\in[d]$. Moreover,
\[
  \dim_{\bbK}\paren*{\bigotimes_{i=1}^d C_i}
    =\prod_{i=1}^d\dim_{\bbK}C_i,\qquad
  \Delta\paren*{\bigotimes_{i=1}^d C_i}
    \ge\prod_{i=1}^d\Delta(C_i).
\]
Equality holds in the latter bound when all factors are nonzero.
\end{lemma}

\subsection{Local Testability and Linear Checks}
\label{subsec:local-testing}

\begin{definition}[Algorithmic Local Testability]
\label[definition]{def:local-testability}
Let $\bbK$ be a finite field and let
$\mcC=\set{C_i\subseteq\bbK^{n_i}}_{i\in\Lambda}$ be a family of
nonempty codes, with block-lengths $n_i \to \infty$ indexed by
$\Lambda\subseteq\bits^*$.
Fix an integer $q\ge 1$ and $\sigma>0$.
A $(q,\sigma)$-local test for $\mcC$ is a randomized
polynomial-time algorithm $T$ which, on input $(i,1^{n_i})$,
makes at most $q$ nonadaptive queries to an oracle string $x\in\bbK^{n_i}$, such that:
\begin{itemize}
  \item[\textnormal{(C)}] If $x\in C_i$, then
  $T^x(i,1^{n_i})$ accepts with probability one.
  \item[\textnormal{(S)}] If $x\notin C_i$, then
  $T^x(i,1^{n_i})$ rejects with probability at least
  $\sigma\, \dist(x,C_i)$.
\end{itemize}
\end{definition}

The unary block-length makes the running-time bound polynomial in
$n_i+|i|$, and the index $i \in \Lambda$ distinguishes codes of the same length.
We will only deal with the case in which $|i| \le \poly(n_i)$. 
By condition (S) of \Cref{def:local-testability}, rejection probability at
most $\varepsilon$ implies $\dist(x,C_i)\le\varepsilon/\sigma$.

A \emph{linear check} on $\bbK^I$ is a functional
$\chi(x)=\ip{\lambda,x}=\sum_{i\in I}\lambda_i x(i)$, for a unique
$\lambda\in\bbK^I$. Its \emph{width} is $|\supp(\lambda)|$.
It is a \emph{parity check} for $C\le\bbK^I$ if
$C\subseteq\ker\chi$, equivalently if $\lambda\in C^\perp$, where
$C^\perp:=\set*{\lambda:\ip{\lambda,c}=0\text{ for all }c\in C}$.
A linear test samples a parity check and accepts if its value is zero.
The checks sampled with positive probability detect every word outside
$C$ exactly when their coefficient vectors span $C^\perp$.

\begin{lemma}\label[lemma]{lem:parity-check-test}
Let $\mcC=\set{C_i}_{i\in\Lambda}$ be a family of linear codes as in
\Cref{def:local-testability}, and let $q\ge 1$ and $\sigma>0$.
Suppose a deterministic polynomial-time algorithm, on input
$(i,1^{n_i})$, outputs a matrix
$H_i\in\bbK^{m_i\times n_i}$ with $m_i\ge 1$.
Assume $C_i = \ker(H_i)$, every row of $H_i$ has width at most $q$, and
\[
  \frac{|\supp(H_i x)|}{m_i}
    \ge\sigma\dist(x,C_i)
  \qquad\text{for every }x\in\bbK^{[n_i]}.
\]
Then $\mcC$ has a $(q,\sigma/2)$-local test using
$\lceil\log_2 m_i\rceil$ random bits at level $i$.
If every $m_i$ is a power of two, the soundness parameter is $\sigma$.
\end{lemma}

Under ideal uniform row sampling the rejection probability is exactly
$|\supp(H_i x)|/m_i$, the factor of two allows a bounded fair-coin
implementation. Later algebraic tests are analyzed under their stated
finite sampling distributions, and the PCP construction accounts for their
implementation using random bits.

\subsection{Group Actions, Group Algebras, and Equivariant Codes}
\label{subsec:group-actions}

Let $G$ be a finite group, $\bbK$ an arbitrary field, and $b\ge 1$.
Write $I:=[b]\times G$. The group acts on $I$ and $\bbK^I$, respectively, by
\[
  g\cdot(j,h):=(j,gh),\qquad
  (g\cdot x)(j,h):=x(j,g^{-1}h).
\]
The orbits $\set{j}\times G$ are called the \emph{fibres}, the first is called the
\emph{data fibre}. Every free action with $b$ orbits has this form after
choosing orbit representatives.
A code $C\le\bbK^I$ is $G$-invariant if $g\cdot C=C$ for all $g\in G$.
Define $\res:\bbK^I\to\bbK^G$ by $\res(x)(g):=x(1,g)$.
For $x,y\in\bbK^I$,
\[
  \dist(\res(x),\res(y))\le b\,\dist(x,y).
\]
The group algebra $\bbK[G]$ is the vector space with basis $G$.
Identify $f\in\bbK^G$ with $\sum_{g\in G}f(g)g$.
Multiplication is convolution:
\[
  (af)(g):=\sum_{s\in G}a(s)f(s^{-1}g).
\]
Thus left multiplication by $s\in G$ is the action $s\cdot f$.
The convolution unit is $\delta_{1_G}$, whereas the pointwise unit is
$\Ind_G$. Extending the group action linearly gives
\[
  a\cdot x:=\sum_{s\in G}a(s)(s\cdot x),
\]
for $a \in \bbK[G]$ and $x \in \bbK^I$.
This identifies $\bbK^I$ with the left module $\bbK[G]^b$, and
$G$-invariant linear codes are precisely its submodules.
Restriction is $\bbK[G]$-linear and preserves pointwise products:
\[
  \res(a\cdot x)=a\,\res(x),\qquad
  \res(x\odot y)=\res(x)\odot\res(y).
\]
Finally, the $\bbK$-linear involution
$\iota:\bbK[G]\to\bbK[G]$ is defined by
\[
  \iota(a):=\sum_{g\in G}a(g)g^{-1}. 
\]
This satisfies $\iota(aa')=\iota(a')\iota(a)$.
Applied to each block of $\bbK^I$, it permutes coordinates and hence
preserves weight and distance.

\begin{lemma}\label[lemma]{lem:equivariant-encoding}
Let $C\le\bbK^I$ be a $G$-invariant linear code.
The following are equivalent:
\begin{itemize}
  \item[\textnormal{(1)}] The restriction
  $\res|_C:C\to\bbK^G$ is surjective.
  \item[\textnormal{(2)}] There is $\eta\in C$ with
  $\res(\eta)=\delta_{1_G}$.
  \item[\textnormal{(3)}] There is a $G$-equivariant linear map
  $\Enc:\bbK^G\to C$ with $\res\circ\Enc=\id_{\bbK^G}$.
\end{itemize}
When these conditions hold, for any $\eta$ as in \textnormal{(2)} one may take
\[
  \Enc(f):=\sum_{g\in G}f(g)(g\cdot\eta)=f\cdot\eta.
\]
Consequently $\dim_{\bbK}C\ge|G|$ and $R(C)\ge 1/b$.
\end{lemma}

The image of the encoder in \Cref{lem:equivariant-encoding} may be a proper
subspace of $C$.
Given a nonempty indexed list
$\mcL=\set{\chi_\rho}_{\rho\in[h]}$ of linear checks on $\bbK^I$, the
\emph{translated tester} chooses $(\rho,s)$ uniformly from
$[h]\times G$ and accepts if $\chi_\rho(s\cdot x)=0$.
Repeated checks are allowed and retain their sampling multiplicity.

\begin{lemma}\label[lemma]{lem:translated-checks}
Suppose every check in $\mcL$ has width at most $w$.
For every $B\subseteq I$,
\[
  \Pr[\textnormal{at least one queried coordinate lies in }B]
    \le\frac{w|B|}{|G|}.
\]
For any $x,y\in\bbK^I$ at relative distance $\delta$, the rejection
probabilities of the translated tester on $x$ and $y$ differ by at
most $wb\delta$.
\end{lemma}

\subsection{PCPs and Computational Conventions}
\label{subsec:pcp-conventions}

\begin{definition}[Probabilistically Checkable Proofs]
\label[definition]{def:pcp}
Let $\ell,\rho,q:\bits^*\to\N$ be polynomial-time computable and
polynomially bounded in the input length.
A $\mathrm{PCP}_\ell(\rho,q)$ verifier for a language
$L\subseteq\bits^*$ is a randomized polynomial-time algorithm $V$
which, on input $x$, uses at most $\rho(x)$ independent unbiased random
bits and makes at most $q(x)$ nonadaptive queries to an oracle string
$\Pi\in\bits^{\ell(x)}$, such that:
\begin{itemize}
  \item[\textnormal{(C)}] If $x\in L$, there exists
  $\Pi\in\bits^{\ell(x)}$ for which $V^\Pi(x)$ accepts with
  probability one.
  \item[\textnormal{(S)}] If $x\notin L$, then for every
  $\Pi\in\bits^{\ell(x)}$ it holds that $V^\Pi(x)$ rejects with probability at
  least $1/2$.
\end{itemize}
\end{definition}

A Boolean circuit consists of a finite DAG with input gates, NAND gates, and one output gate.
Every NAND gate has fan-in two and unrestricted fan-out.
The size is the total number of gates and wires. 
Fix a binary list encoding $\cenc$ recording the gate types, ordered
predecessor indices, and output index. A natural encoding of a circuit of size $n$ has length between $\Omega(n)$ and $O(n\log(n))$.
The language \textnormal{Circuit-SAT} consists of encodings of circuits
having a satisfying assignment. Malformed encodings are rejected. 

\section{Certifying Vanishing on a Free Orbit}
\label{sec:fibre-certificates}

A codeword may restrict to a word supported at one point of the data
fibre. Sampling the restriction then detects nonvanishing with
probability only $1/|G|$. We give a certificate that a codeword vanishes
on the entire fibre. The certificate forces the restrictions of several
words to form a codeword whose distance is known.

\begin{theorem}\label[theorem]{thm:fibre-certificate}
Let $\bbK$ be a finite field, let $G$ be a finite group, and let
$b,h,w,q\ge 1$ be integers. Put $I=[b]\times G$.
Let $C,D\le\bbK^I$ be $G$-invariant linear codes with
$\Delta(C)\ge\Delta_C>0$ and $\Delta(D)\ge\Delta_D>0$.
Suppose there are linear checks
$\mcL=\set{\chi_\rho}_{\rho\in[h]}$, each of width at most $w$, such that
\[
  C=\set*{x\in\bbK^I:
    \chi_\rho(s\cdot x)=0
    \text{ for every }\rho\in[h],\ s\in G},
\]
and there is $\eta\in C$ with $\res(\eta)=\delta_{1_G}$.
Suppose $D$ has a $(q,\sigma)$-local tester $T_D$, where $\sigma>0$. 
Set $D_0:=\set*{p\in D:\res(p)=0}$
and
\begin{equation}\label{eq:fibre-certificate-gap}
  \gamma:=
  \frac{\sigma}{3}
  \min\set*{
    \frac{\Delta_D}{h\sigma+bw},
    \frac{\Delta_C}{\sigma+b}
  }.
\end{equation}
There is a test $V$ with oracle access to an input word $p\in\bbK^I$
and witness words $\mcW=\set{w_j\in\bbK^I}_{j=2}^b$
which makes at most $\max\set{q,w,1}$ nonadaptive queries and has the
following properties:
\begin{itemize}
  \item[\textnormal{(C)}] If $p\in D_0$, then there exists $\mcW$ for
  which $V^{p,\mcW}$ accepts with probability one.
  \item[\textnormal{(S1)}] If $p \in D \setminus D_0$, then for every $\mcW$ it holds that $V^{p,\mcW}$ rejects with probability at least $\gamma$. 
  \item[\textnormal{(S2)}] If $p \notin D$, then for every $\mcW$ it holds that $V^{p,\mcW}$ rejects with probability at least $\gamma \dist(p, D_0)$. 
\end{itemize}
\end{theorem}

\subsection{The Opposite Code}
\label{subsec:opposite-code}

We first express translated linear checks as equations over a group
algebra. This calculation holds over any field.

\begin{lemma}\label[lemma]{lem:opposite-presentation}
Let $\bbK$ be a field, let $G$ be a finite group, and let $b\ge 1$.
Put $I=[b]\times G$. Let
$\mcL=\set{\chi_\rho}_{\rho\in[h]}$ be linear checks on $\bbK^I$, and let
\[
  C=\set*{x\in\bbK^I:
    \chi_\rho(t\cdot x)=0
    \text{ for every }\rho\in[h],\ t\in G}.
\]
Write
\[
  \chi_\rho(x)
    =\sum_{j=1}^b\sum_{s\in G}
      \lambda_{\rho,j,s}x(j,s),
\]
and define $H\in\bbK[G]^{h\times b}$ by $H_{\rho j}:=\sum_{s\in G}\lambda_{\rho,j,s}s$.
Then $C^{\mathrm{op}}:=\iota(C)$
is a right $\bbK[G]$-submodule of $\bbK[G]^b$, such that
$\ker H=C^{\mathrm{op}}$ and
$\Delta(C^{\mathrm{op}})=\Delta(C)$.
If $\eta\in C$ satisfies $\res(\eta)=\delta_{1_G}$, then
$\eta^{\mathrm{op}}:=\iota(\eta)$ satisfies
$H\eta^{\mathrm{op}}=0$ and $\eta^{\mathrm{op}}_1=\delta_{1_G}=1_{\bbK[G]}$.
\end{lemma}

\begin{proof}
The space $C$ is $G$-invariant. Indeed, if $x\in C$ and $u\in G$, then
for every $\rho\in[h]$ and $t\in G$,
$\chi_\rho(t\cdot(u\cdot x))
=\chi_\rho((tu)\cdot x)=0$.
Thus $C$ is a left $\bbK[G]$-submodule. For $x\in C$ and
$a\in\bbK[G]$, the identity
$\iota(x)a=\iota(\iota(a)\cdot x)$
holds in every component. Its right-hand side belongs to $\iota(C)$,
so $C^{\mathrm{op}}$ is a right submodule.
Let $z=\iota(x)$, where $x\in\bbK^I$.
Under the identification $\bbK^I=\bbK[G]^b$, its components are
\[
  z_j=\sum_{g\in G}x(j,g)g^{-1}.
\]
For every $\rho\in[h]$, multiplication in the group algebra gives
\[
\begin{aligned} (Hz)_\rho
    &=\sum_{j=1}^b H_{\rho j}z_j
    =\sum_{j=1}^b\sum_{s,g\in G}
      \lambda_{\rho,j,s}x(j,g)\,sg^{-1}.
  \end{aligned}
\]
The coefficient of $t\in G$ comes from pairs satisfying
$sg^{-1}=t$ iff $g=t^{-1}s$. 
Consequently,
\[
  \begin{aligned}
  (Hz)_\rho(t)
    &=\sum_{j=1}^b\sum_{s\in G}
      \lambda_{\rho,j,s}x(j,t^{-1}s)
    =\chi_\rho(t\cdot x).
  \end{aligned}
\]
It follows that
$Hz=0$ iff $x\in C$ iff $z\in C^{\mathrm{op}}$.
This proves the kernel identity. Since $\iota$ permutes the coordinates
of $\bbK^I$, it preserves relative distance and gives
$\Delta(C^{\mathrm{op}})=\Delta(C)$.
Finally, $\eta\in C$ implies $H\eta^{\mathrm{op}}=0$, and
$\eta^{\mathrm{op}}_1
    =\iota(\res(\eta))
    =\iota(\delta_{1_G})
    =\delta_{1_G}$.
\end{proof}

If every $\chi_\rho$ has width at most $w$, then
\[
  \sum_{j=1}^b|\supp(H_{\rho j})|
    =\abs{\set*{(j,s):\lambda_{\rho,j,s}\ne 0}}
    \le w.
\]
Thus every row of $H$ contains at most $w$ nonzero group monomials.

\subsection{The Certificate Verifier}
\label{subsec:fibre-verifier}

Return to the hypotheses of \Cref{thm:fibre-certificate}, and construct
$H$ as in \Cref{lem:opposite-presentation}.
Given $p$ and $\mcW=\set{w_j}_{j=2}^b$, put $w_1:=p$.
The verifier chooses one of the following tests, each with probability
$1/3$:
\begin{enumerate}[label=\textnormal{(F\arabic*)},leftmargin=*]
  \item Choose $j\in[b]$ uniformly and run $T_D$ on $w_j$.
  \item Choose $\rho\in[h]$ and $i\in I$ independently and uniformly.
  Accept if
  \[
    \left(\sum_{j=1}^b H_{\rho j}\cdot w_j\right)(i)=0.
  \]
  \item Choose $j\in[b]$ and $g\in G$ independently and uniformly.
  Accept if $w_j(1,g)=0$.
\end{enumerate}

In \Cref{lem:opposite-presentation}, the entries of $H$ multiply elements
of $\bbK[G]$. In \textnormal{(F2)}, they act on the words $w_j$.
For $i=(k,g)$, the expression tested in \textnormal{(F2)} is
\[
  \begin{aligned}
  \left(\sum_{j=1}^b H_{\rho j}\cdot w_j\right)(k,g)
    &=\sum_{j=1}^b\sum_{s\in G}
      \lambda_{\rho,j,s}(s\cdot w_j)(k,g)
    =\sum_{j=1}^b\sum_{s\in G}
      \lambda_{\rho,j,s}w_j(k,s^{-1}g).
  \end{aligned}
\]
There are at most $w$ nonzero coefficients in this sum, so
\textnormal{(F2)} makes at most $w$ queries.
The other two tests make at most $q$ queries and one query, respectively.
All query locations are chosen before any answers are read.
The word $p$ is used as $w_1$, so the witness has length
$(b-1)|I|=(b-1)b|G|$.
For a fixed input and witness, write $\varepsilon_1,\varepsilon_2,
\varepsilon_3$ for the rejection probabilities of
\textnormal{(F1)}, \textnormal{(F2)}, and \textnormal{(F3)}, respectively.
The rejection probability of $V$ is
\[
  \varepsilon=\frac{\varepsilon_1+\varepsilon_2+\varepsilon_3}{3}.
\]

\subsection{Completeness and Soundness}
\label{subsec:fibre-certificate-proof}

\begin{proof}[Proof of~\Cref{thm:fibre-certificate}]
The preceding construction gives the query and witness-length bounds.

\emph{Completeness.}
Suppose $p\in D_0$, and put
$w_j:=\eta_j^{\mathrm{op}}\cdot p$ for $j \in [b]$.
By \Cref{lem:opposite-presentation},
$\eta_1^{\mathrm{op}}=1_{\bbK[G]}$ and $H\eta^{\mathrm{op}}=0$.
The first identity gives $w_1=p$.
Since $D$ is a left $\bbK[G]$-submodule, every $w_j$ belongs to $D$,
and \textnormal{(F1)} always accepts.
For every $\rho\in[h]$, associativity gives
\[
  \begin{aligned}
  \sum_{j=1}^b H_{\rho j}\cdot w_j
    &=\sum_{j=1}^b
      H_{\rho j}\cdot(\eta_j^{\mathrm{op}}\cdot p)
    =\left(\sum_{j=1}^b
      H_{\rho j}\eta_j^{\mathrm{op}}\right)\cdot p
    =(H\eta^{\mathrm{op}})_\rho\cdot p\\
    =0.
  \end{aligned}
\]
Hence \textnormal{(F2)} always accepts.
The $\bbK[G]$-linearity of restriction gives
\[
  \res(w_j)
    =\res(\eta_j^{\mathrm{op}}\cdot p)
    =\eta_j^{\mathrm{op}}\res(p)
    =0.
\]
Thus \textnormal{(F3)} always accepts as well.

\emph{Soundness.}
Now let $p\notin D_0$ and $\mcW$ be arbitrary.
If $\varepsilon \ge \gamma$, then we are done, so suppose the
rejection probability satisfies $\varepsilon<\gamma$.
For every $j\in[b]$, choose a nearest codeword $w'_j\in D$, and write
\[
  \delta_j:=\dist(w_j,w'_j)=\dist(w_j,D),
  \qquad
  S:=\sum_{j=1}^b\delta_j.
\]
Soundness of $T_D$ gives
\[
  \begin{aligned}
  \varepsilon_1
    &=\frac1b\sum_{j=1}^b
      \Pr[T_D^{w_j}\text{ rejects}]
    \ge\frac{\sigma}{b}
      \sum_{j=1}^b\dist(w_j,D)
    =\frac{\sigma}{b}S.
  \end{aligned}
\]
In particular,
\begin{equation}\label{eq:fibre-rounding-sum}
  S\le\frac{b\varepsilon_1}{\sigma}
    \le\frac{3b\varepsilon}{\sigma}.
\end{equation}

For $\rho\in[h]$, define
\[
  \widetilde r_\rho
    :=\sum_{j=1}^b H_{\rho j}\cdot w_j,
  \qquad
  r_\rho
    :=\sum_{j=1}^b H_{\rho j}\cdot w'_j.
\]
The word $r_\rho$ belongs to $D$, since every $w'_j\in D$ and $D$
is a left $\bbK[G]$-submodule.
By the definition of \textnormal{(F2)},
\[
  \varepsilon_2
    =\frac1h\sum_{\rho=1}^h\weight(\widetilde r_\rho),
\]
which implies $\weight(\wt{r}_\rho) \le h\eps_2$.
For a fixed $(j,s)$ with $\lambda_{\rho,j,s}\ne 0$, the map
$i\mapsto s^{-1}\cdot i$ is a permutation of $I$.
Consequently, for uniform $i\in I$,
$\Pr\bigl[w_j(s^{-1}\cdot i)\ne
    w'_j(s^{-1}\cdot i)\bigr]=\delta_j$.
If all queried entries agree, the two row expressions agree.
A union bound therefore gives
\[
  \begin{aligned}
  \dist(r_\rho,\widetilde r_\rho)
    &\le
      \sum_{\substack{j\in[b],\,s\in G \\
        \lambda_{\rho,j,s}\ne 0}}\delta_j
    =\sum_{j=1}^b|\supp(H_{\rho j})|\delta_j
    \le wS.
  \end{aligned}
\]
Using \eqref{eq:fibre-rounding-sum}, we obtain
\[
  \begin{aligned}
  \weight(r_\rho)
    \le\weight(\widetilde r_\rho)
      +\dist(r_\rho,\widetilde r_\rho)
    \le h\varepsilon_2+wS
    \le 3\varepsilon
      \left(h+\frac{bw}{\sigma}\right)
    <\Delta_D,
  \end{aligned}
\]
where the last inequality follows from
$\varepsilon<\gamma$ and \eqref{eq:fibre-certificate-gap}.
Since $r_\rho\in D$ and $\Delta(D)\ge\Delta_D$,
we have $r_\rho=0$ for all $\rho\in[h]$.
Now, define $z,\widetilde z\in\bbK[G]^b$ by
$z_j:=\res(w'_j)$ and $\widetilde z_j:=\res(w_j)$.
For every $\rho\in[h]$, restriction of the row equation gives
\[
  \begin{aligned}
  (Hz)_\rho
    &=\sum_{j=1}^b H_{\rho j}\res(w'_j)
    =\res\left(\sum_{j=1}^b H_{\rho j}\cdot w'_j\right)
    =\res(r_\rho)
    =0.
  \end{aligned}
\]
By \Cref{lem:opposite-presentation},
$z\in\ker H=C^{\mathrm{op}}$.
Here $z$ is a word on $[b]\times G$, with
$z(j,g)=w'_j(1,g)$.
The third test gives
\[
  \varepsilon_3
    =\frac1{b|G|}
      \sum_{j=1}^b
      \abs{\set*{g\in G:w_j(1,g)\ne 0}}
    =\weight(\widetilde z).
\]
Moreover,
\[
  \begin{aligned}
  \dist(z,\widetilde z)
    &=\frac1b\sum_{j=1}^b
      \dist(\res(w'_j),\res(w_j))
    \le\frac1b\sum_{j=1}^b b\,\dist(w'_j,w_j)
    =S.
  \end{aligned}
\]
Together with \eqref{eq:fibre-rounding-sum}, this yields
\[
  \begin{aligned}
  \weight(z)
    &\le\weight(\widetilde z)+\dist(z,\widetilde z)
    \le\varepsilon_3+S
    \le 3\varepsilon\left(1+\frac b\sigma\right)
    <\Delta_C.
  \end{aligned}
\]
The last inequality again follows from
\eqref{eq:fibre-certificate-gap}.
Since $z\in C^{\mathrm{op}}$ and
$\Delta(C^{\mathrm{op}})\ge\Delta_C$, we conclude that
$z=0$.
In particular, $w'_1\in D_0$.
If $p\in D$, we chose $w'_1=p$.
Thus $\varepsilon<\gamma$ implies $\res(p)=0$, proving the constant
rejection bound for $p\in D\setminus D_0$.
For arbitrary $p$, we have
\[
  \dist(p,D_0)
    \le\dist(p,w'_1)
    =\delta_1
    \le S
    \le\frac{3b\varepsilon}{\sigma}.
\]
Since $\Delta_C\le 1$, \eqref{eq:fibre-certificate-gap} gives
\[
  \gamma
    \le\frac{\sigma\Delta_C}{3(\sigma+b)}
    \le\frac{\sigma}{3b}.
\]
Therefore
\[
  \varepsilon
    \ge\frac{\sigma}{3b}\dist(p,D_0)
    \ge\gamma\,\dist(p,D_0)
\]
whenever $\varepsilon<\gamma$.
If $\varepsilon\ge\gamma$, the same conclusion follows from
$\dist(p,D_0)\le 1$.
\end{proof}

\subsection{Replacing the Input by a Codeword}
\label{subsec:fibre-stability}

Later we will apply the certificate after replacing $p$ by a nearby
codeword. The other witness words can be kept fixed.

\begin{lemma}\label[lemma]{lem:fibre-stability}
Let $V$ be the verifier above, let $p,p'\in\bbK^I$, and fix
$\mcW=\set{w_j}_{j=2}^b$.
Put $\delta:=\dist(p,p')$, and let $\varepsilon_a,\varepsilon'_a$
be the rejection probabilities of test \textnormal{(F$a$)} on
$(p,\mcW)$ and $(p',\mcW)$, respectively.
Then
$|\varepsilon'_2-\varepsilon_2|\le w\delta$ and 
$|\varepsilon'_3-\varepsilon_3|\le\delta$.
If $p'\in D$, then $\varepsilon'_1\le\varepsilon_1$, and hence
\begin{equation}\label{eq:fibre-input-replacement}
  \Pr[V^{p',\mcW}\text{ rejects}]
    \le
  \Pr[V^{p,\mcW}\text{ rejects}]
    +\frac{w+1}{3}\dist(p,p').
\end{equation}
In particular, if $p'\in D$ and $\res(p')\ne 0$, then
\[
  \Pr[V^{p,\mcW}\text{ rejects}]
    \ge\gamma-\frac{w+1}{3}\dist(p,p').
\]
\end{lemma}

\begin{proof}
Couple the two runs by using the same choices in each test.
In \textnormal{(F2)}, a queried entry of the first word has the form
$p(s^{-1}\cdot i)$,
where $i$ is uniform in $I$ and $s$ is fixed once the row and the
queried term are fixed.
Thus the probability that this entry changes is $\delta$.
At most $w$ queried entries belong to the first word, so
\[
  |\varepsilon'_2-\varepsilon_2|
    \le\Pr[\text{some queried entry changes}]
    \le w\delta.
\]

In \textnormal{(F3)}, an answer can change only when $j=1$.
The probability of this event and a disagreement on the data fibre is
\[
  \begin{aligned}
  \Pr[j=1\text{ and }p(1,g)\ne p'(1,g)]
    &=\frac1b\,\dist(\res(p),\res(p'))\\
    &\le\frac1b\,b\,\dist(p,p')\\
    &=\delta.
  \end{aligned}
\]
This gives $|\varepsilon'_3-\varepsilon_3|\le\delta$.
For \textnormal{(F1)}, the tests agree whenever $j\ne 1$.
If $p'\in D$, perfect completeness of $T_D$ gives
$\Pr[T_D^{p'}\text{ rejects}]=0$,
so replacing $p$ by $p'$ cannot increase the membership rejection
probability. Averaging the three bounds proves
\eqref{eq:fibre-input-replacement}.
If also $\res(p')\ne 0$, then \Cref{thm:fibre-certificate} gives
\[
  \gamma
    \le\Pr[V^{p',\mcW}\text{ rejects}]
    \le\Pr[V^{p,\mcW}\text{ rejects}]
      +\frac{w+1}{3}\dist(p,p'),
\]
which proves the last assertion.
\end{proof}
\section{Tensor Products and Certification of Algebraic Identities}
\label{sec:algebraic-identities}

We combine the certificate of \Cref{sec:fibre-certificates} with
local tests for tensor products to certify quadratic equations on
data fibres.

\begin{definition}
\label[definition]{def:multiplication-pair}
Let $\bbK$ be a finite field and let $G$ be a finite group.
Fix integers $b,w,h\ge 1$ and reals $\Delta,\sigma\in(0,1]$,
and put $I=[b]\times G$.
An \emph{equivariant multiplication pair} with parameters
$(b,\Delta,\sigma,w,h)$ consists of linear codes
$C_1,C_2\le\bbK^I$, a word $\eta\in C_1$, and lists
$\mcL_a=\set{\chi_{a,\rho}}_{\rho\in[h]}$ for $a\in\set{1,2}$,
such that:
\begin{itemize}
  \item[\textnormal{(E1)}] Both codes are $G$-invariant and satisfy
  $\Delta(C_a)\ge\Delta$.
  \item[\textnormal{(E2)}] $C_1*C_1\le C_2$.
  \item[\textnormal{(E3)}] $\res(\eta)=\delta_{1_G}$.
  \item[\textnormal{(E4)}] Every check in $\mcL_a$ has width at most
  $w$, and the translated tester associated to $\mcL_a$ is a
  $(w,\sigma)$-local test for $C_a$.
\end{itemize}
\end{definition}

Fix two such pairs, indexed by $v\in\set{\mathrm s,\mathrm t}$, over
the same field and with the same parameters. Write
$I_v=[b]\times G_v$ and $I_\otimes=I_{\mathrm s}\times I_{\mathrm t}$.
For $a\in\set{1,2}$, put
$C_a(I_{\mathrm s}):=C_{\mathrm s,a}$ and $C_a(I_\otimes):=C_{\mathrm s,a}\otimes C_{\mathrm t,a}$.
The product group $G_{\mathrm s}\times G_{\mathrm t}$ acts on
$I_\otimes$, whose data fibre is the product of the two data fibres.
Write $\res_E$ for the data restriction on
$E\in\set{I_{\mathrm s},I_\otimes}$, and put
$b_E=b$ or $b^2$, respectively.
Write $G_E=G_{\mathrm s}$ or $G_{\mathrm s}\times G_{\mathrm t}$
according to $E=I_{\mathrm s}$ or $I_\otimes$.
Put
\[
  u_{\mathrm t}:=\sum_{g\in G_{\mathrm t}}g\cdot\eta_{\mathrm t}
    \in C_{\mathrm t,1}.
\]
Then $\res_{\mathrm t}(u_{\mathrm t})=\Ind_{G_{\mathrm t}}$.

\begin{definition}
\label[definition]{def:quadratic-fibre-identities}
Let $\mathbf z=\set{z_r\in\bbK^{D_r}}_{r\in[N]}$, where
$D_r\in\set{I_{\mathrm s},I_\otimes}$.
An \emph{atom} on $E\in\set{I_{\mathrm s},I_\otimes}$ is one of
the following words:
\begin{itemize}
  \item[\textnormal{(A1)}] a fixed codeword in $C_1(E)$,
  \item[\textnormal{(A2)}] $g\cdot z_r$, where $D_r=E$ and $g\in G_E$,
  \item[\textnormal{(A3)}] $g\cdot(z_r\otimes u_{\mathrm t})$, where
  $D_r=I_{\mathrm s}$, $E=I_\otimes$, and
  $g\in G_{\mathrm s}\times G_{\mathrm t}$.
\end{itemize}
A \emph{quadratic expression} on $E$ is a sum
\[
  \Psi(\mathbf z)
    =\sum_{\ell=1}^{n'}c_\ell
      A_\ell(\mathbf z)\odot B_\ell(\mathbf z),
\]
where $c_\ell\in\bbK$ and $A_\ell,B_\ell$ are atoms on $E$.
A \emph{quadratic fibre identity} is the equation
$\res_E(\Psi(\mathbf z))=0$.
A system consists of the coordinate sets $D_r$ and a family
$\set{\res_{E_\alpha}(\Psi_\alpha(\mathbf z))=0}_{\alpha\in[J]}$.
A solution is a family with $z_r\in C_1(D_r)$ satisfying every
identity. All coefficients, group elements, and fixed codewords are
specified as part of the system.
\end{definition}

\begin{theorem}\label[theorem]{thm:quadratic-fibre-identities}
Let $N,J,n\ge 0$ be integers with $N+J\ge 1$.
Consider a system of quadratic fibre identities as in
\Cref{def:quadratic-fibre-identities}, with words
$\set{z_r\in\bbK^{D_r}}_{r\in[N]}$, equation domains
$\set{E_\alpha}_{\alpha\in[J]}$, and at most $n$ products in each
expression. Here $D_r,E_\alpha\in\set{I_{\mathrm s},I_\otimes}$,
and a solution requires $z_r\in C_1(D_r)$ for every $r$.
Put $M:=N+3J$ and 
\[
  \sigma_*:=\min\set*{\sigma,
    \frac{\Delta^2\sigma^2}{2w(2+bw)}}.
\]
There is a nonadaptive verifier using at most
$\max\set{w,2n+1}$ queries to a proof of
\[
  \sum_{r=1}^N|D_r|
    +\sum_{\alpha=1}^J b_{E_\alpha}|E_\alpha|
\]
field symbols, and there is $\gamma_0>0$ depending only on
$b,\Delta,\sigma,w,h,N,J,n$, with the following properties:
\begin{itemize}
  \item[\textnormal{(C)}] If the system has a solution, then there
  exists a proof which the verifier accepts with probability one.
  \item[\textnormal{(S)}] If the system has no solution, then every
  proof is rejected with probability at least $\gamma_0$.
\end{itemize}
Every solution can be extended to an accepting proof. Moreover, if a
proof is rejected with probability $\varepsilon<\gamma_0$, there are
nearest codewords $z'_r\in C_1(D_r)$ satisfying every identity, with
$\dist(z_r,z'_r)\le M\varepsilon/\sigma_*$ for every $r$.
In particular, $z'_r=z_r$ whenever $z_r\in C_1(D_r)$.
\end{theorem}

\subsection{Local Testability of Tensor Products}
\label{subsec:tensor-testing}

The tensor tester applies a local test to a uniform column or row.
We prove its soundness for linear testers on arbitrary coordinate
sets. The only additional hypothesis is a bound on how much one
factor's rejection probability changes when its input changes.

\begin{lemma}\label[lemma]{lem:tensor-local-testability}
Let $I_{\mathrm s},I_{\mathrm t}$ be finite nonempty sets and let
$A_v\le\bbK^{I_v}$ have relative distance at least
$\Delta_v\in(0,1]$, for $v\in\set{\mathrm s,\mathrm t}$.
Suppose $T_v$ is a $(q_v,\sigma_v)$-local test which samples a linear
parity check of width at most $q_v$. Write
$\pi_v(x):=\Pr[T_v^x\text{ rejects}]$.
Suppose $\kappa\ge 0$ satisfies
\begin{equation}\label{eq:tensor-tester-perturbation}
  |\pi_{\mathrm t}(x)-\pi_{\mathrm t}(y)|
    \le\kappa\dist(x,y)
\end{equation}
for all $x,y\in\bbK^{I_{\mathrm t}}$.
The test which chooses, with probability $1/2$ each, to run
$T_{\mathrm s}$ on a uniform column or $T_{\mathrm t}$ on a uniform
row is a $(\max\set{q_{\mathrm s},q_{\mathrm t}},\tau)$-local test
for $A_{\mathrm s}\otimes A_{\mathrm t}$, where
\begin{equation}\label{eq:tensor-tester-soundness}
  \tau:=
  \frac{\Delta_{\mathrm s}\Delta_{\mathrm t}
    \sigma_{\mathrm s}\sigma_{\mathrm t}}
    {2q_{\mathrm s}
      (\Delta_{\mathrm s}\sigma_{\mathrm t}
        +\sigma_{\mathrm s}+\kappa)}.
\end{equation}
\end{lemma}

\begin{proof}
A column is a word $Y(\cdot,j)$ on $I_{\mathrm s}$, and a row is a
word $Y(i,\cdot)$ on $I_{\mathrm t}$.
Perfect completeness follows from \Cref{lem:tensor-codes}, and the
query bound follows from those of the two testers.
Let $\varepsilon$ be the rejection probability on $Y$.
If $\varepsilon\ge\tau$, soundness follows from
$\dist(Y,A_{\mathrm s}\otimes A_{\mathrm t})\le 1$, so suppose
$\varepsilon<\tau$.
Replace each column by a nearest codeword of $A_{\mathrm s}$,
obtaining $Y^{(1)}$.
The column branch rejects with probability at most $2\varepsilon$,
so local testability gives
\[
  \dist(Y,Y^{(1)})
    =\frac1{|I_{\mathrm t}|}\sum_{j\in I_{\mathrm t}}
      \dist(Y(\cdot,j),A_{\mathrm s})
    \le\frac{2\varepsilon}{\sigma_{\mathrm s}}.
\]
Apply \eqref{eq:tensor-tester-perturbation} to each row and average.
The row branch on $Y^{(1)}$ rejects with probability at most
\[
  a:=2\varepsilon+\kappa\dist(Y,Y^{(1)})
    \le\frac{2\varepsilon(\sigma_{\mathrm s}+\kappa)}
      {\sigma_{\mathrm s}}.
\]
For a check $\chi(x)=\sum_{j\in I_{\mathrm t}}\lambda_jx(j)$
sampled by $T_{\mathrm t}$, put
$r_\chi:=\sum_j\lambda_jY^{(1)}(\cdot,j)$.
This word belongs to $A_{\mathrm s}$, and its value at $i$ is the
value of $\chi$ on row $i$. Consequently,
\[
  \Delta_{\mathrm s}\Pr_\chi[r_\chi\ne 0]
    \le\mathbb E_\chi\weight(r_\chi)
    =\frac1{|I_{\mathrm s}|}\sum_{i\in I_{\mathrm s}}
      \pi_{\mathrm t}(Y^{(1)}(i,\cdot))
    \le a.
\]
For each $i$, a check that rejects row $i$ has $r_\chi\ne 0$.
Thus every row has rejection probability at most
$a/\Delta_{\mathrm s}$, and local testability implies
\[
  \dist(Y^{(1)}(i,\cdot),A_{\mathrm t})
    \le\beta:=\frac{a}{\Delta_{\mathrm s}\sigma_{\mathrm t}}
    \le\frac{2\varepsilon(\sigma_{\mathrm s}+\kappa)}
      {\Delta_{\mathrm s}\sigma_{\mathrm s}\sigma_{\mathrm t}}.
\]
Replace each row by a nearest codeword of $A_{\mathrm t}$,
obtaining $Y^{(2)}$. Every row changes in at most a $\beta$ fraction
of its coordinates.
We show that the columns of $Y^{(2)}$ still belong to
$A_{\mathrm s}$.
Fix a check $\psi(x)=\sum_{i\in I_{\mathrm s}}\mu_ix(i)$
sampled with positive probability by $T_{\mathrm s}$, and put
$v_\psi:=\sum_i\mu_iY^{(2)}(i,\cdot)\in A_{\mathrm t}$.
Every column of $Y^{(1)}$ belongs to $A_{\mathrm s}$, so
$\sum_i\mu_iY^{(1)}(i,\cdot)=0$. Hence
\[
  \weight(v_\psi)
    \le\sum_{i:\mu_i\ne 0}
      \dist(Y^{(1)}(i,\cdot),Y^{(2)}(i,\cdot))
    \le q_{\mathrm s}\beta
    <\Delta_{\mathrm t}.
\]
The last inequality follows from $\varepsilon<\tau$ and
\eqref{eq:tensor-tester-soundness}, since
\[
  q_{\mathrm s}\beta
    <\Delta_{\mathrm t}
      \frac{\sigma_{\mathrm s}+\kappa}
      {\Delta_{\mathrm s}\sigma_{\mathrm t}
        +\sigma_{\mathrm s}+\kappa}
    <\Delta_{\mathrm t}.
\]
Distance forces $v_\psi=0$. Thus every column of $Y^{(2)}$ is
accepted with probability one by $T_{\mathrm s}$, and soundness
places it in $A_{\mathrm s}$.
The rows already belong to $A_{\mathrm t}$, so
\Cref{lem:tensor-codes} gives
$Y^{(2)}\in A_{\mathrm s}\otimes A_{\mathrm t}$.
Finally,
\[
  \dist(Y,A_{\mathrm s}\otimes A_{\mathrm t})
    \le\frac{2\varepsilon}{\sigma_{\mathrm s}}+\beta
    \le\frac{2(\Delta_{\mathrm s}\sigma_{\mathrm t}
        +\sigma_{\mathrm s}+\kappa)}
      {\Delta_{\mathrm s}\sigma_{\mathrm s}\sigma_{\mathrm t}}
      \varepsilon
      =\frac{\Delta_{\mathrm t}}{q_{\mathrm s}}
      \frac{\varepsilon}{\tau}
    \le\frac{\varepsilon}{\tau}. \qedhere
\]
\end{proof}

\subsection{Tensor Products of Multiplication Pairs}
\label{subsec:tensor-multiplication-pairs}

We first record the relation between tensor products and Schur
products. This identity holds over any field.

\begin{lemma}\label[lemma]{lem:tensor-schur-products}
For linear codes $A_v,B_v\le\bbK^{I_v}$, with
$v\in\set{\mathrm s,\mathrm t}$,
\[
  (A_{\mathrm s}\otimes A_{\mathrm t})*
    (B_{\mathrm s}\otimes B_{\mathrm t})
    =(A_{\mathrm s}*B_{\mathrm s})\otimes
      (A_{\mathrm t}*B_{\mathrm t}).
\]
\end{lemma}

\begin{proof}
For $a_v\in A_v$ and $b_v\in B_v$, pointwise multiplication gives
\[
  (a_{\mathrm s}\otimes a_{\mathrm t})\odot
    (b_{\mathrm s}\otimes b_{\mathrm t})
    =(a_{\mathrm s}\odot b_{\mathrm s})\otimes
      (a_{\mathrm t}\odot b_{\mathrm t}).
\]
The left sides span the left-hand code by bilinearity, and the right
sides span the right-hand code by the definition of Schur product.
\end{proof}

\begin{lemma}\label[lemma]{lem:tensor-multiplication-pairs}
For $v\in\set{\mathrm s,\mathrm t}$, let
$(C_{v,1},C_{v,2})$ be an equivariant multiplication pair with
parameters $(b_v,\Delta_v,\sigma_v,w_v,h_v)$ over the same field.
Their tensor products form an equivariant multiplication pair for
$G_{\mathrm s}\times G_{\mathrm t}$, with parameters
\[
  \begin{gathered}
  b_\otimes=b_{\mathrm s}b_{\mathrm t},\qquad
  \Delta_\otimes=\Delta_{\mathrm s}\Delta_{\mathrm t},\qquad
  w_\otimes=\max\set{w_{\mathrm s},w_{\mathrm t}},\\
  h_\otimes=2\operatorname{lcm}(b_{\mathrm t}h_{\mathrm s},
    b_{\mathrm s}h_{\mathrm t}),\qquad
  \sigma_\otimes=
    \frac{\Delta_{\mathrm s}\Delta_{\mathrm t}
      \sigma_{\mathrm s}\sigma_{\mathrm t}}
      {2w_{\mathrm s}(\Delta_{\mathrm s}\sigma_{\mathrm t}
        +\sigma_{\mathrm s}+b_{\mathrm t}w_{\mathrm t})}.
  \end{gathered}
\]
One may take $\eta_\otimes=\eta_{\mathrm s}\otimes\eta_{\mathrm t}$.
The translated tester chooses a uniform column or row with
probability $1/2$ each and applies the corresponding factor tester.
\end{lemma}

\begin{proof}
The product action has $b_{\mathrm s}b_{\mathrm t}$ fibres, indexed
by $[b_{\mathrm s}]\times[b_{\mathrm t}]$, with $(1,1)$ indexing
the data fibre. Tensor products of invariant codes are invariant.
Their distance bound follows from \Cref{lem:tensor-codes}, and
\Cref{lem:tensor-schur-products} gives the multiplication inclusion.
Moreover,
\[
  \res_\otimes(\eta_{\mathrm s}\otimes\eta_{\mathrm t})
    =\res_{\mathrm s}(\eta_{\mathrm s})\otimes
      \res_{\mathrm t}(\eta_{\mathrm t})
    =\delta_{(1_{G_{\mathrm s}},1_{G_{\mathrm t}})}.
\]
By \Cref{lem:translated-checks}, the testers for
$C_{\mathrm t,1}$ and $C_{\mathrm t,2}$ satisfy
\eqref{eq:tensor-tester-perturbation} with
$\kappa=b_{\mathrm t}w_{\mathrm t}$.
Applying \Cref{lem:tensor-local-testability} for each
$a\in\set{1,2}$ gives the stated soundness and query bounds.

It remains to represent this test by a list of translated checks.
Fix $a\in\set{1,2}$. The base column and row checks are
\[
  \begin{aligned}
  \chi^{\mathrm{col}}_{\rho,j_{\mathrm t}}(Y)
    &:=\chi_{\mathrm s,a,\rho}
      (Y(\cdot,(j_{\mathrm t},1_{G_{\mathrm t}}))), \qquad
  \chi^{\mathrm{row}}_{\rho,j_{\mathrm s}}(Y)
    :=\chi_{\mathrm t,a,\rho}
      (Y((j_{\mathrm s},1_{G_{\mathrm s}}),\cdot)).
  \end{aligned}
\]
There are $a_{\mathrm s}:=b_{\mathrm t}h_{\mathrm s}$ column
checks and $a_{\mathrm t}:=b_{\mathrm s}h_{\mathrm t}$ row checks.
If $\ell=\operatorname{lcm}(a_{\mathrm s},a_{\mathrm t})$, repeat
each column check $\ell/a_{\mathrm s}$ times and each row check
$\ell/a_{\mathrm t}$ times. The resulting list has $2\ell$ checks
and chooses each orientation with probability $1/2$.
For a column check, translation by $(s,t)$ gives
\[
  \chi^{\mathrm{col}}_{\rho,j_{\mathrm t}}((s,t)\cdot Y)
    =\chi_{\mathrm s,a,\rho}
      \bigl(s\cdot Y(\cdot,(j_{\mathrm t},t^{-1}))\bigr).
\]
Uniform $t\in G_{\mathrm t}$ chooses a uniform column within its
fibre. The uniform list index chooses $j_{\mathrm t}$ and the
check $\chi_{\mathrm s,a,\rho}$, while $s\in G_{\mathrm s}$
independently chooses the translate.
Thus this orientation is exactly the column test. The row case is
the same calculation with the factors interchanged.
Every check has width at most $w_\otimes$.
\end{proof}

When the factor parameters are both $(b,\Delta,\sigma,w,h)$,
we may therefore take
\[
  (b_\otimes,\Delta_\otimes,\sigma_\otimes,w_\otimes,h_\otimes)
    =\left(b^2,\Delta^2,
      \frac{\Delta^2\sigma^2}{2w(2+bw)},w,2bh\right),
\]
since $\Delta\sigma+\sigma\le 2$.
These bounds impose no relation between $|G_{\mathrm s}|$ and
$|G_{\mathrm t}|$.

By \Cref{lem:equivariant-encoding}, write
$\Enc_v(f)=f\cdot\eta_v$ and
$\Enc_\otimes(f)=f\cdot(\eta_{\mathrm s}\otimes\eta_{\mathrm t})$.
Set $u_{\mathrm s}:=\Enc_{\mathrm s}(\Ind_{G_{\mathrm s}})$.
For $v\in\set{\mathrm s,\mathrm t}$, we then have
$u_v=\Enc_v(\Ind_{G_v})$ and $\res_v(u_v)=\Ind_{G_v}$.
The map $B:\bbK^{I_{\mathrm s}}\to\bbK^{I_\otimes}$ given by
$Bx=x\otimes u_{\mathrm t}$ sends $C_{\mathrm s,1}$ into
$C_1(I_\otimes)$ and satisfies
$\res_\otimes(Bx)(g,t)=\res_{\mathrm s}(x)(g)$ and $\dist(Bx,By)=\dist(x,y)\weight(u_{\mathrm t})
    \le\dist(x,y)$.
The distance identity follows because a coordinate of
$(x-y)\otimes u_{\mathrm t}$ is nonzero exactly when both factors
are nonzero. Evaluating $Bx$ at $(i,j)$ reads $x(i)$ and multiplies
it by the fixed value $u_{\mathrm t}(j)$.

\subsection{Algebraic Expressions}
\label{subsec:algebraic-expressions}

The comparison of an expression before and after rounding depends
only on the distribution of its queried coordinates. We state this
fact without assumptions on codes or on the algebraic form of the
expression.

\begin{lemma}\label[lemma]{lem:coordinate-expression-stability}
Let $E,D_1,\ldots,D_N$ be finite nonempty sets, and let
$\Psi:\prod_{r=1}^N\bbK^{D_r}\to\bbK^E$.
Suppose there are indices $r_1,\ldots,r_L$ and maps
$\phi_\ell:E\to D_{r_\ell}$ such that $\Psi(\mathbf z)(i)$ is
determined by $i$ and
$\set{z_{r_\ell}(\phi_\ell(i))}_{\ell\in[L]}$.
Assume each $\phi_\ell$ sends the uniform distribution on $E$ to
the uniform distribution on $D_{r_\ell}$.
Then, for all $\mathbf z,\mathbf z'$,
\[
  \dist(\Psi(\mathbf z),\Psi(\mathbf z'))
    \le\sum_{\ell=1}^L
      \dist(z_{r_\ell},z'_{r_\ell}).
\]
\end{lemma}

\begin{proof}
For uniform $i\in E$, the hypothesis on $\phi_\ell$ gives
$\Pr[z_{r_\ell}(\phi_\ell(i))
    \ne z'_{r_\ell}(\phi_\ell(i))]
    =\dist(z_{r_\ell},z'_{r_\ell})$.
If the two outputs differ at $i$, at least one of these events
occurs. The union bound proves the claim.
\end{proof}

Return to the two multiplication pairs fixed above.

\begin{lemma}\label[lemma]{lem:quadratic-expression-properties}
If every $z_r\in C_1(D_r)$, then a quadratic expression $\Psi$ on
$E$ belongs to $C_2(E)$.
If $\Psi$ has at most $n$ products, each coordinate can be evaluated
using at most $2n$ nonadaptive queries to words $z_r$. Moreover, if
$\dist(z_r,z'_r)\le\delta$ for every $r$, then
$\dist(\Psi(\mathbf z),\Psi(\mathbf z'))\le 2n\delta$.
\end{lemma}

\begin{proof}
If $z_r\in C_1(D_r)$ for every $r$, then every atom belongs
to $C_1(E)$, by invariance and the tensor-product construction.
Each product therefore belongs to $C_2(E)$, and linearity gives
the first assertion.

A fixed atom requires no query to the words $z_r$. A translated atom evaluated
at $i$ reads $z_r(g^{-1}\cdot i)$.
For $g=(s,t)$, a translated tensor extension evaluated at $(i,j)$
has value
$z_r(s^{-1}\cdot i)u_{\mathrm t}(t^{-1}\cdot j)$, and also requires
one query. Thus each product uses at most two queries.
All their addresses are determined by the output coordinate and the
specified group elements.
For a uniform output coordinate, a translated coordinate is uniform,
and the projection of a uniform coordinate of $I_\otimes$ onto
$I_{\mathrm s}$ is uniform. \Cref{lem:coordinate-expression-stability} therefore
bounds the change by a sum of at most $2n$ terms of the form
$\dist(z_r,z'_r)$, and hence by $2n\delta$.
\end{proof}

These identities include the Boolean constraints needed later.
On either domain $E$, take
$u_E=\Enc_E(\Ind_{G_E})$, where $\Enc_E=\Enc_{\mathrm s}$ for $E=I_{\mathrm s}$ and
$\Enc_E=\Enc_\otimes$ for $E=I_\otimes$. Then $\res_E(u_E)=\Ind_{G_E}$, and the equation
$\res_E(x\odot x-u_E\odot x)=0$
forces every value $a$ of $\res_E(x)$ to satisfy $a(a-1)=0$, hence
to lie in $\set{0,1}$. When the input restrictions are Boolean,
the NAND equation is
$\res_E(a\odot b+u_E\odot z-u_E\odot u_E)=0$.
For a subset $S\subseteq G_E$, let $m_S:=\Enc_E(\Ind_S)$.
Equality of $\res_E(x)$ and $\res_E(y)$ on $S$ is equivalent to
$\res_E(m_S\odot x-m_S\odot y)
    =\Ind_S\odot(\res_E(x)-\res_E(y))=0$.
Taking $y=u_E$ and $S=\set{g_0}$ requires
$\res_E(x)(g_0)=1$. These expressions use only the prescribed
restriction of $m_S$ on the data fibre.

\subsection{The Verifier and Its Analysis}
\label{subsec:identity-verifier}

We prove the composition step for arbitrary expression maps with
the two properties established above: they take codeword inputs to
codewords, and changing the inputs changes their values by a
controlled amount.

\begin{proposition}\label[proposition]{prop:identity-certification}
Let $N,J\ge 0$ with $N+J\ge 1$.
For each $r\in[N]$, let $A_r\le\bbK^{I_r}$ have a
$(q,\sigma)$-local tester.
For each $\alpha\in[J]$, let
$E_\alpha=[b_\alpha]\times G_\alpha$ and suppose
$C_\alpha,D_\alpha\le\bbK^{E_\alpha}$ satisfy the hypotheses of
\Cref{thm:fibre-certificate}, with check width at most $w$,
$\Delta(D_\alpha)\ge\Delta>0$, a $(q,\sigma)$-local tester for
$D_\alpha$, and a certificate rejection bound at least $\gamma>0$.
Let
$\Psi_\alpha:\prod_{r=1}^N\bbK^{I_r}\to\bbK^{E_\alpha}$
satisfy the following conditions:
\begin{itemize}
  \item[\textnormal{(1)}] If $z_r\in A_r$ for every $r$, then
  $\Psi_\alpha(\mathbf z)\in D_\alpha$.
  \item[\textnormal{(2)}] Each coordinate of $\Psi_\alpha$ can be evaluated using at
  most $\ell$ nonadaptive queries to the words $z_r$.
  \item[\textnormal{(3)}] For every $\delta\ge 0$, if $\dist(z_r,z'_r)\le\delta$
  for every $r$, then
  $\dist(\Psi_\alpha(\mathbf z),\Psi_\alpha(\mathbf z'))\le L\delta$.
\end{itemize}
Here $L\ge 0$ and $\ell\ge 0$ is an integer.
Put $M=N+3J$ and
\begin{equation}\label{eq:identity-certifier-gap}
  \beta:=\frac1M\min\set*{
    \frac{\Delta}{1+(L+1)/\sigma},
    \frac{\gamma}{1+(w+1)/(3\sigma)}}.
\end{equation}
There is a verifier for the equations
$\res_{E_\alpha}(\Psi_\alpha(\mathbf z))=0$ which uses at most
$\max\set{q,w,\ell+1}$ nonadaptive queries and proof length
\[
  \sum_{r=1}^N|I_r|+\sum_{\alpha=1}^J b_\alpha|E_\alpha|.
\]
Every solution in $\prod_r A_r$ extends to an accepting proof.
If a proof is rejected with probability $\varepsilon<\beta$, there
are nearest codewords $z'_r\in A_r$ satisfying all the equations,
with $\dist(z_r,z'_r)\le M\varepsilon/\sigma$ for all $r$.
\end{proposition}

\begin{proof}
For each equation, the proof contains
$p_\alpha,w_{\alpha,2},\ldots,w_{\alpha,b_\alpha}
\in\bbK^{E_\alpha}$ in addition to the input words.
Let $V_\alpha$ be the verifier of \Cref{thm:fibre-certificate} for
$(C_\alpha,D_\alpha)$, and write
$\mcW_\alpha=\set{w_{\alpha,j}}_{j=2}^{b_\alpha}$.
Choose uniformly among the following $M=N+3J$ tests:
\begin{enumerate}[label=\textnormal{(T\arabic*)},leftmargin=*]
  \item For each $r\in[N]$, test membership of $z_r$ in $A_r$.
  \item For each $\alpha\in[J]$, test membership of $p_\alpha$
  in $D_\alpha$.
  \item For each $\alpha\in[J]$, choose $i\in E_\alpha$
  uniformly and accept if $p_\alpha(i)=\Psi_\alpha(\mathbf z)(i)$.
  \item For each $\alpha\in[J]$, run
  $V_\alpha^{p_\alpha,\mcW_\alpha}$.
\end{enumerate}
Thus each input word contributes one choice and each equation contributes
three choices. The last choice includes the three branches of
$V_\alpha$ with their probabilities from Section~\ref{sec:fibre-certificates}.
The proof-length formula counts the $b_\alpha$ words for each
equation, including $p_\alpha$.
The membership tests use at most $q$ queries, the comparison uses
at most $\ell+1$, and $V_\alpha$ uses at most
$\max\set{q,w}$. Their nonadaptivity gives that of the verifier.

\emph{Completeness.}
Suppose $\mathbf z$ is a solution, and set
$p_\alpha=\Psi_\alpha(\mathbf z)$.
Then $p_\alpha\in D_\alpha$ and
$\res_{E_\alpha}(p_\alpha)=0$.
Choose the witnesses supplied by \Cref{thm:fibre-certificate}.
Every membership test, comparison, and fibre certificate accepts.

\emph{Soundness.}
Suppose a proof is rejected with probability
$\varepsilon<\beta$, and put
$\widehat\varepsilon=M\varepsilon$ and
$\delta=\widehat\varepsilon/\sigma$.
Each of the $M$ tests rejects with probability at most
$\widehat\varepsilon$.
Choose nearest codewords $z'_r\in A_r$ and
$p'_\alpha\in D_\alpha$. The membership tests give
$\dist(z_r,z'_r)\le\delta$ and
$\dist(p_\alpha,p'_\alpha)\le\delta$.

Fix $\alpha$ and put $\Psi'_\alpha:=\Psi_\alpha(\mathbf z')$.
The comparison test gives
$\dist(p_\alpha,\Psi_\alpha(\mathbf z))\le\widehat\varepsilon$.
The hypothesis on $\Psi_\alpha$ then yields
\[
  \begin{aligned}
  \dist(p'_\alpha,\Psi'_\alpha)
    &\le\dist(p'_\alpha,p_\alpha)
      +\dist(p_\alpha,\Psi_\alpha(\mathbf z))
      +\dist(\Psi_\alpha(\mathbf z),\Psi_\alpha(\mathbf z'))\\
    &\le\widehat\varepsilon+(L+1)\delta
    =M\varepsilon\left(1+\frac{L+1}{\sigma}\right)
    <\Delta.
  \end{aligned}
\]
The final inequality follows from the first term in
\eqref{eq:identity-certifier-gap}.
Both $p'_\alpha$ and $\Psi'_\alpha$ belong to $D_\alpha$, so its
distance implies $p'_\alpha=\Psi'_\alpha$.

Now keep $\mcW_\alpha$ fixed and replace $p_\alpha$ by
$p'_\alpha$ in the fibre certificate.
Since $p'_\alpha\in D_\alpha$, \Cref{lem:fibre-stability} gives
\[
  \begin{aligned}
  \Pr[V_\alpha^{p'_\alpha,\mcW_\alpha}\text{ rejects}]
    &\le\Pr[V_\alpha^{p_\alpha,\mcW_\alpha}\text{ rejects}]
      +\frac{w+1}{3}\dist(p_\alpha,p'_\alpha)\\
    &\le\widehat\varepsilon+\frac{w+1}{3}\delta
    =M\varepsilon\left(1+\frac{w+1}{3\sigma}\right)
    <\gamma.
  \end{aligned}
\]
The second term in \eqref{eq:identity-certifier-gap} gives the
strict inequality.
Condition \textnormal{(S1)} of \Cref{thm:fibre-certificate} now
forces $\res_{E_\alpha}(p'_\alpha)=0$.
Since $p'_\alpha=\Psi_\alpha(\mathbf z')$, the words $z'_r$
satisfy this identity. The same argument applies to every $\alpha$.
\end{proof}

\begin{proof}[Proof of~\Cref{thm:quadratic-fibre-identities}]
By \Cref{lem:tensor-multiplication-pairs}, the tensor pair has
parameters $(b^2,\Delta^2,\sigma_\otimes,w,2bh)$, where
$\sigma_\otimes=\Delta^2\sigma^2/(2w(2+bw))$.
Thus, on either domain $E$, both $C_1(E)$ and $C_2(E)$ have
distance at least $\Delta^2$ and local tests with soundness
at least $\sigma_*$.
The translated checks present each first code as their common
kernel, since acceptance with probability one implies membership.
Consequently \Cref{thm:fibre-certificate} applies on either domain
with $C=C_1(E)$ and $D=C_2(E)$.
Let $\gamma_{\mathrm s}$ and $\gamma_\otimes$ be its bounds:
\[
  \begin{aligned}
  \gamma_{\mathrm s}
    &:=\frac{\sigma}{3}\min\set*{
      \frac{\Delta}{h\sigma+bw},\frac{\Delta}{\sigma+b}},\\
  \gamma_\otimes
    &:=\frac{\sigma_\otimes}{3}\min\set*{
      \frac{\Delta^2}{2bh\sigma_\otimes+b^2w},
      \frac{\Delta^2}{\sigma_\otimes+b^2}}.
  \end{aligned}
\]
Put $\gamma_*:=\min\set{\gamma_{\mathrm s},\gamma_\otimes}$.
By \Cref{lem:quadratic-expression-properties}, the expression maps
satisfy \Cref{prop:identity-certification} with query bound $w$,
soundness $\sigma_*$, output distance $\Delta^2$, certificate bound
$\gamma_*$, and $L=\ell=2n$.
Apply that proposition, taking
\[
  \gamma_0:=\frac1M\min\set*{
    \frac{\Delta^2}{1+(2n+1)/\sigma_*},
    \frac{\gamma_*}{1+(w+1)/(3\sigma_*)}}.
\]
It gives the query and proof-length bounds, completeness for every
solution, and the asserted rounding conclusion.
If the system has no solution, this conclusion rules out rejection
probability below $\gamma_0$, proving \textnormal{(S)}.
Finally, a nearest codeword to a word already in its code is that
word itself, so $z'_r=z_r$ whenever $z_r\in C_1(D_r)$.
\end{proof}
\section{Certifying Permutation Constraints}
\label{sec:permutation-certificates}

We encode the history of a routing schedule in the tensor code.
Grouping the transition equations by their translations and word
indices gives a number of quadratic fibre identities independent
of the length of the schedule.

\begin{definition}\label[definition]{def:routing-schedule}
Let $G$ be a finite group and let $T\subseteq G\setminus\set{1_G}$
be symmetric. A \emph{$T$-schedule} of length $L\ge 0$ is a family
$\set{\mathcal M_i}_{i\in[L]}$, where each $\mathcal M_i$ is a
matching in the graph with edges $\set{g,tg}$ for $g\in G$ and
$t\in T$.
Let $\omega_i$ exchange the endpoints of every edge in
$\mathcal M_i$ and fix all unmatched vertices.
The schedule \emph{realizes} the permutation
$\pi=\omega_L\circ\cdots\circ\omega_1$.
The empty schedule realizes the identity.
\end{definition}

Retain the two multiplication pairs from
\Cref{sec:algebraic-identities}, with parameters
$(b,\Delta,\sigma,w,h)$.
Let $T_{\mathrm s}\subseteq G_{\mathrm s}\setminus\set{1}$ be
symmetric and let
$T_{\mathrm t}\subseteq G_{\mathrm t}\setminus\set{1}$ be a
symmetric generating set. Write
$d_{\mathrm s}=|T_{\mathrm s}|$, $d_{\mathrm t}=|T_{\mathrm t}|$,
and $k=d_{\mathrm t}+1$.

\begin{theorem}\label[theorem]{thm:permutation-certificate}
Let $\pi$ be a permutation of $G_{\mathrm s}$, let
$J\subseteq G_{\mathrm s}$, and suppose a $T_{\mathrm s}$-schedule
of length $L<|G_{\mathrm t}|$ realizing $\pi$ is given.
There is a system of $k+2$ quadratic fibre identities on
$I_\otimes$, involving words $x,y\in\bbK^{I_{\mathrm s}}$ and
$k$ additional words $\set{z_\nu\in\bbK^{I_\otimes}}_{\nu\in[k]}$,
each expression having at most
$n_{\mathrm{perm}}:=
    \max\set{2,2(d_{\mathrm s}+1)d_{\mathrm t}k}$
products, with the following properties:
\begin{itemize}
  \item[\textnormal{(C)}] If for all $g \in J$,
  \begin{equation}\label{eq:permutation-constraint}
    \res_{\mathrm s}(y)(g)
      =\res_{\mathrm s}(x)(\pi^{-1}(g)),
  \end{equation}
  then there are words $z_\nu\in C_1(I_\otimes)$ satisfying every
  identity.
  \item[\textnormal{(S)}] If \eqref{eq:permutation-constraint}
  fails, then every choice of the words $z_\nu\in\bbK^{I_\otimes}$
  violates at least one identity.
\end{itemize}
The fixed codewords depend only on the schedule, $J$, the sets
$T_{\mathrm s},T_{\mathrm t}$, and the data of the two multiplication pairs.
From the group multiplication tables, the words
$\eta_{\mathrm s},\eta_{\mathrm t}$, the sets
$T_{\mathrm s},T_{\mathrm t}$, the schedule, and $J$, the system
can be constructed in time polynomial in $|I_{\mathrm s}||I_{\mathrm t}|$.
Given $x,y$ satisfying \eqref{eq:permutation-constraint}, the words
in \textnormal{(C)} can be constructed in the same time bound.
\end{theorem}

\subsection{Routing Histories}
\label{subsec:routing-histories}

Fix the schedule in \Cref{thm:permutation-certificate}.
For $f_0:G_{\mathrm s}\to\bbK$, let $f_i$ record the values after
round $i$. The value arriving at $g$ was at $\omega_i(g)$ before
the round, so for all $i \in [L]$,
\begin{equation}\label{eq:routing-history}
  f_i=f_{i-1}\circ\omega_i.
\end{equation}
Every $\omega_i$ is an involution. Induction therefore gives
$f_i=f_0\circ\omega_1\circ\cdots\circ\omega_i$ and 
$f_L=f_0\circ\pi^{-1}$.
For every $i$ and $g$, there is a unique
$t\in T_{\mathrm s}\cup\set{1}$ with $\omega_i(g)=tg$.
For an unmatched vertex, take $t=1$. For a matched vertex, this
follows from symmetry of $T_{\mathrm s}$, and uniqueness follows
from $t=\omega_i(g)g^{-1}$.

The proof will encode the functions $\set{f_i}_{i=0}^L$ on the
data fibre $G_{\mathrm s}\times G_{\mathrm t}$, using several
words to distinguish repeated visits to the same element of $G_{\mathrm t}$.

\subsection{Encoding the Time Coordinate}
\label{subsec:time-coordinate}

\begin{lemma}\label[lemma]{lem:time-indexing}
Let $\Gamma$ be a finite connected graph on $N\ge 1$ vertices,
with maximum degree at most $d$.
There is a walk $\set{\theta_i}_{i=0}^{2N-2}$ visiting every vertex and indices
$\set{\nu_i\in[d+1]}_{i=0}^{2N-2}$ such that all pairs
$(\theta_i,\nu_i)$ are distinct.
The walk and indices can be computed in polynomial time from
$\Gamma$, starting at any prescribed vertex.
\end{lemma}

\begin{proof}
Choose a spanning tree rooted at the prescribed vertex and traverse
it depth first, traversing each edge once in each direction.
Record the initial vertex and every arrival along an edge.
There are $2(N-1)$ moves and $2N-1$ recorded vertices.
The root appears initially and after the return from each of its
children, so it appears at most $d+1$ times.
A nonroot vertex appears upon arrival from its parent and upon
return from each child, so it appears at most $d$ times.
For each $i$, put
$\nu_i:=|\set{j:0\le j\le i,\ \theta_j=\theta_i}|$.
Thus $\nu_i\in[d+1]$, and two visits to the same vertex have
different indices. A spanning tree and its traversal can both be
computed in polynomial time.
\end{proof}

Apply \Cref{lem:time-indexing} to the Cayley graph of
$G_{\mathrm t}$ generated by $T_{\mathrm t}$, and retain the first
$L+1$ visits.
This is possible because $L<|G_{\mathrm t}|$.
We obtain $\theta_0,\ldots,\theta_L\in G_{\mathrm t}$ and
$\nu_0,\ldots,\nu_L\in[k]$ with distinct pairs
$(\theta_i,\nu_i)$ and
$\theta_{i-1}=t'_i\theta_i$ and $t'_i:=\theta_{i-1}\theta_i^{-1}\in T_{\mathrm t}$ for $i \in [L]$.
For arbitrary words $\set{z_\nu}_{\nu\in[k]}$ on $I_\otimes$,
write $F_\nu:=\res_\otimes(z_\nu)$ and define, for $0\le i\le L$ and $g\in G_{\mathrm s}$,
\begin{equation}\label{eq:history-restrictions}
  f_i(g):=F_{\nu_i}(g,\theta_i).
\end{equation}
Conversely, any family $\set{f_i}_{i=0}^L$ can be stored this way.
Indeed, prescribe $F_{\nu_i}(g,\theta_i)=f_i(g)$ and set all other
values to zero. Distinctness of the pairs $(\theta_i,\nu_i)$ makes
these prescriptions consistent, and the words
$z_\nu:=\Enc_\otimes(F_\nu)$ have the required restrictions.

\subsection{Transition Identities}
\label{subsec:routing-transitions}

Let
$\mathcal E:=(T_{\mathrm s}\cup\set{1})\times
T_{\mathrm t}\times[k]$.
For $\nu\in[k]$ and $e=(t,t',\nu')\in\mathcal E$, define
\[
  A_{\nu,e}:=\set*{(g,\theta_i):
    \begin{array}{l}
    g\in G_{\mathrm s},\ i\in[L],\ \nu_i=\nu,\\
    \omega_i(g)=tg,\ \theta_{i-1}=t'\theta_i,
      \ \nu_{i-1}=\nu'
    \end{array}}.
\]
For each fixed $\nu$, these sets partition the positions
$(g,\theta_i)$ with $i\in[L]$ and $\nu_i=\nu$.
The pair $(\theta_i,\nu)$ determines $i$, after which
$t=\omega_i(g)g^{-1}$, $t'=\theta_{i-1}\theta_i^{-1}$, and
$\nu'=\nu_{i-1}$ are determined.
Put
$m_{\nu,e}:=\Enc_\otimes(\Ind_{A_{\nu,e}})$ and
$s_e:=(t^{-1},t'^{-1})$.
Then $m_{\nu,e}\in C_1(I_\otimes)$ and
$\res_\otimes(m_{\nu,e})=\Ind_{A_{\nu,e}}$.
The transition identity for $\nu$ is
\begin{equation}\label{eq:routing-transitions}
  \res_\otimes\left(
    \sum_{e=(t,t',\nu')\in\mathcal E}
      \bigl(m_{\nu,e}\odot z_\nu
        -m_{\nu,e}\odot(s_e\cdot z_{\nu'})\bigr)
    \right)=0.
\end{equation}

\begin{proposition}\label[proposition]{prop:routing-transitions}
For arbitrary words $z_\nu\in\bbK^{I_\otimes}$, the identities
\eqref{eq:routing-transitions}, for all $\nu\in[k]$, hold if and
only if the functions defined by \eqref{eq:history-restrictions}
satisfy \eqref{eq:routing-history}.
Each expression in \eqref{eq:routing-transitions} is quadratic and
has at most $2(d_{\mathrm s}+1)d_{\mathrm t}k$ products.
\end{proposition}

\begin{proof}
For $e=(t,t',\nu')$, the group-action convention gives
$\res_\otimes(s_e\cdot z_{\nu'})(g,\theta)
    =F_{\nu'}(tg,t'\theta)$.
Fix $i\in[L]$, $g\in G_{\mathrm s}$, and put $\nu=\nu_i$.
Exactly one set $A_{\nu,e}$ contains $(g,\theta_i)$.
At this position, the restricted expression in
\eqref{eq:routing-transitions} is
\[
  \begin{aligned}
  F_\nu(g,\theta_i)-F_{\nu'}(tg,t'\theta_i)
    &=F_{\nu_i}(g,\theta_i)
      -F_{\nu_{i-1}}(\omega_i(g),\theta_{i-1})
    =f_i(g)-f_{i-1}(\omega_i(g)).
  \end{aligned}
\]
At a position outside all the sets $A_{\nu,e}$, every mask
restriction is zero, so the restricted expression is zero.
This proves the equivalence.
The two factors in each product are atoms from
\Cref{def:quadratic-fibre-identities}, and
$|\mathcal E|=(d_{\mathrm s}+1)d_{\mathrm t}k$.
Thus there are at most twice this many products.
\end{proof}

\subsection{Endpoints and Local Verification}
\label{subsec:routing-endpoints}

Let
$m_{\mathrm{in}}:=\Enc_\otimes(\Ind_{G_{\mathrm s}\times\set{\theta_0}})$
and
$m_{\mathrm{out}}:=\Enc_\otimes(\Ind_{J\times\set{\theta_L}})$.
Recall that $Bx=x\otimes u_{\mathrm t}$ satisfies
$\res_\otimes(Bx)(g,\theta)=\res_{\mathrm s}(x)(g)$.
Add the two endpoint identities
\begin{equation}\label{eq:routing-endpoints}
  \begin{aligned}
  \res_\otimes(m_{\mathrm{in}}\odot z_{\nu_0}
    -m_{\mathrm{in}}\odot Bx)&=0, \qquad
  \res_\otimes(m_{\mathrm{out}}\odot z_{\nu_L}
    -m_{\mathrm{out}}\odot By)=0.
  \end{aligned}
\end{equation}
Each is a quadratic fibre identity with two products.

\begin{proof}[Proof of~\Cref{thm:permutation-certificate}]
Use \eqref{eq:routing-transitions} for every $\nu\in[k]$ and the
two identities in \eqref{eq:routing-endpoints}.
The counts of identities and products follow from
\Cref{prop:routing-transitions}.

\emph{Completeness.}
Suppose \eqref{eq:permutation-constraint} holds.
Take $f_0=\res_{\mathrm s}(x)$ and define
$f_i=f_{i-1}\circ\omega_i$.
Then $f_L=f_0\circ\pi^{-1}$, so
$f_L|_J=\res_{\mathrm s}(y)|_J$.
Construct the functions $F_\nu$ from this history as above and put
$z_\nu=\Enc_\otimes(F_\nu)\in C_1(I_\otimes)$.
The transition identities hold by \Cref{prop:routing-transitions}.
For the endpoint identities, restriction gives
\[
  \res_\otimes(m_{\mathrm{in}}\odot z_{\nu_0}
    -m_{\mathrm{in}}\odot Bx)(g,\theta)
    =\Ind_{\set{\theta_0}}(\theta)
      \bigl(F_{\nu_0}(g,\theta)-f_0(g)\bigr),\]
and 
\[
  \res_\otimes(m_{\mathrm{out}}\odot z_{\nu_L}
    -m_{\mathrm{out}}\odot By)(g,\theta)
    =\Ind_J(g)\Ind_{\set{\theta_L}}(\theta)
      \bigl(F_{\nu_L}(g,\theta)-\res_{\mathrm s}(y)(g)\bigr).
\]
Both expressions vanish, by the prescribed initial and final values.

\emph{Soundness.}
Suppose arbitrary words $\set{z_\nu}_{\nu\in[k]}$ satisfy all the
identities, and define $f_i$ by \eqref{eq:history-restrictions}.
The same endpoint calculations show that
$f_0=\res_{\mathrm s}(x)$ and
$f_L|_J=\res_{\mathrm s}(y)|_J$.
By \Cref{prop:routing-transitions}, the history satisfies
\eqref{eq:routing-history}. Hence, for every $g\in J$,
\[
  \res_{\mathrm s}(y)(g)
    =f_L(g)=f_0(\pi^{-1}(g))
    =\res_{\mathrm s}(x)(\pi^{-1}(g)).
\]
This proves \textnormal{(S)}.
To construct the system, compute the walk in $G_{\mathrm t}$ and the sets
$A_{\nu,e}$ by enumerating the $L|G_{\mathrm s}|$ transitions.
All masks are obtained by applying $\Enc_\otimes$ to their
indicators. Explicitly, for a function $F$ on
$G_{\mathrm s}\times G_{\mathrm t}$,
\[
  \Enc_\otimes(F)
    =\sum_{(s,t)\in G_{\mathrm s}\times G_{\mathrm t}}
      F(s,t)\bigl((s,t)\cdot
        (\eta_{\mathrm s}\otimes\eta_{\mathrm t})\bigr).
\]
This uses polynomially many field operations and group lookups in
the stated size. The fixed words depend only on the schedule,
$J$, the sets $T_{\mathrm s},T_{\mathrm t}$, and the multiplication-pair data.
For completeness, the same computation constructs the words
$z_\nu$ after the history is evaluated.
\end{proof}

\begin{corollary}\label[corollary]{cor:permutation-verifier}
Under the hypotheses of \Cref{thm:permutation-certificate}, there
is a nonadaptive verifier with oracle access to
$x,y\in\bbK^{I_{\mathrm s}}$ and an auxiliary proof of
$\bigl(k+b^2(k+2)\bigr)|I_\otimes|$
field symbols, using at most $\max\set{w,2n_{\mathrm{perm}}+1}$
queries. There is $\gamma_{\mathrm{perm}}>0$, depending only on
$b,\Delta,\sigma,w,h,d_{\mathrm s},d_{\mathrm t}$, such that:
\begin{itemize}
  \item[\textnormal{(C)}] If $x,y\in C_{\mathrm s,1}$ satisfy
  \eqref{eq:permutation-constraint}, then there is an auxiliary
  proof accepted with probability one.
  \item[\textnormal{(S)}] If $x,y\in C_{\mathrm s,1}$ do not
  satisfy \eqref{eq:permutation-constraint}, then every auxiliary
  proof is rejected with probability at least $\gamma_{\mathrm{perm}}$.
\end{itemize}
For arbitrary $x,y$, rejection probability
$\varepsilon<\gamma_{\mathrm{perm}}$ implies that there are
nearest codewords $x',y'\in C_{\mathrm s,1}$ satisfying
\eqref{eq:permutation-constraint}, with $x,y$ replaced by $x',y'$,
and
\[
  \dist(x,x'),\dist(y,y')
    \le\frac{(4k+8)\varepsilon}{\sigma_*},
  \qquad
  \sigma_*:=\min\set*{\sigma,
    \frac{\Delta^2\sigma^2}{2w(2+bw)}}.
\]
\end{corollary}

\begin{proof}
Apply \Cref{thm:quadratic-fibre-identities} to the $k+2$ input
words $x,y,\set{z_\nu}_{\nu\in[k]}$ and the $k+2$ identities of
\Cref{thm:permutation-certificate}.
The verifier has $M=(k+2)+3(k+2)=4k+8$ choices.
Its auxiliary proof contains the $k$ history words and $b^2$
words on $I_\otimes$ for each identity, giving the stated length.
The bound on the products gives the query count, and the
completeness constructions of the two theorems give
\textnormal{(C)}.

If the rejection probability is below the bound supplied by
\Cref{thm:quadratic-fibre-identities}, nearest codewords for all
input words satisfy every identity. The words $x',y'$
therefore satisfy \eqref{eq:permutation-constraint}, by
\Cref{thm:permutation-certificate}, and have the stated distances.
When $x,y$ are already codewords, their nearest codewords are
$x,y$ themselves. Thus a failure of
\eqref{eq:permutation-constraint} rules out rejection probability
below $\gamma_{\mathrm{perm}}$, proving \textnormal{(S)}.
\end{proof}
\section{Code-to-PCP Holography}
\label{sec:code-to-pcp}

We express a circuit computation by quadratic fibre identities and
permutation constraints. We first state the properties of a code
family needed for this construction.

\begin{definition}\label[definition]{def:admissible-family}
A family of equivariant multiplication pairs is \emph{admissible}
if its members are over a fixed finite field $\bbK$, have common
parameters $(b,\Delta,\sigma,w,h)$, and satisfy the following
conditions for an integer $c\ge 1$:
\begin{itemize}
  \item[\textnormal{(A1)}] Given $u\ge 1$, a deterministic algorithm
  constructs a member with group $G$ satisfying
  $u\le |G|\le cu$, in time polynomial in $u$.
  It outputs the multiplication table of $G$, the word $\eta$, the
  check lists $\mcL_1,\mcL_2$, and a symmetric generating set
  $T\subseteq G\setminus\set{1_G}$ with $|T|\le c$.
  \item[\textnormal{(A2)}] Given the data of a member and a
  permutation $\pi$ of $G$, a deterministic algorithm constructs
  a $T$-schedule realizing $\pi$, of length at most
  $c\lceil\log_2|G|\rceil$, in time polynomial in $|G|$.
\end{itemize}
\end{definition}

\begin{theorem}\label[theorem]{thm:code-to-pcp}
Suppose an admissible family exists. There are polynomial-time
computable and polynomially bounded functions
$\ell,\rho,q:\bits^*\to\N$ and a
$\mathrm{PCP}_\ell(\rho,q)$ verifier for \textnormal{Circuit-SAT}
such that, for every circuit $C$ of size $n$,
\[
  \ell(\cenc(C))=O(n\log n),\qquad
  \rho(\cenc(C))=O(\log n),\qquad
  q(\cenc(C))=O(1).
\]
An accepting proof can be constructed deterministically from
$\cenc(C)$ and a satisfying assignment in time polynomial in $n$.
\end{theorem}

\subsection{Circuit Occurrences and Permutation Constraints}
\label{subsec:circuit-occurrences}

We use five types of occurrences, indexed by
$\mathcal P:=\set{\mathrm A,\mathrm B,\mathrm Z,
\mathrm{in},\mathrm{out}}$.
The first three record the two inputs and the output of a NAND
gate. The last two record circuit inputs and the designated output.
Cyclic equalities identify all occurrences of each wire value.
The following reduction does not require a group structure on the
coordinate set or a bound on fan-out.

\begin{lemma}\label[lemma]{lem:circuit-arithmetization}
Let $\bbK$ be a field, let $C$ be a circuit of size $n$ with NAND gates
$\Gamma$ and input gates $U$, and let $\Omega$ be a finite set
with $|\Omega|\ge\max\set{|\Gamma|,|U|,1}$.
There are $g_0\in\Omega$, permutations
$\set{\pi_{\kappa,\kappa'}}_{\kappa,\kappa'\in\mathcal P}$ of
$\Omega$, and subsets
$\set{J_{\kappa,\kappa'}\subseteq\Omega}_{\kappa,\kappa'\in\mathcal P}$
such that $C$ is satisfiable if and only if there are functions
$\set{f_\kappa\in\bbK^\Omega}_{\kappa\in\mathcal P}$ satisfying, for every $\kappa\in\mathcal P$,
\begin{equation}\label{eq:circuit-value-constraints}
  \begin{aligned}
  f_\kappa\odot f_\kappa&=f_\kappa, \qquad
  f_{\mathrm A}\odot f_{\mathrm B}+f_{\mathrm Z}
    =\Ind_\Omega,\qquad
  f_{\mathrm{out}}(g_0)=1,
  \end{aligned}
\end{equation}
and
\begin{equation}\label{eq:circuit-wiring-constraints}
  f_{\kappa'}(g)
    =f_\kappa(\pi_{\kappa,\kappa'}^{-1}(g))
  \quad\text{for all }\kappa,\kappa'\in\mathcal P,
    \ g\in J_{\kappa,\kappa'}.
\end{equation}
The permutations and subsets can be constructed from $\cenc(C)$
and an enumeration of $\Omega$ in time polynomial in $n+|\Omega|$.
A satisfying assignment gives such functions, and such functions
give a satisfying assignment, in the same time bound.
\end{lemma}

\begin{proof}
Choose injections $\alpha:\Gamma\to\Omega$ and
$\beta:U\to\Omega$, and choose $g_0\in\Omega$.
Their images need not be disjoint. Let $o\in\Gamma\cup U$ be
the output gate, and let $p_1(v),p_2(v)\in\Gamma\cup U$ be
the predecessors of $v\in\Gamma$.
The occupied occurrences form the set
\[
  \mathcal O:=
  \bigl(\set{\mathrm A,\mathrm B,\mathrm Z}\times\alpha(\Gamma)\bigr)
  \cup\bigl(\set{\mathrm{in}}\times\beta(U)\bigr)
  \cup\set{(\mathrm{out},g_0)}.
\]
Define $\zeta:\mathcal O\to\Gamma\cup U$ by
\[
  \begin{aligned}
  \zeta(\mathrm A,\alpha(v))&=p_1(v),&
  \zeta(\mathrm B,\alpha(v))&=p_2(v),&
  \zeta(\mathrm Z,\alpha(v))&=v,\\
  \zeta(\mathrm{in},\beta(u))&=u,&
  \zeta(\mathrm{out},g_0)&=o.
  \end{aligned}
\]
Thus $\zeta$ assigns to an occurrence the gate whose value it
records. Every gate has a defining occurrence, of type
$\mathrm Z$ or $\mathrm{in}$, so every fibre of $\zeta$ is
nonempty. Choose a cyclic ordering of each fibre and let
$\operatorname{succ}$ be the resulting permutation of
$\mathcal O$. A fibre with one occurrence gives a fixed point.

For $\kappa,\kappa'\in\mathcal P$, let
\[
  D_{\kappa,\kappa'}:=\set*{g\in\Omega:
    (\kappa,g)\in\mathcal O,\ 
    \operatorname{succ}(\kappa,g)\in\set{\kappa'}\times\Omega}.
\]
Write
$\operatorname{succ}(\kappa,g)=(\kappa',\psi_{\kappa,\kappa'}(g))$
for $g\in D_{\kappa,\kappa'}$, and put
$J_{\kappa,\kappa'}:=\psi_{\kappa,\kappa'}(D_{\kappa,\kappa'})$.
Since $\operatorname{succ}$ is a permutation,
$\psi_{\kappa,\kappa'}$ is injective.
Extend it to a permutation $\pi_{\kappa,\kappa'}$ of $\Omega$
by any bijection between the complements of
$D_{\kappa,\kappa'}$ and $J_{\kappa,\kappa'}$.
These complements have the same size.
For every $g\in D_{\kappa,\kappa'}$, the corresponding equation
in \eqref{eq:circuit-wiring-constraints} is exactly
$f_{\kappa'}(\psi_{\kappa,\kappa'}(g))=f_\kappa(g)$.
Consequently, the wiring constraints say that the values are
constant on every cycle of $\operatorname{succ}$, hence on every
fibre of $\zeta$. There are $|\mathcal P|^2=25$ permutations,
regardless of the number of occurrences of any gate value.

Suppose first that $C$ has a satisfying assignment, and evaluate
every gate. At an occupied occurrence $(\kappa,g)$, set
$f_\kappa(g)$ to the value of $\zeta(\kappa,g)$.
For $g\notin\alpha(\Gamma)$, set
$f_{\mathrm A}(g)=f_{\mathrm B}(g)=0$ and
$f_{\mathrm Z}(g)=1$.
Set all other unoccupied values to zero.
All functions are Boolean, the NAND identity holds at gate
positions and at the remaining positions, and
$f_{\mathrm{out}}(g_0)=1$.
Values agree along the cycles of $\operatorname{succ}$, so both
\eqref{eq:circuit-value-constraints} and
\eqref{eq:circuit-wiring-constraints} hold.

Conversely, suppose the functions satisfy the constraints.
Over a field, $a^2=a$ implies $a\in\set{0,1}$, so all their
values are Boolean. The wiring constraints assign a common value
$a_v$ to the occurrences in $\zeta^{-1}(v)$.
At $g=\alpha(v)$, the NAND identity gives
$a_v=1-a_{p_1(v)}a_{p_2(v)}$ for $v \in \Gamma$. 
Thus the values $\set{a_u}_{u\in U}$ extend to the circuit
evaluation $\set{a_v}_{v\in\Gamma\cup U}$.
The output constraint gives $a_o=1$, so this assignment satisfies
$C$.
The injections, cycles, and extensions can all be computed by
enumerating the gates and coordinates. The two conversions of
assignments use the same enumerations and a circuit evaluation.
\end{proof}

\subsection{The Quadratic Identity System}
\label{subsec:compiler-identities}

Fix an admissible family and a circuit $C$ of size $n$.
Use \textnormal{(A1)} to choose a member with
$n\le N_{\mathrm s}:=|G_{\mathrm s}|\le cn$.
Apply \Cref{lem:circuit-arithmetization} with
$\Omega=G_{\mathrm s}$.
By \textnormal{(A2)}, compute a $T_{\mathrm s}$-schedule for
each of the $25$ permutations.
Pad the schedules with empty matchings so that they have a common
length $L\le c\lceil\log_2 N_{\mathrm s}\rceil$.
Choose a member with
$L+1\le N_{\mathrm t}:=|G_{\mathrm t}|\le c(L+1)$, and put
$k:=|T_{\mathrm t}|+1\le c+1$.
In particular, $N_{\mathrm s}=O(n)$ and
$N_{\mathrm t}=O(\log n)$.

Introduce five words
$\set{x_\kappa\in\bbK^{I_{\mathrm s}}}_{\kappa\in\mathcal P}$.
Recall $u_{\mathrm s}=\Enc_{\mathrm s}(\Ind_{G_{\mathrm s}})$,
and put $m_{\mathrm{acc}}:=\Enc_{\mathrm s}(\delta_{g_0})$.
The seven identities on $I_{\mathrm s}$ are, for $\kappa\in\mathcal P$,
\begin{equation}\label{eq:compiler-spatial-identities}
  \begin{aligned}
  \res_{\mathrm s}(x_\kappa\odot x_\kappa
    -u_{\mathrm s}\odot x_\kappa)&=0, \\
  \res_{\mathrm s}(x_{\mathrm A}\odot x_{\mathrm B}
    +u_{\mathrm s}\odot x_{\mathrm Z}
    -u_{\mathrm s}\odot u_{\mathrm s})&=0,\\
  \res_{\mathrm s}(m_{\mathrm{acc}}\odot x_{\mathrm{out}}
    -m_{\mathrm{acc}}\odot u_{\mathrm s})&=0.
  \end{aligned}
\end{equation}
Their restrictions are exactly the conditions in
\eqref{eq:circuit-value-constraints}.
For each $(\kappa,\kappa')\in\mathcal P^2$, apply
\Cref{thm:permutation-certificate} to
$x=x_\kappa$, $y=x_{\kappa'}$,
$\pi=\pi_{\kappa,\kappa'}$, and
$J=J_{\kappa,\kappa'}$.
This adds $k$ history words on $I_\otimes$ and $k+2$ identities
certifying the corresponding wiring constraint.
Let $\mathcal S_C$ be the union of these identities and
\eqref{eq:compiler-spatial-identities}, with the five words
$x_\kappa$ shared by every occurrence of the relevant type.

\begin{lemma}\label[lemma]{lem:compiler-identity-system}
The system $\mathcal S_C$ has a solution if and only if $C$ is
satisfiable. It has
\begin{equation}\label{eq:compiler-system-counts}
  N=5+25k\quad\text{input words},\qquad
  J=7+25(k+2)\quad\text{identities},
\end{equation}
and each expression has at most
$n_0:=\max\set{3,2c(c+1)^2}$ products.
The system can be constructed from $\cenc(C)$ in time polynomial
in $n$. A satisfying assignment gives a solution in the same
time bound.
\end{lemma}

\begin{proof}
Suppose $C$ is satisfiable and take the functions from
\Cref{lem:circuit-arithmetization}.
Set $x_\kappa=\Enc_{\mathrm s}(f_\kappa)$, so
$x_\kappa\in C_{\mathrm s,1}$ and
$\res_{\mathrm s}(x_\kappa)=f_\kappa$.
The value constraints give
\eqref{eq:compiler-spatial-identities}.
The wiring constraints and completeness of
\Cref{thm:permutation-certificate} give codeword histories for
all $25$ schedules. Together these words solve $\mathcal S_C$.

Conversely, let the input words be a solution of $\mathcal S_C$
and put $f_\kappa=\res_{\mathrm s}(x_\kappa)$.
The identities on $I_{\mathrm s}$ imply
\eqref{eq:circuit-value-constraints}.
For every pair of types, soundness of
\Cref{thm:permutation-certificate} implies the corresponding
equations in \eqref{eq:circuit-wiring-constraints}.
Thus \Cref{lem:circuit-arithmetization} gives a satisfying
assignment for $C$.

The counts in \eqref{eq:compiler-system-counts} follow from the
five words $x_\kappa$, the seven identities in
\eqref{eq:compiler-spatial-identities}, and the counts for each
schedule. Those seven expressions use at most three products. For the routing expressions,
\[
  2(|T_{\mathrm s}|+1)|T_{\mathrm t}|k
    \le 2(c+1)c(c+1)=2c(c+1)^2.
\]
The group data and schedules are constructed by admissibility,
and \Cref{thm:permutation-certificate} constructs their masks
and history words in polynomial time.
For words on $I_{\mathrm s}$, 
the encoder is explicitly
\[
  \Enc_{\mathrm s}(f)(j,g)
    =\sum_{a\in G_{\mathrm s}}
      f(a)\eta_{\mathrm s}(j,a^{-1}g).
\]
It is computable in polynomial time from the multiplication table
and $\eta_{\mathrm s}$. The tensor encoder has the same formula
over $G_{\mathrm s}\times G_{\mathrm t}$.
All relevant groups and words have size polynomial in $n$.
\end{proof}

\subsection{Sampling with Random Bits}
\label{subsec:compiler-randomness}

The algebraic verifiers use uniform choices from finite sets whose
sizes need not be powers of two. We implement their combined
sampling distribution with a bounded number of unbiased random
bits.

\begin{lemma}\label[lemma]{lem:finite-sampling}
Let $\Omega$ be a finite nonempty set with an enumeration, and put
$R=|\Omega|$.
There is a sampler using $\lceil\log_2 R\rceil$ independent
unbiased random bits whose output $Y$ satisfies
$\Pr[Y=\omega]\ge\frac{1}{2R}$ for all $\omega \in \Omega$.
Consequently, for every $F:\Omega\to\mathbb R_{\ge 0}$,
\[
  \mathbb E[F(Y)]\ge
    \frac12\mathbb E_{\omega\in\Omega}[F(\omega)],
\]
where the expectation on the right is uniform.
The index of $Y$ is computable in time polynomial in $\log_2 R$.
\end{lemma}

\begin{proof}
Put $a=\lceil\log_2 R\rceil$ and choose $U$ uniformly from
$\set{0,\ldots,2^a-1}$.
Return the element with index $1+\lfloor RU/2^a\rfloor$.
Since $2^a\ge R$, every index has at least one preimage, so its
probability is at least $2^{-a}\ge 1/(2R)$.
Multiplying these inequalities by $F(\omega)$ and summing proves
the expectation bound. The calculation of the index uses integer
arithmetic on $O(\log R)$ bits.
\end{proof}

Apply \Cref{thm:quadratic-fibre-identities} to $\mathcal S_C$,
and write $M=N+3J$.
The quantities $N,J,n_0$ are bounded by constants, so the theorem
gives a common rejection bound $\gamma_0>0$ for unsatisfiable
systems. Use the translated check testers for the codes on
$I_{\mathrm s}$ and $I_\otimes$. Their check lists have lengths
$h$ and $2bh$, respectively. All random choices of this verifier
can be generated from a uniform element of
\begin{equation}\label{eq:compiler-random-seed}
  \mathcal R:=[M]\times[3]\times[b^2]\times[2bh]
    \times G_{\mathrm s}\times G_{\mathrm t},
  \qquad
  R:=|\mathcal R|=6Mb^3hN_{\mathrm s}N_{\mathrm t}.
\end{equation}
The first factor chooses one of the $M$ tests in
\Cref{prop:identity-certification}.
Write $j\in[b^2]$ and $r\in[2bh]$ for the third and fourth
coordinates. On $I_{\mathrm s}$, use
\[
  j_{\mathrm s}:=1+\left\lfloor\frac{j-1}{b}\right\rfloor,
  \qquad
  r_{\mathrm s}:=1+\left\lfloor\frac{r-1}{2b}\right\rfloor.
\]
These are uniform in $[b]$ and $[h]$, respectively.
Use the coordinate in $G_{\mathrm s}$ and ignore the one in
$G_{\mathrm t}$. On $I_\otimes$, use $j$, $r$, and the group
pair directly, identifying the $b^2$ fibres with $[b]^2$.

For a membership test, the check index and group element specify
the translated check. For a comparison test, the fibre index and
group element specify the coordinate.
When a fibre certificate is selected, the second factor chooses
among \textnormal{(F1)}, \textnormal{(F2)}, and
\textnormal{(F3)} from \Cref{subsec:fibre-verifier}.
In \textnormal{(F1)}, the fibre index chooses the word, and the
check index and group element give its membership test.
In \textnormal{(F2)}, the check index chooses the row, and the
fibre index and group element choose the coordinate.
In \textnormal{(F3)}, the fibre index chooses the word, and the
group element chooses a coordinate of its data fibre.
Thus every test has its prescribed distribution.
The choices that must be independent come from different factors,
and unused coordinates of the seed are ignored.

Apply \Cref{lem:finite-sampling} to $\mathcal R$.
For any fixed proof, rejection is a nonnegative function of the
seed, so the implemented verifier has rejection probability at
least half that of the ideal verifier.
It retains perfect completeness and has rejection bound
$\gamma:=\gamma_0/2$ on unsatisfiable circuits, using
$\lceil\log_2 R\rceil=O(\log n)$ random bits.
All query addresses are determined by $\cenc(C)$ and the seed,
so the verifier is nonadaptive.

\subsection{Proof of the Holography Theorem}
\label{subsec:compiler-proof}

\begin{proof}[Proof of~\Cref{thm:code-to-pcp}]
Given a valid circuit encoding, construct $\mathcal S_C$ and use
the verifier just described. Before amplification, its query
bound is $\max\set{w,2n_0+1}=O(1)$.
Each input word has length at most
$|I_\otimes|=b^2N_{\mathrm s}N_{\mathrm t}$, and each identity
uses at most $b^2$ auxiliary words of this length.
By \eqref{eq:compiler-system-counts}, its proof length in field
symbols is at most
$(N+Jb^2)b^2N_{\mathrm s}N_{\mathrm t}
    =O(n\log n)$.
The field $\bbK$ is fixed, so binary encoding changes the
proof length and query bound by constant factors.

\emph{Completeness.}
If $C$ is satisfiable, \Cref{lem:compiler-identity-system}
constructs a solution of $\mathcal S_C$ from any satisfying
assignment. By \Cref{thm:quadratic-fibre-identities}, this
solution extends to a proof accepted with probability one.
The implementation of the sampler preserves this property.

\emph{Soundness.}
If $C$ is unsatisfiable, \Cref{lem:compiler-identity-system}
shows that $\mathcal S_C$ has no solution.
Thus \Cref{thm:quadratic-fibre-identities} rejects every proof
with probability at least $\gamma_0$ under ideal sampling, and
the implemented verifier rejects every proof with probability
at least $\gamma$.

Choose a constant integer $t\ge 1$ with
$(1-\gamma)^t\le 1/2$.
Run $t$ independent copies on the same proof and accept if every
copy accepts. Choose all seeds before reading the proof.
Completeness remains perfect, and for every proof of an
unsatisfiable circuit the rejection probability is at least
$1-(1-\gamma)^t\ge 1/2$.
The proof length is unchanged, the number of queries remains
constant, and the number of random bits is
$t\lceil\log_2 R\rceil=O(\log n)$ by
\eqref{eq:compiler-random-seed}.
The repeated verifier is still nonadaptive.

All preprocessing is polynomial in $n$ by admissibility and
\Cref{lem:compiler-identity-system}.
The tensor check lists are constructed as in
\Cref{lem:tensor-multiplication-pairs}, and the remaining
verification steps use field operations, group operations, and
the explicitly specified expressions.
An accepting proof is also computable in polynomial time.
After constructing its input codewords, set
$p_\alpha=\Psi_\alpha(\mathbf z)$ for every identity.
The witnesses from \Cref{thm:fibre-certificate} are
$w_{\alpha,j}=(\eta_{E_\alpha}^{\mathrm{op}})_j\cdot p_\alpha$,
where $\eta_{E_\alpha}$ is $\eta_{\mathrm s}$ or
$\eta_{\mathrm s}\otimes\eta_{\mathrm t}$ according to the
domain. These group-algebra products are computable from the
same data in polynomial time.

The proof layout, the seed cardinality $R$, and the query bound
are computable from $\cenc(C)$.
Take $\ell$ to be the binary proof length, $\rho$ to be
$t\lceil\log_2 R\rceil$, and $q$ to be the resulting constant
upper bound on binary queries.
For malformed encodings, use fixed values of these functions
and reject immediately.
Since a natural encoding has length $\Omega(n)$, the resource
functions are polynomial-time computable and polynomially
bounded in the input length, and the verifier has polynomial
running time. It therefore satisfies \Cref{def:pcp}.
\end{proof}
\section{Multiplicative Chamber Codes}
\label{sec:chamber-codes}

We construct codes whose coordinates are the chambers of a quotient
of a product of trees. The code constraints prescribe a linear code
on the chambers incident to each vertex. Compatible choices of these
local codes give multiplication of global codewords and an action
of the covering group.

Let $d\ge 1$ and $n_i\ge 2$, and let
$\mathcal T=\prod_{i=1}^d T_i$, where $T_i$ is an
$n_i$-regular tree with a chosen bipartition.
A cell is a product of vertices and edges of the factors.
Its direction set $\dir(f)\subseteq[d]$ consists of its edge
factors, and its dimension is $|\dir(f)|$.
The cells of dimension $d$ are called \emph{chambers}.
The bipartitions give each vertex a type in $\bits^d$.
Let $\Lambda$ act on $\mathcal T$ preserving the factors and
types, freely on vertices, and with finite quotient
$X_0=\Lambda\backslash\mathcal T$.
For a quotient $X$, write $X(j)$ for its $j$-cells and put
$\operatorname{St}(f):=\set{Q\in X(d):f\le Q}$.

\begin{definition}\label[definition]{def:chamber-sheaf}
Let $\bbK$ be a field.
A \emph{chamber sheaf} $\mathcal F$ on $X$ assigns a subspace
$\mathcal F_f\le\bbK^{\operatorname{St}(f)}$ to each cell $f$,
with $\mathcal F_Q=\bbK$ for chambers $Q$, such that ordinary
restriction maps $\mathcal F_f$ into $\mathcal F_e$ whenever
$f\le e$.
Its \emph{chamber code} is
\[
  C(X,\mathcal F):=
  \set*{z\in\bbK^{X(d)}:
    z|_{\operatorname{St}(v)}\in\mathcal F_v
    \text{ for every }v\in X(0)}.
\]
For an integer $A\ge 0$, a family
$\set{\mathcal F^a}_{a=0}^A$ is \emph{multiplicative}
if $\mathcal F_f^a*\mathcal F_f^{a'}\le\mathcal F_f^{a+a'}$
for every $f$ and $a+a'\le A$.
Write $C_a(X):=C(X,\mathcal F^a)$.
\end{definition}

\begin{theorem}\label[theorem]{thm:multiplicative-chamber-codes}
Let $\set{\mathcal F^a}_{a=0}^A$ be a multiplicative family of
chamber sheaves on $X_0$, and pull it back to a finite cover
$X=\Lambda'\backslash\mathcal T$, where $\Lambda'\le\Lambda$.
Then:
\begin{itemize}
  \item[\textnormal{(1)}] A family
  $\set{x_v\in\mathcal F_v^a}_{v\in X(0)}$ is the family of
  vertex restrictions of a codeword in $C_a(X)$ if and only if
  $x_v|_{\operatorname{St}(e)}=x_w|_{\operatorname{St}(e)}$
  for every edge $e$ with endpoints $v,w$. The codeword is unique.
  \item[\textnormal{(2)}] If $a+a'\le A$, then
  $C_a(X)*C_{a'}(X)\le C_{a+a'}(X)$.
  \item[\textnormal{(3)}] Pullback along a cover $Y\to X$
  sends $C_a(X)$ injectively into $C_a(Y)$.
  \item[\textnormal{(4)}] If $\Lambda'\triangleleft\Lambda$,
  put $G=\Lambda/\Lambda'$ and $b=|X_0(d)|$.
  Choosing a lift of each chamber of $X_0$ identifies $X(d)$
  with $[b]\times G$, under which every $C_a(X)$ is
  $G$-invariant for the action
  $(s\cdot z)(j,g)=z(j,s^{-1}g)$.
\end{itemize}
\end{theorem}

\subsection{Cubical Quotients and Chamber Stars}
\label{subsec:cubical-chamber-stars}

The quotient retains the faces of each product cell.
Types distinguish the vertices of a cell, so its vertices and faces
remain distinct in the quotient.
We say that \emph{vertex stars embed} if, for every vertex
$\widetilde v$ of $\mathcal T$, projection is injective on the
cells in the union of the chambers containing $\widetilde v$,
including all their faces.
This condition is stronger than the bijection on incident chambers
established below.

\begin{lemma}\label[lemma]{lem:cubical-incidence}
For a cell $f$ of $X$ and a chosen lift $\widetilde f$, the map
$\operatorname{St}(\widetilde f)\to\operatorname{St}(f)$ is a
bijection.
At a vertex $v$, choosing labels $S_{v,i}$ for the $n_i$ edges
in each direction identifies
$\operatorname{St}(v)$ with $\prod_{i=1}^d S_{v,i}$.
For $v\le f$, the subset $\operatorname{St}(f)$ fixes one label
in each direction in $\dir(f)$.
Consequently, with $N_*:=\prod_i n_i$ and
$n_\Sigma:=\sum_i n_i$,
\[
  |\operatorname{St}(f)|=\prod_{i\notin\dir(f)}n_i,
  \qquad
  |X(d)|=\frac{N_*}{2^d}|X(0)|,
  \qquad
  |X(1)|=\frac{n_\Sigma}{2}|X(0)|.
\]
If vertex stars embed, two incident edges of different directions
determine a unique square, as does a path of two edges in different
directions.
\end{lemma}

\begin{proof}
Any chamber containing $f$ has a lift containing some translate
of $\widetilde f$, which can be translated to contain
$\widetilde f$ itself.
For injectivity, suppose chambers $Q,Q'$ containing
$\widetilde f$ satisfy $Q'=\gamma Q$ for $\gamma\in\Lambda'$.
Choose a vertex $\widetilde v\le\widetilde f$.
There is exactly one vertex of its type in $Q'$, so
$\gamma\widetilde v=\widetilde v$.
The action is free on vertices, giving $\gamma=1$ and $Q=Q'$.

In the product of trees, a chamber containing a vertex is specified
by one incident edge in each direction. A face containing that
vertex fixes the edges in its directions and leaves the others
free. The bijection just proved transports this description to
$X$. It gives the first cardinality formula.
A vertex belongs to $N_*$ chambers and $n_\Sigma$ edges, a
chamber has $2^d$ vertices, and an edge has two endpoints.
Counting these incidences gives the remaining formulas.
The two edge choices in distinct tree factors determine their
product square. Embedded vertex stars preserve this square and
its uniqueness. Applying this at the intermediate vertex gives
the assertion for a path of two edges.
\end{proof}

For a face $f$ and a vertex $v\le f$, the preceding identification
will be used without changing the chamber coordinates.
In particular, the chambers incident to an edge at $v$ form a
coordinate hyperplane, and the chambers incident to a square at
$v$ form the intersection of two perpendicular hyperplanes.

\subsection{Projective Reed--Solomon Codes}
\label{subsec:projective-reed-solomon}

Let $S\subseteq\bbP^1(\bbK)$ be finite and nonempty, and choose
a representative $r_p\in\bbK^2\setminus\set{0}$ for each
$p\in S$.
For $0\le t<|S|$, let $\mathcal H_t=\bbK[X,Y]_t$ be the
space of homogeneous binary forms of degree $t$, and define
$\operatorname{ev}_t:\mathcal H_t\to\bbK^S$ by 
$\operatorname{ev}_t(F)(p):=F(r_p)$ and put
$\operatorname{PRS}_S(t):=\operatorname{im}(\operatorname{ev}_t)$.
When $\bbF_q\subseteq\bbK$ and $S=\bbP^1(\bbF_q)$, we use
$r_p=(p,1)^{\mathsf T}$ for $p\in\bbF_q$ and
$r_\infty=(1,0)^{\mathsf T}$, and write
$\operatorname{PRS}_q(t)$ for this code.

\begin{lemma}\label[lemma]{lem:prs-properties}
For $0\le t<|S|$, evaluation is injective and we have
$\dim\operatorname{PRS}_S(t)=t+1$ and $\Delta(\operatorname{PRS}_S(t))=1-\frac{t}{|S|}$.
Restriction to any $t+1$ coordinates is an isomorphism onto
$\bbK^{t+1}$.
If $t+t'<|S|$, then
$\operatorname{PRS}_S(t)*\operatorname{PRS}_S(t')
    =\operatorname{PRS}_S(t+t')$.
\end{lemma}

\begin{proof}
A nonzero form $F\in\mathcal H_t$ has at most $t$ projective
zeros. Indeed, its finite zeros are roots of $F(X,1)$.
If infinity is also a zero, the coefficient of $X^t$ vanishes,
so $\deg F(X,1)\le t-1$.
Thus evaluation on $S$, or on any $t+1$ points of $S$, is
injective. Since $\dim\mathcal H_t=t+1$, this proves the
dimension and restriction assertions and the lower bound on
distance. A product of linear forms vanishing at $t$ chosen
points has exactly those zeros in $S$, giving equality.

Pointwise multiplication satisfies
$\operatorname{ev}_t(F)\odot\operatorname{ev}_{t'}(F')
=\operatorname{ev}_{t+t'}(FF')$.
Every monomial of degree $t+t'$ is a product of monomials of
degrees $t$ and $t'$, so such products span
$\mathcal H_{t+t'}$. This proves the Schur-product identity.
\end{proof}

Changing a projective frame changes both the coordinate labels
and the representatives at which a form is evaluated.
The scalar in the next lemma records the latter change.

\begin{lemma}\label[lemma]{lem:prs-coordinate-change}
Let $g\in\operatorname{GL}_2(\bbK)$ preserve $S$ projectively.
Define $c(g,p)\in\bbK^\times$ by
$gr_p=c(g,p)r_{gp}$.
Then, for $F\in\mathcal H_t$ and $p\in S$,
\begin{equation}\label{eq:prs-frame-change}
  \operatorname{ev}_t(F\circ g)(p)
    =c(g,p)^t\operatorname{ev}_t(F)(gp).
\end{equation}
For $g,h$ preserving $S$, the scalars satisfy
$c(gh,p)=c(g,hp)c(h,p)$.
The coordinate map
$M_g^{(t)}x(p):=c(g,p)^t x(gp)$ preserves
$\operatorname{PRS}_S(t)$ and obeys
$(M_g^{(t)}x)\odot(M_g^{(t')}y)
    =M_g^{(t+t')}(x\odot y)$.
\end{lemma}

\begin{proof}
Homogeneity gives $F(gr_p)=c(g,p)^tF(r_{gp})$, proving
\eqref{eq:prs-frame-change}.
Also
$ghr_p=c(h,p)gr_{hp}=c(h,p)c(g,hp)r_{ghp}$,
so uniqueness of the scalar gives its composition law.
Substitution by $g$ is an invertible map of $\mathcal H_t$,
which proves preservation of the code.
The product identity follows by adding the exponents of
$c(g,p)$ at each coordinate.
\end{proof}

A permutation of coordinates followed by multiplication by
nonzero coordinate scalars preserves dimension and Hamming
distance. We use such coordinate maps in the local charts.
The multiplication identity requires the exponents of the
coordinate scalars to add with the degrees.

\subsection{Compatible Tensor Charts and Cochains}
\label{subsec:compatible-tensor-charts}

\begin{definition}\label[definition]{def:compatible-tensor-charts}
A chamber sheaf $\mathcal F$ has \emph{compatible tensor charts}
if, for every vertex $v$, there are linear codes
$B_{v,i}\le\bbK^{S_{v,i}}$ and a function
$\mu_v:\operatorname{St}(v)\to\bbK^\times$ such that, for
every $f\ge v$,
\[
  \mathcal F_f=
  \mu_v|_{\operatorname{St}(f)}\odot
  \bigotimes_{i\notin\dir(f)}B_{v,i}.
\]
Here the tensor code is viewed as a space of functions on
$\operatorname{St}(f)$ by fixing the labels in $\dir(f)$.
The same multiplier $\mu_v$ is used for every face containing
$v$, and the displayed equality is required at every vertex
of $f$.
\end{definition}

Thus a chart identifies a vertex stalk and all its face
restrictions at once. Dividing an array and its restrictions by
$\mu_v$ gives the corresponding tensor codes and ordinary
coordinate restrictions. Agreement of arrays on intersections is
preserved under this division.
By \Cref{lem:tensor-codes}, if $B_{v,i}$ has relative distance
at least $\theta_i$, then
$\Delta(\mathcal F_f)\ge\prod_{i\notin\dir(f)}\theta_i$.

\begin{lemma}\label[lemma]{lem:graded-tensor-charts}
Suppose each $\mathcal F^a$ has compatible tensor charts with
factors $B_{v,i}^a$ and multipliers $\mu_v^a$.
Assume that, for $a+a'\le A$,
we have $B_{v,i}^a*B_{v,i}^{a'}\le B_{v,i}^{a+a'}$ and
$\mu_v^a\odot\mu_v^{a'}=\mu_v^{a+a'}$.
Then $\set{\mathcal F^a}_{a=0}^A$ is multiplicative.
In particular, the hypotheses hold when the factors are
$\operatorname{PRS}_{q_i}(ae_i)$, with a fixed permutation of
each factor's coordinates for all grades, and the multiplier at
grade $a$ is the $a$-th power of a fixed nowhere-zero function,
provided $Ae_i\le q_i$ for every $i$.
\end{lemma}

\begin{proof}
Fix $f$ and a vertex $v\le f$.
In its chart, the product of two stalk words is
\[
  (\mu_v^a\odot u)\odot(\mu_v^{a'}\odot u')
    =\mu_v^{a+a'}\odot(u\odot u').
\]
By \Cref{lem:tensor-schur-products}, or its iterated form,
$u\odot u'$ belongs to the tensor product of the factor
Schur products. The assumed inclusions place this in the
tensor code for grade $a+a'$, so the product belongs to
$\mathcal F_f^{a+a'}$.
The Reed--Solomon case follows from \Cref{lem:prs-properties}.
Using the same permutation in each grade preserves the product
identity, as does taking powers of the fixed multiplier.
\end{proof}

The compatibility in \Cref{def:compatible-tensor-charts} is part
of the construction. Choosing Reed--Solomon parameters at each
vertex alone does not establish equality of the two descriptions
of an edge stalk. The arithmetic construction will supply these
compatible charts.

For the cochain statements, assume that $\bbK$ has
characteristic two. For $0\le j\le d$, put
\[
  C^j(X,\mathcal F):=
    \bigoplus_{f\in X(j)}\mathcal F_f.
\]
For $0\le j<d$, define
$\delta^j:C^j(X,\mathcal F)\to C^{j+1}(X,\mathcal F)$ by
\[
  (\delta^j x)(e):=
    \sum_{\substack{f\le e\\\dim f=j}}
      x(f)|_{\operatorname{St}(e)}
\]
for $e \in X(j+1)$.
The maps are ordinary restriction, so their compositions agree
along any two face chains with the same endpoints.
Each face of codimension two lies in exactly two intermediate
faces. Its contribution to $\delta^{j+1}\delta^j x$ therefore
appears twice and vanishes. Thus $\delta^{j+1}\delta^j=0$.
An element of $\ker\delta^j$ is a \emph{$j$-cocycle}.
The support of a cochain is
$\supp(x):=\set{f:x(f)\ne 0}$, counting a stalk value as one
entry.

\begin{lemma}\label[lemma]{lem:chamber-cocycles}
The map
$C(X,\mathcal F)\to C^0(X,\mathcal F)$ defined by
$z\mapsto\set{z|_{\operatorname{St}(v)}}_{v\in X(0)}$
is an isomorphism onto $\ker\delta^0$.
\end{lemma}

\begin{proof}
If $z\in C(X,\mathcal F)$, its restrictions at the two
endpoints of an edge agree on that edge's chamber set.
Their sum is zero in characteristic two, so the resulting
zero-cochain is a cocycle.

Conversely, let $x\in\ker\delta^0$.
For every edge $e=[v,w]$,
$x(v)|_{\operatorname{St}(e)}=x(w)|_{\operatorname{St}(e)}$.
Fix a chamber $Q$. Its vertices are connected by its edges,
so the values $x(v)(Q)$ are equal for every $v\le Q$.
Let $z(Q)$ be this common value.
Then $z|_{\operatorname{St}(v)}=x(v)\in\mathcal F_v$ for
every vertex, giving $z\in C(X,\mathcal F)$.
This construction is inverse to restriction, proving the claim.
The gluing argument itself uses equality on edge stars and is
valid over any field.
\end{proof}

\subsection{Covers and Group Coordinates}
\label{subsec:chamber-cover-coordinates}

Let $p:Y\to X$ arise from an inclusion of the acting groups.
For each cell $f$ of $Y$, \Cref{lem:cubical-incidence} gives a
bijection
$p:\operatorname{St}(f)\to\operatorname{St}(p(f))$.
Define the pulled-back stalk by
\[
  (p^*\mathcal F)_f:=
  \set*{u\circ p|_{\operatorname{St}(f)}:
    u\in\mathcal F_{p(f)}}.
\]
For $f\le e$, restriction commutes with this pullback, since
both evaluate the same function on the chambers containing $e$.
Thus $p^*\mathcal F$ is a chamber sheaf.
Pullback also commutes with pointwise multiplication, so it
preserves multiplicative families and compatible tensor charts.

\begin{proof}[Proof of~\Cref{thm:multiplicative-chamber-codes}]
For \textnormal{(1)}, the gluing argument in
\Cref{lem:chamber-cocycles} gives existence and uniqueness.
For \textnormal{(2)}, take $z\in C_a(X)$ and
$z'\in C_{a'}(X)$.
At every vertex $v$,
\[
  (z\odot z')|_{\operatorname{St}(v)}
    =z|_{\operatorname{St}(v)}\odot
      z'|_{\operatorname{St}(v)}
    \in\mathcal F_v^{a+a'}.
\]
Hence $z\odot z'\in C_{a+a'}(X)$, and linearity gives the
Schur-product inclusion.

For \textnormal{(3)}, let $z\in C_a(X)$ and define
$p^*z=z\circ p$ on the chambers of $Y$.
At a vertex $v$ of $Y$, its restriction is the pullback of
$z|_{\operatorname{St}(p(v))}$, so it belongs to
$(p^*\mathcal F^a)_v$.
Thus $p^*z\in C_a(Y)$.
Projection is surjective on chambers, making the pullback map
injective.

For \textnormal{(4)}, normality defines the deck action of
$G$ on $X=\Lambda'\backslash\mathcal T$.
The stabilizer in $\Lambda$ of a cell is trivial: an element
stabilizing the cell fixes its vertex of each type and hence is
the identity.
Consequently, $G$ acts freely and transitively on the cells above
each cell of $X_0$.
Choose lifts $\widehat Q_1,\ldots,\widehat Q_b$ of the base
chambers. The map
$[b]\times G\to X(d)$ defined by $(j,g)\mapsto g\widehat Q_j$
is therefore a bijection.
The pulled-back stalk at $sv$ is the image under $s$ of the
stalk at $v$. For $z\in C_a(X)$, the restriction of $s\cdot z$
at $sv$ consequently belongs to $\mathcal F_{sv}^a$.
Thus $C_a(X)$ is $G$-invariant.
Finally,
$ (s\cdot z)(g\widehat Q_j)=z(s^{-1}g\widehat Q_j)$,
which gives the stated action in the coordinates $[b]\times G$.
\end{proof}
\section{Distance and Local Testability from Local Agreement}
\label{sec:chamber-testability}

We prove distance and local testability for chamber codes with
compatible tensor charts. The proof combines agreement of the
patches around a vertex with a quadratic-form bound for the walk
on the vertices.

Let $X$ be a finite quotient of a product of $d\ge 3$ regular
trees as in \Cref{sec:chamber-codes}, and assume that its vertex
stars embed. Let $\mathcal F$ be a chamber sheaf over a finite
field $\bbK$ of characteristic two, with compatible tensor charts
as in \Cref{def:compatible-tensor-charts}.
Write $V=X(0)$, $E=X(1)$, $Q=X(d)$, and
$C=C(X,\mathcal F)\le\bbK^Q$.
If the tree degrees are $\set{n_i}_{i\in[d]}$, put
\[
  n_{\min}:=\min_i n_i,\qquad
  \rho:=\frac{\max_i n_i}{n_{\min}},\qquad
  n_\Sigma:=\sum_{i=1}^d n_i,\qquad
  N_*:=\prod_{i=1}^d n_i.
\]
Let $W$ choose an incident edge uniformly, so
$(Wf)(v)=n_\Sigma^{-1}\sum_{w\sim v}f(w)$.
Each incident edge contributes once to this sum.
For real functions on $V$, write
$\mathbb E f=|V|^{-1}\sum_{v\in V}f(v)$ and
$\langle f,g\rangle=\mathbb E(fg)$.

\begin{theorem}\label[theorem]{thm:chamber-distance-testability}
Suppose every factor code in every vertex chart has relative
distance at least $\theta\in(0,1]$.
Put $\kappa:=\frac{\theta^2}{4d\rho}$ and $\alpha:=\frac{\kappa n_{\min}}{n_\Sigma}$.
Suppose $\lambda\ge 0$ and $\beta>0$ satisfy $\lambda<\alpha$
and
\begin{equation}\label{eq:chamber-mixing}
  \langle f,Wf\rangle
    \le\lambda\mathbb E(f^2)+\beta(\mathbb E f)^2
\end{equation}
for $f : V \to \R_{\ge 0}$.
Set $c:=(\alpha-\lambda)/\beta$ and
$c_{\mathrm{rep}}:=\max\set{1,2/c}$.
Then
\[
  \Delta(C)\ge\frac{\theta^d(\theta-\lambda)}{\beta}.
\]
The test that chooses a uniform $v\in V$, reads
$y|_{\operatorname{St}(v)}$, and checks membership in
$\mathcal F_v$ uses $N_*$ queries and satisfies:
\begin{itemize}
  \item[\textnormal{(C)}] If $y\in C$, the test accepts with
  probability one.
  \item[\textnormal{(S)}] If $y\notin C$, the test rejects with
  probability at least
  $\sigma_*\dist(y,C)$, where
  $\sigma_*:=1/(1+c_{\mathrm{rep}}n_\Sigma)$.
\end{itemize}
\end{theorem}

\subsection{Agreement of Tensor Patches}
\label{subsec:tensor-patch-agreement}

The local agreement statement holds over any field and for
arbitrary linear factor codes.
Let $S_1,\ldots,S_d$ be finite nonempty sets, let
$n_i=|S_i|$, and put $S=\prod_i S_i$.
For $i\in[d]$ and $a\in S_i$, write
$H(i,a)=\set{x\in S:x_i=a}$.
Given codes $B_i\le\bbK^{S_i}$, a \emph{patch} on $H(i,a)$ is
a word in $\bigotimes_{j\ne i}B_j$, viewed as a function on
that hyperplane.
Hyperplanes in different directions are perpendicular.
Two perpendicular patches disagree when their restrictions to the
intersection are unequal.

\begin{lemma}\label[lemma]{lem:patch-disagreement-witnesses}
Suppose $d\ge 3$ and $\Delta(B_i)\ge\theta$ for every $i$.
Prescribe a patch on every hyperplane, and form the graph whose
vertices are the indices $(i,a)$ and whose edges are disagreeing
perpendicular pairs.
If $H(i,a)$ and $H(j,b)$ form an edge and
$k\notin\set{i,j}$, their degrees in this graph have sum at
least $\theta n_k$.
\end{lemma}

\begin{proof}
The difference of the two patches on their intersection is a
nonzero word of $\bigotimes_{r\notin\set{i,j}}B_r$.
Some line in direction $k$ is nonzero, and its restriction is a
word of $B_k$. There are therefore at least $\theta n_k$ labels
$a_k\in S_k$ at which the two original patches have different
values on this line.
The patch on $H(k,a_k)$ cannot agree with both original patches
on their respective intersections. Thus it is adjacent to at
least one of $H(i,a)$ and $H(j,b)$ in the disagreement graph.
Different labels $a_k$ give different neighboring vertices,
proving the claim.
\end{proof}

\begin{lemma}\label[lemma]{lem:tensor-patch-agreement}
Let $d\ge 3$ and let $B_i\le\bbK^{S_i}$ have relative distance
at least $\theta\in(0,1]$.
Prescribe a patch $p_{i,a}$ on every $H(i,a)$, and let $D$ be
the number of unordered disagreeing perpendicular pairs.
There is $u\in\bigotimes_iB_i$ whose restrictions disagree with
at most $D/(\kappa n_{\min})$ patches, where
$n_{\min}=\min_i n_i$, $\rho=\max_i n_i/n_{\min}$, and
$\kappa=\theta^2/(4d\rho)$.
\end{lemma}

\begin{proof}
First suppose a factor code is zero. If at least two factor
codes are zero, every patch is zero, so take $u=0$.
If only $B_i$ is zero, only patches in direction $i$ can be
nonzero. Fix $j\ne i$. Every nonzero patch in direction $i$
has a nonzero direction-$j$ line, hence disagrees with at least
$\theta n_j$ of the zero patches in direction $j$.
These edges of the disagreement graph are different for
different direction-$i$ patches. Thus there are at most
$D/(\theta n_j)\le D/(\kappa n_{\min})$ nonzero patches,
and again $u=0$ works.

We may therefore assume all factor codes are nonzero.
Call a hyperplane bad if its degree in the disagreement graph
is at least $\theta n_{\min}/2$, and good otherwise.
By the degree sum, the number of bad hyperplanes is at most
$4D/(\theta n_{\min})$.
\Cref{lem:patch-disagreement-witnesses} shows that every
disagreeing pair contains a bad hyperplane. Consequently,
perpendicular good patches agree on their intersection.

Suppose first that $4D/(\theta n_{\min})<\theta n_{\min}$.
In each direction $i$, fewer than $\theta n_i$ labels belong
to bad hyperplanes. Restriction of $B_i$ to the remaining
labels is injective, since a nonzero word supported on the
deleted labels would have weight below its distance.
By \Cref{lem:information-sets}, choose an information set $A_i$
among the remaining labels.
At each point of $\prod_i A_i$, the good patches through that
point have a common value.
Since each map $B_i\to\bbK^{A_i}$ is an isomorphism, their
tensor product gives an isomorphism
\[
  \bigotimes_{i=1}^dB_i
    \longrightarrow\bbK^{\prod_i A_i}.
\]
Let $u$ be the codeword with these prescribed values.

If $a\in A_i$, then $u|_{H(i,a)}$ and $p_{i,a}$ are words
of $\bigotimes_{j\ne i}B_j$ agreeing on its information set
$\prod_{j\ne i}A_j$. Hence they are equal everywhere.
Now let $H(i,a)$ be any good hyperplane.
At a point of its information set, choose $j\ne i$.
The point lies on a good hyperplane $H(j,a_j)$ with
$a_j\in A_j$, where $u$ equals the prescribed patch by the
preceding argument. Agreement of the two good patches shows
that $u$ equals $p_{i,a}$ at this point.
Injectivity of restriction to $\prod_{j\ne i}A_j$ now gives
$u|_{H(i,a)}=p_{i,a}$.
Thus $u$ disagrees only on bad hyperplanes, and their number
is at most $4D/(\theta n_{\min})\le D/(\kappa n_{\min})$.

If $4D/(\theta n_{\min})\ge\theta n_{\min}$, take $u=0$.
There are $n_\Sigma\le d\rho n_{\min}$ hyperplanes, and
\[
  \frac{D}{\kappa n_{\min}}
    =\frac{4d\rho D}{\theta^2n_{\min}}
    \ge d\rho n_{\min}\ge n_\Sigma.
\]
This proves the bound in the remaining case.
\end{proof}

Return to a vertex $v$ of $X$.
The edges at $v$ correspond to the coordinate hyperplanes of its
chart. Dividing all edge patches by the restriction of the same
nonzero chart multiplier preserves equality on intersections.
Applying \Cref{lem:tensor-patch-agreement} and restoring the
multiplier therefore gives the following conclusion: if $D_v$
perpendicular pairs of edge patches disagree, some
$u\in\mathcal F_v$ disagrees with at most
$D_v/(\kappa n_{\min})$ of those patches.

\subsection{Locally Minimal Cocycles}
\label{subsec:locally-minimal-cocycles}

For $v\in V$ and $u\in\mathcal F_v$, let $J_vu$ be the
zero-cochain equal to $u$ at $v$ and zero elsewhere.
Supports of cochains count the cells on which the stalk value
is nonzero.

\begin{definition}\label[definition]{def:locally-minimal-cocycle}
A one-cochain $z\in C^1(X,\mathcal F)$ is \emph{locally minimal}
if $|\supp(z+\delta^0J_vu)|\ge|\supp(z)|$
for $v\in V$ and $u\in\mathcal F_v$.
\end{definition}

\begin{lemma}\label[lemma]{lem:locally-minimal-support}
Under the hypotheses of \Cref{thm:chamber-distance-testability},
every nonzero locally minimal $z\in\ker\delta^1$ satisfies
$|\supp(z)|\ge c|V|/2$.
\end{lemma}

\begin{proof}
For $v\in V$, put
$d_z(v):=|\set{e\in E:v\le e,\ z_e\ne 0}|$.
Regard the values of $z$ on all edges at $v$, including the
zero values, as prescribed patches.
Let $D_v$ count their disagreeing perpendicular pairs.
Adding $\delta^0J_vu$ replaces each incident patch by
$z_e+u|_{\operatorname{St}(e)}$ and leaves all other edges
unchanged. In characteristic two, the new patch is zero
exactly when the old patch equals the restriction of $u$.
Local minimality and the agreement conclusion above therefore
give
\begin{equation}\label{eq:local-support-agreement}
  \kappa n_{\min}d_z(v)\le D_v.
\end{equation}

A disagreeing perpendicular pair at $v$ spans a square $f$.
The two edge restrictions have nonzero sum on
$\operatorname{St}(f)$.
Since $(\delta^1z)_f=0$, at least one of the other two edges
has nonzero restriction there, hence belongs to $\supp(z)$.
Choose such an edge and the two-edge path that reaches it from
$v$. This path determines $f$: its edges, viewed from their
common vertex, have different directions and determine a unique
square. It therefore determines the original pair at $v$.
The chosen paths for different disagreeing pairs are distinct,
so
\[
  D_v\le\sum_{w\sim v}d_z(w).
\]
Combining this with \eqref{eq:local-support-agreement} gives
$\alpha d_z(v)\le(Wd_z)(v)$.
Multiply by $d_z(v)$ and average. By
\eqref{eq:chamber-mixing},
\[
  \alpha\mathbb E(d_z^2)
    \le\langle d_z,Wd_z\rangle
    \le\lambda\mathbb E(d_z^2)
      +\beta(\mathbb E d_z)^2.
\]
The function $d_z$ is nonnegative and integer-valued, so
$\mathbb E(d_z^2)\ge\mathbb E d_z$.
Since $z\ne 0$, its mean is positive. Dividing yields
$\mathbb E d_z\ge(\alpha-\lambda)/\beta=c$.
Each supported edge is counted at its two endpoints, and hence
$2|\supp(z)|=|V|\mathbb E d_z\ge c|V|$.
\end{proof}

\subsection{Repair and the Vertex Test}
\label{subsec:chamber-repair}

The repair argument applies to a complex of vector spaces with
any choice of blocks in its first two spaces.

\begin{lemma}\label[lemma]{lem:cochain-repair}
Let $C^0=\bigoplus_{v\in V}U_v$ and
$C^1=\bigoplus_{e\in E}U_e$ be vector spaces over a field,
with $V,E$ finite. Let
$\delta^0:C^0\to C^1$ and $\delta^1:C^1\to C^2$ be linear
maps satisfying $\delta^1\delta^0=0$.
Call $z\in C^1$ locally minimal if adding $\delta^0u$, for
$u$ supported on one block of $C^0$, never reduces its number
of nonzero blocks.
Suppose every nonzero locally minimal element of
$\ker\delta^1$ has at least $s>0$ nonzero blocks.
For every $x\in C^0$, there is $x'\in\ker\delta^0$ with
\[
  |\supp(x-x')|
    \le\max\set{1,|V|/s}\,|\supp(\delta^0x)|.
\]
\end{lemma}

\begin{proof}
If $|\supp(\delta^0x)|\ge s$, take $x'=0$.
Then $|\supp(x-x')|\le|V|\le(|V|/s)|\supp(\delta^0x)|$.
Otherwise, $z=\delta^0x$ is a cocycle with fewer than $s$
nonzero blocks. If $z\ne 0$, it is not locally minimal.
Thus changing $x$ on one block reduces
$|\supp(\delta^0x)|$ by at least one.
The new coboundary is again a cocycle below the same threshold.
After at most the original $|\supp(\delta^0x)|$ changes,
the coboundary is zero. The number of blocks on which the final
word differs from the original is at most the number of changes.
\end{proof}

Apply \Cref{lem:cochain-repair} to the complex of
$\mathcal F$ with $s=c|V|/2$.
By \Cref{lem:locally-minimal-support}, every
$x\in C^0(X,\mathcal F)$ has $x'\in\ker\delta^0$ such that
\begin{equation}\label{eq:chamber-repair-bound}
  |\supp(x-x')|
    \le c_{\mathrm{rep}}|\supp(\delta^0x)|.
\end{equation}

\begin{lemma}\label[lemma]{lem:chamber-star-test}
Suppose the chamber sheaf $\mathcal F$ satisfies
\eqref{eq:chamber-repair-bound} for some
$c_{\mathrm{rep}}>0$.
For every word $y\in\bbK^Q$, the fraction of vertices where
$y|_{\operatorname{St}(v)}\notin\mathcal F_v$ is at least
$\dist(y,C)/(1+c_{\mathrm{rep}}n_\Sigma)$.
\end{lemma}

\begin{proof}
Let $B$ be the set of failing vertices, and define a zero-cochain
$x$ by $x_v=y|_{\operatorname{St}(v)}$ for $v\notin B$ and
$x_v=0$ for $v\in B$.
If neither endpoint of an edge is in $B$, its two arrays are
restrictions of the same word $y$ and agree on its chamber star.
Thus $\delta^0x$ is supported on edges incident to $B$, giving
$|\supp(\delta^0x)|\le n_\Sigma|B|$.
Repair to $x'\in\ker\delta^0$, and let
$A=\supp(x-x')$.
Then $|A|\le c_{\mathrm{rep}}n_\Sigma|B|$.
By \Cref{lem:chamber-cocycles}, there is $y'\in C$ whose
vertex restrictions are the entries of $x'$.

If $y(q)\ne y'(q)$, every vertex of the chamber $q$ lies in
$A\cup B$. Indeed, at a vertex outside this union the two
chamber values belong to equal vertex restrictions.
Write $D=\set{q\in Q:y(q)\ne y'(q)}$.
By \Cref{lem:cubical-incidence}, each chamber has $2^d$ vertices
and each vertex belongs to $N_*$ chambers. Counting incidences gives
$2^d|D|\le N_*|A\cup B|$ and $2^d|Q|=N_*|V|$.
Consequently,
\[
  \dist(y,C)\le\frac{|D|}{|Q|}
    \le\frac{|A\cup B|}{|V|}
    \le(1+c_{\mathrm{rep}}n_\Sigma)\frac{|B|}{|V|}.
\]
This is the required rejection bound.
\end{proof}

\subsection{Distance and Translated Checks}
\label{subsec:chamber-distance-checks}

\begin{lemma}\label[lemma]{lem:chamber-distance}
Suppose every vertex chart has factor distances at least
$\theta\in(0,1]$, and suppose \eqref{eq:chamber-mixing} holds
with $\lambda<\theta$ and $\beta>0$.
Every nonzero $y\in C$ has
$\weight(y)\ge\theta^d(\theta-\lambda)/\beta$.
\end{lemma}

\begin{proof}
Let $S=\set{v\in V:y|_{\operatorname{St}(v)}\ne 0}$ and
put $p=|S|/|V|>0$.
Fix $v\in S$ and a direction $i$.
In the tensor chart at $v$, some line in direction $i$ is
nonzero. Its entries are a codeword of the $i$th factor,
so at least $\theta n_i$ of its entries are nonzero.
The corresponding direction-$i$ hyperplanes therefore have
nonzero restrictions of $y$, and the other endpoints of their
edges belong to $S$.
It follows that $(W\Ind_S)(v)\ge\theta$ for every $v\in S$.
Applying \eqref{eq:chamber-mixing} to $\Ind_S$ gives
$\theta p\le\langle\Ind_S,W\Ind_S\rangle
    \le\lambda p+\beta p^2$ and $p\ge\frac{\theta-\lambda}{\beta}$.
By \Cref{lem:tensor-codes}, every nonzero vertex restriction has
at least $\theta^dN_*$ nonzero chamber entries.
Count pairs of a vertex and a nonzero chamber containing it:
\[
  2^d|\supp(y)|
    =\sum_{v\in S}
      |\supp(y|_{\operatorname{St}(v)})|
    \ge\theta^dN_*|S|.
\]
Since $2^d|Q|=N_*|V|$, division gives
$\weight(y)\ge\theta^dp\ge\theta^d(\theta-\lambda)/\beta$.
\end{proof}

\begin{proof}[Proof of~\Cref{thm:chamber-distance-testability}]
The inequalities $n_\Sigma\ge n_{\min}$ and
$\kappa\le\theta$ give $\alpha\le\theta$.
Thus $\lambda<\alpha$ implies $\lambda<\theta$, and
\Cref{lem:chamber-distance} gives the distance bound.
The repair bound \eqref{eq:chamber-repair-bound} and
\Cref{lem:chamber-star-test} give soundness.
Perfect completeness follows from the definition of $C$,
and the test reads the $N_*$ coordinates of one vertex star.
\end{proof}

\begin{corollary}\label[corollary]{cor:chamber-translated-tester}
Under the hypotheses of \Cref{thm:chamber-distance-testability},
suppose a finite group $G$ acts on $X$ freely on vertices and
chambers and preserves $\mathcal F$.
Let $v_0=|V/G|$ and identify $Q=[b]\times G$ by choosing
representatives of its $b$ chamber orbits.
There is a list of $h=v_0N_*$ linear checks, each of width at
most $w=N_*$, whose translated tester satisfies:
\begin{itemize}
  \item[\textnormal{(C)}] If $y\in C$, the tester accepts with
  probability one.
  \item[\textnormal{(S)}] If $y\notin C$, the tester rejects
  with probability at least $\sigma\dist(y,C)$, where
  $\sigma:=1/(N_*(1+c_{\mathrm{rep}}n_\Sigma))$.
\end{itemize}
The list can be computed by linear algebra from the vertex
orbits, their chamber stars, and their stalk spaces.
\end{corollary}

\begin{proof}
Choose one vertex $\widehat v$ from each orbit.
Take a basis of the linear conditions defining
$\mathcal F_{\widehat v}$ inside
$\bbK^{\operatorname{St}(\widehat v)}$, and pad it with zero
conditions to length $N_*$.
Write these conditions as
$\set{\lambda_{\widehat v,r}}_{r\in[N_*]}$ and extend them to
checks on chamber words by
\[
  \chi_{\widehat v,r}(y)
    :=\lambda_{\widehat v,r}
      (y|_{\operatorname{St}(\widehat v)}).
\]
Their combined list has $v_0N_*$ entries and width at most
$N_*$.
The experiment chooses a uniform list entry and $g\in G$ and
tests $\chi_{\widehat v,r}(g\cdot y)=0$.
By equivariance, this is one of the defining conditions at
$g^{-1}\widehat v$.
Freeness on vertices implies that choosing an orbit uniformly
and then $g$ uniformly gives a uniform vertex of $V$.
Every failing vertex restriction violates at least one of its
$N_*$ conditions, while every codeword satisfies them all.
The rejection probability is therefore at least $1/N_*$ times
that of the vertex test.
\Cref{thm:chamber-distance-testability} gives the stated bound.
\end{proof}
\section{Arithmetic Covers with Directional Mixing}
\label{sec:arithmetic-covers}

We construct the covers and compatible tensor charts to which
\Cref{thm:chamber-distance-testability} will apply.
The principal cover supplies the group action on the chamber
coordinates. An intermediate cover will be used to construct
codewords with prescribed values on a chamber fibre.

Fix a sufficiently large integer $h$ divisible by four, and for $i \in [6]$ put
\begin{equation}\label{eq:arithmetic-parameters}
  \mu_i:=2h(64+i),\qquad s_i:=2^{\mu_i},\qquad
  n_i:=s_i+1,\qquad e_i:=\frac{s_i-1}{2^h+1},
  \qquad \ell:=\frac{2^h-4}{6}. 
\end{equation}
The numbers $e_i$ and $\ell$ are integers.
We take $h>407$ and $\ell\ge 2\cdot140\cdot2^{20}$.
The latter inequality leaves the grades needed for interpolation
below $\ell$.
Fix distinct monic irreducible polynomials
$\set{\mathfrak p_i}_{i\in[6]}$ in $\bbF_2[t]$, with
$\deg\mathfrak p_i=\mu_i$.
The construction in this section works for every such choice.
Write $\mathcal T=\prod_{i=1}^6T_i$, where $T_i$ is the
$n_i$-regular tree.

\begin{theorem}\label[theorem]{thm:arithmetic-covers}
There are a finite field $\bbK$ of characteristic two, a
type-preserving group $\Lambda$ acting freely on the vertices
of $\mathcal T$, and a finite quotient
$X_0=\Lambda\backslash\mathcal T$ with the following properties.
For every sufficiently large even $r$, every monic irreducible
$\mathfrak r\in\bbF_2[t]$ of degree $r$ gives covers
\[
  \widehat X_R\longrightarrow X_R\longrightarrow X_0,
  \qquad
  R=2^r,\qquad
  G_R=\operatorname{SL}_2(\bbF_R),\qquad
  U_R=\operatorname{Stab}_{G_R}(e_1).
\]
\begin{itemize}
  \item[\textnormal{(1)}]
  The cover $\widehat X_R\to X_0$ is regular with deck group
  $G_R$, and $X_R=U_R\backslash\widehat X_R$.
  Above each cell of $X_0$, the two fibres are indexed by
  $G_R$ and $U_R\backslash G_R$, respectively.
  Their sizes are $R(R^2-1)$ and $R^2-1$.
  \item[\textnormal{(2)}]
  There is a multiplicative family
  $\set{\mathcal F^a}_{a=0}^{2^h+1}$ on $X_0$ and its covers.
  It has compatible tensor charts whose direction-$i$ factor
  at grade $a$ is a permutation of
  $\operatorname{PRS}_{s_i}(ae_i)$.
  \item[\textnormal{(3)}]
  There is $\beta>0$ such that every convex combination $W$
  of the six directional vertex walks on either cover satisfies
  \[
    \langle f,Wf\rangle
      \le\lambda\mathbb E(f^2)+\beta(\mathbb E f)^2
      \quad(f\ge0),
    \qquad
    \lambda:=\max_i\frac{2\sqrt{s_i}}{s_i+1}.
  \]
  The measure is uniform on vertices, and $\beta$ is independent
  of $\mathfrak r$.
  \item[\textnormal{(4)}]
  For every fixed $L$, all sufficiently large $r$ have the
  property that no nonidentity element of either covering
  subgroup moves a vertex of $\mathcal T$ by distance at most
  $L$. In particular, vertex stars in both covers embed.
  \item[\textnormal{(5)}]
  The cells, incidences, group action, and stalk matrices of
  the covers can be computed deterministically in time
  polynomial in $R$.
  There is a fixed integer $k_0$ such that, when $2k_0\mid r$,
  one can also compute a symmetric generating set $T$ of $G_R$
  of size $2^{k_0}+1$ and, for any permutation of $G_R$, a
  $T$-schedule of length $O(\log|G_R|)$ in polynomial time.
\end{itemize}
For $a\in\set{\ell,2\ell}$, the chamber codes
$C_a(\widehat X_R)$ and $C_a(X_R)$ have positive relative
distance and local-test soundness independent of $R$.
\end{theorem}

\subsection{The Quaternion Lattice and Local Codes}
\label{subsec:quaternion-lattice}

Put $\bbF=\bbF_2(t)$ and $\bbE=\bbF_4(t)$.
Let $\sigma$ square the coefficients in $\bbF_4$ and fix $t$.
Consider the algebra
$D=\bbE\oplus\bbE u$, where $u^2=t$ and $ua=\sigma(a)u$ for all $a\in\bbE$.
For $a,b,c,d\in\bbE$, its multiplication and reduced norm are
\[
  (a+bu)(c+du)
    =ac+tb\sigma(d)+(ad+b\sigma(c))u,
  \qquad
  \operatorname{Nrd}(a+bu)=a\sigma(a)+tb\sigma(b).
\]
Conjugation sends $a+bu$ to $\sigma(a)+bu$, so
$\operatorname{Trd}(a+bu)=a+\sigma(a)$.
The two terms in the norm have even and odd valuations at
$t=0$, respectively. They cannot cancel.
Thus every nonzero element has nonzero norm and is invertible.
An element commuting with both $\bbE$ and $u$ belongs to
$\bbF$, so $D$ is a quaternion division algebra over $\bbF$.

Let $\mathcal O_D$ be the skew polynomial ring consisting of
the finite sums $\sum_{k\ge0}a_ku^k$, with $a_k\in\bbF_4$
and $ua=a^2u$.
Its centre is $\bbF_2[t]$, and its fraction algebra is $D$.
For a finite place $\mathfrak p$, write
$D_{\mathfrak p}=D\otimes_{\bbF}\bbF_{\mathfrak p}$ and
$\mathcal O_{D,\mathfrak p}
 =\mathcal O_D\otimes_{\bbF_2[t]}\mathcal O_{\mathfrak p}$.
Since $\mu_i$ is even,
$\bbF_4\otimes_{\bbF_2}\mathcal O_{\mathfrak p_i}
 \simeq\mathcal O_{\mathfrak p_i}\times
       \mathcal O_{\mathfrak p_i}$.
The element $u$ exchanges the two idempotents and satisfies
$u^2=t\in\mathcal O_{\mathfrak p_i}^{\times}$.
Consequently,
$\mathcal O_{D,\mathfrak p_i}
    \simeq M_2(\mathcal O_{\mathfrak p_i})$ and 
$D_{\mathfrak p_i}\simeq M_2(\bbF_{\mathfrak p_i})$.
Use these identifications to realize $T_i$ as the tree of
homothety classes of rank-two
$\mathcal O_{\mathfrak p_i}$-lattices in
$\bbF_{\mathfrak p_i}^2$.
The neighbors of a vertex correspond to lines in its residue
plane, giving the degree $s_i+1$.
Let $v_{0,i}$ be the standard lattice vertex and put
$v_0=\{v_{0,i}\}_{i\in[6]}$.
Choose its type to be zero.

Write $\mathsf G=\operatorname{PGL}_1(D)$, so
$\mathsf G(\bbF)=D^\times/\bbF^\times$.
Let $\Gamma$ consist of the classes with representatives
$\gamma$ satisfying
$\gamma\in
    \bbF_{\mathfrak p}^{\times}
      \mathcal O_{D,\mathfrak p}^{\times}
  \quad\text{for every finite }\mathfrak p
    \notin\set{t,\mathfrak p_1,\ldots,\mathfrak p_6}$.
This is the quaternionic construction of
\cite[\S2]{RSV19}. We give the lattice argument for the
fixed places used here.

\begin{proposition}\label[proposition]{prop:quaternion-lattice}
The group $\Gamma$ is transitive on the vertices of
$\mathcal T$, with
$\operatorname{Stab}_{\Gamma}(v_0)
 \simeq\bbF_4^\times\rtimes\bbZ/2$ of order six.
There is a homomorphism
$\operatorname{lt}_0:\Gamma\to
 \bbF_4^\times\rtimes\bbZ/2$
whose restriction to this stabilizer is an isomorphism.
Its kernel $\Lambda_1$ acts simply transitively on vertices.
The subgroup $\Lambda\le\Lambda_1$ preserving all six types
and having leading coefficient one at infinity has finite
index and acts freely on cells.
\end{proposition}

\begin{proof}
Represent a vertex of $\mathcal T$ by lattices
$L_i=g_i\mathcal O_{\mathfrak p_i}^2$.
The matrices with columns in $L_i$ form the right
$M_2(\mathcal O_{\mathfrak p_i})$-module
$g_iM_2(\mathcal O_{\mathfrak p_i})$.
There is a fractional right $\mathcal O_D$-ideal $I$ with
these localizations at $\mathfrak p_i$ and with
$I_{\mathfrak p}=\mathcal O_{D,\mathfrak p}$ at the other
finite places.
To see this directly, multiply the prescribed modules by
a central polynomial so that each lies between
$b\mathcal O_{D,\mathfrak p}$ and
$\mathcal O_{D,\mathfrak p}$ for a common nonzero polynomial
$b$. The Chinese remainder decomposition of
$\mathcal O_D/b\mathcal O_D$ assembles their images into a
right submodule. Its inverse image, followed by undoing the
central scaling, has the prescribed localizations.

Every right ideal of $\mathcal O_D$ is principal.
Indeed, choose a nonzero element of least $u$-degree.
The skew division algorithm cancels the leading term of any
other element by a right multiple of it. The remainder lies
in the ideal and has smaller degree, so it vanishes.
Clearing a central denominator gives the same assertion for
fractional right ideals.
Thus $I=\gamma\mathcal O_D$ for some $\gamma\in D^\times$.
Its class belongs to $\Gamma$, and
$\gamma M_2(\mathcal O_{\mathfrak p_i})
 =g_iM_2(\mathcal O_{\mathfrak p_i})$
shows that it sends $v_0$ to the prescribed vertex.

Now let $\gamma$ represent an element fixing $v_0$.
At every finite place other than $t$, it is a unit up to a
scalar. Multiplying by a product of powers of the corresponding
irreducible polynomials makes it a unit at all these places.
It is then a unit of $\mathcal O_D[1/t]$.
In a product of skew Laurent polynomials, both the highest
and the lowest $u$-degrees add. A unit therefore has the
form $au^k$, with $a\in\bbF_4^\times$ and $k\in\bbZ$.
Conversely, every such element fixes $v_0$.
Quotienting by the central units $t^\bbZ=u^{2\bbZ}$
gives $\bbF_4^\times\rtimes\bbZ/2$.

The completion at $t=0$ consists of skew Laurent series.
For a nonzero series with lowest term $au^k$, put
$\operatorname{lt}_0(au^k+\cdots)=(a,k\bmod2)$.
Multiplication of lowest terms is the multiplication in
$\bbF_4^\times\rtimes\bbZ/2$, where the nontrivial element
of $\bbZ/2$ acts by squaring.
A scalar in $\bbF^\times$ has lowest coefficient one and
even $u$-degree. Hence this map is well-defined on projective
classes. On the stabilizer just computed it is an isomorphism.
It follows that
$\Gamma=\Lambda_1\operatorname{Stab}_{\Gamma}(v_0)$
and that $\Lambda_1$ has trivial vertex stabilizers.
Together with transitivity, this proves simple transitivity.

At infinity, write a nonzero quaternion as a skew Laurent
series with highest term $au^k$.
For $\gamma\in\Gamma$, the norm has even valuation at every
finite place outside $t,\mathfrak p_1,\ldots,\mathfrak p_6$.
Since all $\mu_i$ are even, the degree formula for the
principal divisor of $\operatorname{Nrd}(\gamma)$ gives
$\operatorname{ord}_{\infty}\operatorname{Nrd}(\gamma)
    \equiv
  \operatorname{ord}_{0}\operatorname{Nrd}(\gamma)
    \pmod2$.
For $\gamma\in\Lambda_1$, the right side is even.
Thus its highest $u$-degree at infinity is even.
Multiplication of leading coefficients on $\Lambda_1$ is
therefore ordinary multiplication in $\bbF_4^\times$.
Scalars again have leading coefficient one, giving a
homomorphism
$\operatorname{lc}_{\infty}:\Lambda_1\to\bbF_4^\times$.
The six type changes give a homomorphism
$\tau:\Lambda_1\to(\bbZ/2)^6$.
Set $\Lambda:=\ker\tau\cap\ker\operatorname{lc}_{\infty}
    \le\Lambda_1$.
Its index is at most $3\cdot2^6$, so its quotient is finite.
An element stabilizing a cell fixes its vertex of each type.
It therefore fixes a vertex and is the identity.
\end{proof}

Choose a finite field $\bbK$ containing every $\bbF_{s_i^2}$.
For a vector $x$, let $x^{[2^j]}$ mean entrywise Frobenius.

\begin{theorem}[{Adapted from
\cite[\S\S3.1.2 and~3.1.4, Propositions~3.3 and~3.4]{LLHL26}}]
\label[theorem]{thm:arithmetic-chart-input}
On $\mathcal T$, choose the lattice and chamber frames of
\cite[\S3.1.2]{LLHL26}.
For a vertex $v$ of type
$\{b_i(v)\}_{i\in[6]}$ and a chamber $Q\ge v$, these frames
give projective labels
$a_i(v,Q)\in\bbP^1(\bbF_{s_i})$ and nonzero scalars
$c_i(v,Q)\in\bbF_{s_i}^\times$.
Let $0\le d_i\le s_i-1$ satisfy
$(s_i-1)\mid(2^h+1)d_i$.
For every $v\le f$, the image of the map
\begin{equation}\label{eq:arithmetic-chart}
  \Phi_{f,v}(\psi)(Q)
   :=
   \left(\prod_{i=1}^6c_i(v,Q)^{2^{hb_i(v)}d_i}\right)
   \psi\left(
     \{r_{a_i(v,Q)}^{[2^{hb_i(v)}]}\}_{i\notin\dir(f)}
   \right)
\end{equation}
is independent of $v$.
Here $\psi$ is a polynomial over $\bbK$, homogeneous of degree
$d_i$ in the variable
pair for each $i\notin\dir(f)$.
The map is injective, and restriction from $f$ to a containing
face is obtained by evaluating the newly fixed variable pairs.
The resulting stalks form a chamber sheaf.
If two degree profiles and their sum satisfy these conditions,
the product of their chart values is the chart value of the
product polynomial at the summed profile.
The normalized changes of frame are unchanged by simultaneous
left multiplication of the two frames.
\end{theorem}

All the fixed algebraic constants needed later can also be
included in $\bbK$.
We explain the descent to $X_0$, since the group action on
chamber words will use these same charts on every cover.
Choose frames on representatives of the $\Lambda$-orbits
of vertices and chambers and extend them
$\Lambda$-equivariantly on $\mathcal T$.
The free action makes these choices consistent.
The scalar ambiguity in a projective matrix disappears in
the determinant-normalized change of frame.
Explicitly, if $U,V$ are frames of homothetic lattices, then
$[U:V]_i
    :=\det\!\left(\operatorname{red}_i(U^{-1}\xi V)\right)^{-1/2}
       \operatorname{red}_i(U^{-1}\xi V)$ and $[AU:AV]_i=[U:V]_i$,
where $\xi$ makes $U^{-1}\xi V$ integral and invertible.
The square root is unique in $\bbF_{s_i}$.
The equality follows by cancelling $A$ before reduction.
It gives
$a_i(\gamma v,\gamma Q)=a_i(v,Q)$ and
$c_i(\gamma v,\gamma Q)=c_i(v,Q)$.
Types are also preserved. Hence
$\Phi_{\gamma f,\gamma v}(\psi)(\gamma Q)
    =\Phi_{f,v}(\psi)(Q)$ for all $\gamma\in\Lambda$.
Thus the image stalks descend to $X_0$ and pull back to its
covers. A deck transformation acts by permuting chamber
coordinates, with no further scalar in that action.

At grade $a\in\set{0,\ldots,2^h+1}$, take $d_i=ae_i$.
The required divisibility holds by
\eqref{eq:arithmetic-parameters}.
In \eqref{eq:arithmetic-chart}, Frobenius permutes the
projective labels and does not act on the coefficients of
$\psi$, so the charts are $\bbK$-linear.
The multiplier at a vertex is the $a$-th power of
\[
  \mu_v(Q):=
    \prod_{i=1}^6c_i(v,Q)^{2^{hb_i(v)}e_i}.
\]
It is the same function restricted to every face through $v$.
Thus these are compatible tensor charts in the sense of
\Cref{def:compatible-tensor-charts}, and
\Cref{lem:graded-tensor-charts} proves multiplicativity.
Scalar normalizations depending only on $(f,v)$ do not
change the image stalks.

\subsection{Congruence Covers and One-Direction Graphs}
\label{subsec:congruence-components}

Let $\mathfrak r$ have even degree $r$ and be different from
$t,\mathfrak p_1,\ldots,\mathfrak p_6$, and put $R=2^r$.
The same splitting calculation as above gives
$\mathcal O_D/\mathfrak r\mathcal O_D\simeq M_2(\bbF_R)$.
Elements of $\Gamma$ have representatives that are units
at $\mathfrak r$, so reduction gives
\[
  \operatorname{red}_{\mathfrak r}:\Gamma
    \longrightarrow\operatorname{PGL}_2(\bbF_R)
    \simeq\operatorname{SL}_2(\bbF_R)=G_R.
\]
The last identification sends a projective matrix $A$ to
$\det(A)^{-1/2}A$.
Define
\begin{equation}\label{eq:arithmetic-covers}
  \begin{aligned}
  \Lambda(\mathfrak r)&:=\ker
    (\operatorname{red}_{\mathfrak r}|_\Lambda),
  &\widehat X_R&:=\Lambda(\mathfrak r)\backslash\mathcal T,\\
  \Lambda_U(\mathfrak r)&:=
    (\operatorname{red}_{\mathfrak r}|_\Lambda)^{-1}(U_R),
  &X_R&:=\Lambda_U(\mathfrak r)\backslash\mathcal T.
  \end{aligned}
\end{equation}
We next prove that the reduction of $\Lambda$ is all of
$G_R$ and that the directional components have fixed
positive vertex proportions.

Fix a direction $j$.
The components of the direction-$j$ graph on $\mathcal T$
are $T_j\times\{y\}$, where $y$ is a tuple of vertices
in the other factors.
Put $\Lambda_y=\operatorname{Stab}_\Lambda(y)$.
Its quotient $\Lambda_y\backslash T_j$ is a component of
the direction-$j$ graph on $X_0$.
There are finitely many such components.
For each $j$, choose one representative $y$ for each of them.

Let
$R_j:=\set*{z\in\bbF:
       \operatorname{ord}_{\mathfrak p}(z)\ge0
       \text{ for every place }\mathfrak p\ne\mathfrak p_j}$ and 
    $\Gamma_j:=\mathsf G(R_j)$,
where the integral group is defined by a fixed faithful
representation of $\mathsf G$.
We choose its integral model as in
\cite[(1.2)--(1.3)]{LSV05}.

\begin{lemma}\label[lemma]{lem:one-place-commensurability}
For each chosen $y$, the groups $\Lambda_y$ and $\Gamma_j$
are commensurable.
\end{lemma}

\begin{proof}
In each factor $i\ne j$, fixing $y_i$ imposes a compact
open condition on $\mathsf G(\bbF_{\mathfrak p_i})$.
Away from these places and from $t,\infty,\mathfrak p_j$,
the definition of $\Gamma$ imposes an integral compact
open condition.
At $t$ and at infinity the algebra is a local division
algebra, as follows from the same norm-valuation argument
used for $D$. Its projective multiplicative group is compact:
modulo central powers of the uniformizer, the valuation
has only two possibilities and the remaining units form
a compact group.

The integral model defining $\Gamma_j$ also imposes compact
open conditions at every place other than $\mathfrak p_j$.
The two systems agree outside a finite set.
At each exceptional place the intersection of two compact
open subgroups has finite index in both.
The product of these local indices bounds the index of
the rational-point intersection in each group.
Finally, imposing $\Lambda\le\Gamma$ changes the index by
at most the fixed number $[\Gamma:\Lambda]$.
This proves the claim.
\end{proof}

\begin{theorem}[{Adapted from
\cite[Theorem~1.2 and Proposition~1.5]{LSV05}
and \cite[Theorem~3.2]{BR17}}]
\label[theorem]{thm:congruence-ramanujan-input}
Let $D'$ be a quaternion division algebra over a global
function field $F'$, split at a place $\mathfrak p$ whose
residue field has size $q$.
Let $\Gamma'$ be the integral arithmetic group of
$\operatorname{PGL}_1(D')$ away from $\mathfrak p$.
If a principal congruence subgroup $\Gamma'(\mathfrak a)$
acts freely and without inversions on the Bruhat--Tits
tree at $\mathfrak p$, its finite quotient graph is
$(q+1)$-regular and its normalized adjacency eigenvalues lie in
\[
  \set{-1,1}\ \cup\
  \left[-\frac{2\sqrt q}{q+1},\frac{2\sqrt q}{q+1}\right].
\]
\end{theorem}

We also write $\mathfrak r$ for the corresponding prime
ideal of $R_j$.
By \Cref{lem:one-place-commensurability}, $\Gamma_j$ has
finitely many vertex and edge orbits, with finite stabilizers.
Choose representatives of these stabilizers.
Every nonidentity element is nonidentity modulo all but
finitely many primes in the faithful representation.
We can therefore choose a prime ideal $\mathfrak a_j$ of
$R_j$ such that $\Gamma_j(\mathfrak a_j)$ has trivial
vertex and edge stabilizers. Normality of the congruence
kernel excludes all their conjugates as well.
Let $\Gamma_j^c$ be its subgroup preserving the two colors
of $T_j$.
For $\mathfrak r$ outside a fixed finite set, put
\[
  Z_{j,\mathfrak r}:=
    (\Gamma_j^c\cap\ker\operatorname{red}_{\mathfrak r})
       \backslash T_j.
\]
Here and below the two reduction maps agree on common
domains after choosing the same local splitting.

\begin{lemma}\label[lemma]{lem:one-direction-ramanujan}
The graph $Z_{j,\mathfrak r}$ is connected, bipartite, and
$n_j$-regular. Its eigenvalues $1,-1$ are simple, and
all other normalized adjacency eigenvalues have absolute
value at most $\lambda_j:=2\sqrt{s_j}/(s_j+1)$.
\end{lemma}

\begin{proof}
The subgroup acts freely and preserves the colors, so the
quotient of the connected tree is connected, bipartite,
and $n_j$-regular.
For coprime levels,
$\Gamma_j(\mathfrak a_j)\cap
 \ker\operatorname{red}_{\mathfrak r}
 =\Gamma_j(\mathfrak a_j\mathfrak r)$.
Its quotient is Ramanujan by
\Cref{thm:congruence-ramanujan-input}.
Intersecting this subgroup with the color kernel either
leaves it unchanged or replaces its graph by the
bipartite double cover.
The eigenvalues of the double cover are the eigenvalues
of the original graph and their negatives.
This gives the asserted interval in both cases.
Connectedness makes the eigenvalue $1$ simple, and in a
connected bipartite graph the eigenvalue $-1$ is simple too.
\end{proof}

\begin{theorem}[{Adapted from \cite{Pra77}}]
\label[theorem]{thm:strong-approximation-input}
Let $H$ be an absolutely almost simple simply connected
group over a global function field $F'$.
If $H(F'_{\mathfrak p})$ is noncompact, then $H(F')$
is dense in the restricted product of $H(F'_{\mathfrak q})$
over $\mathfrak q\ne\mathfrak p$.
In particular, one may prescribe nonempty open conditions
at finitely many of these places and integral membership
at all the remaining places simultaneously.
\end{theorem}

The passage from one fixed finite-index subgroup to another
uses the following elementary fact.

\begin{lemma}\label[lemma]{lem:sl2-bounded-index}
Let $K$ be a finite field of characteristic two and let
$H\le\operatorname{SL}_2(K)$ have index at most $k$.
If $|K|>k!$, then $H=\operatorname{SL}_2(K)$.
\end{lemma}

\begin{proof}
Let $N$ be the kernel of the action on the cosets of $H$.
Then $N$ is normal and
$|\operatorname{SL}_2(K)/N|\le k!$.
The subgroup
$U=\set*{u(a):a\in K}$, where
$u(a)=\left(\begin{smallmatrix}1&a\\0&1\end{smallmatrix}\right)$,
has more than $k!$ elements.
Thus $u(a)\in N$ for some $a\ne0$.
For $t\in K^\times$,
\[
  \begin{pmatrix}t&0\\0&t^{-1}\end{pmatrix}
  u(a)
  \begin{pmatrix}t^{-1}&0\\0&t\end{pmatrix}
    =u(t^2a)\in N.
\]
Squaring permutes $K^\times$, so $U\le N$.
Conjugating by
$\left(\begin{smallmatrix}0&1\\1&0\end{smallmatrix}\right)$
also puts the lower unipotent subgroup in $N$.
These two subgroups generate $\operatorname{SL}_2(K)$
by elementary row operations.
Consequently $N=\operatorname{SL}_2(K)$ and $H=N$.
\end{proof}

\begin{lemma}\label[lemma]{lem:arithmetic-surjectivity}
For all sufficiently large even $r$, reduction at every
degree-$r$ prime $\mathfrak r$ maps each of
$\Gamma_j^c$,
$\Gamma_{j,y}:=\Gamma_j^c\cap\Lambda_y$,
$\Lambda_y$, and $\Lambda$ onto $G_R$.
The direction-$j$ graph of $\widehat X_R$ has exactly one
component above each component $\Lambda_y\backslash T_j$
of $X_0$. Its fraction of the vertices equals the fraction
occupied by that base component.
\end{lemma}

\begin{proof}
Apply \Cref{thm:strong-approximation-input} to
$H=\operatorname{SL}_1(D)$ and the omitted place
$\mathfrak p_j$.
The group is an inner form of $\operatorname{SL}_2$, so
it is absolutely almost simple and simply connected.
Since $D$ splits at $\mathfrak p_j$,
$H(\bbF_{\mathfrak p_j})\simeq
 \operatorname{SL}_2(\bbF_{\mathfrak p_j})$ is noncompact.

Given $g\in G_R$, prescribe reduction $g$ in the local
norm-one group at $\mathfrak r$.
At the finitely many exceptional places, including the
level $\mathfrak a_j$, prescribe a sufficiently small
neighborhood of the identity whose projective image
belongs to the required integral or congruence subgroup.
At all other places except $\mathfrak p_j$, prescribe
integral membership.
Strong approximation gives a rational norm-one element
meeting these conditions.
Its projective class belongs to
$\Gamma_j(\mathfrak a_j)$ and reduces to $g$.
Norm one also implies that it preserves the colors of
$T_j$. Thus $\operatorname{red}_{\mathfrak r}
(\Gamma_j^c)=G_R$.

By \Cref{lem:one-place-commensurability},
$k_{j,y}:=[\Gamma_j^c:\Gamma_{j,y}]$ is finite.
Surjectivity for $\Gamma_j^c$ gives
$[G_R:\operatorname{red}_{\mathfrak r}(\Gamma_{j,y})]
 \le k_{j,y}$.
For $R>k_{j,y}!$, \Cref{lem:sl2-bounded-index} makes this
image all of $G_R$.
There are only finitely many pairs $(j,y)$, so one
threshold works for all of them.
The inclusions
$\Gamma_{j,y}\le\Lambda_y\le\Lambda$
give the remaining surjectivity assertions.

Components above $\Lambda_y\backslash T_j$ are indexed by
$\Lambda(\mathfrak r)\backslash\Lambda/\Lambda_y
    \simeq G_R/\operatorname{red}_{\mathfrak r}(\Lambda_y)$.
This set has one element.
The component is
$(\Lambda_y\cap\ker\operatorname{red}_{\mathfrak r})
 \backslash T_j$ and has $|G_R|$ vertices above each base
vertex. The entire cover has the same number above
every base vertex, proving the proportion assertion.
\end{proof}

It follows that the covers in \eqref{eq:arithmetic-covers}
have the deck group and fibres claimed in
\Cref{thm:arithmetic-covers}.
In characteristic two,
\[
  U_R=\set*{\begin{pmatrix}1&z\\0&1\end{pmatrix}:z\in\bbF_R},
  \qquad
  U_R\backslash G_R\longrightarrow\bbF_R^2\setminus\set{0},
  \quad U_Rg\longmapsto g^{-1}e_1
\]
is a bijection.
Indeed, $G_R$ is transitive on nonzero vectors and $U_R$
is the stabilizer of $e_1$.
There are $R^2-1$ choices for a first column and $R$ choices
for a second column giving determinant one.
Hence $|G_R|=R(R^2-1)$ and $[G_R:U_R]=R^2-1$.

\subsection{Mixing under Finite Covers}
\label{subsec:mixing-finite-covers}

The directional components share covers of bounded degree
with the graphs $Z_{j,\mathfrak r}$.
The following lemma transfers the inequality needed for
local testability.
All graph measures in this subsection are uniform.
Graphs may have parallel edges, counted with multiplicity.

\begin{lemma}\label[lemma]{lem:nonnegative-mixing-covers}
Let $Z$ be a finite regular graph, and suppose
$\lambda,\beta\ge0$ satisfy
$\langle f,W_Zf\rangle
    \le\lambda\mathbb E(f^2)+\beta(\mathbb E f)^2$ for $f \ge 0$. 
Every $k$-sheeted graph cover $\widetilde Z\to Z$ satisfies
this inequality with parameters $\lambda,k\beta$.
If $Z$ covers a regular graph $Z'$, then $Z'$ satisfies
the original inequality with parameters $\lambda,\beta$.
\end{lemma}

\begin{proof}
Given $f\ge0$ on $\widetilde Z$, sort its values over
each vertex $z$ of $Z$ in decreasing order, and let
$f_a(z)$ be the $a$th value, for $a\in[k]$.
An oriented edge of $Z$ pairs the two endpoint fibres.
The sum of products over its lifted edges is at most
the sum obtained by pairing the two decreasing lists.
Indeed, replacing an inverted pair changes the sum by
$(x-x')(y-y')\ge0$.
Averaging over oriented edges gives
\[
  \langle f,W_{\widetilde Z}f\rangle
    \le\frac1k\sum_{a=1}^k\langle f_a,W_Zf_a\rangle.
\]
Sorting preserves the sum of squares and the sum of values:
\[
  \frac1k\sum_{a=1}^k\mathbb E(f_a^2)=\mathbb E(f^2),
  \qquad
  \sum_{a=1}^k\mathbb E f_a=k\mathbb E f.
\]
Because all the means are nonnegative,
$\sum_a(\mathbb E f_a)^2\le(\sum_a\mathbb E f_a)^2$.
Applying the hypothesis to each $f_a$ therefore gives
\[
  \langle f,W_{\widetilde Z}f\rangle
    \le\lambda\mathbb E(f^2)
      +\frac{\beta}{k}\sum_{a=1}^k(\mathbb E f_a)^2
    \le\lambda\mathbb E(f^2)+k\beta(\mathbb E f)^2.
\]
For the quotient assertion, pull a function on $Z'$ back
to $Z$. A graph cover preserves its mean, second moment,
and adjacency quadratic form.
\end{proof}

\begin{lemma}\label[lemma]{lem:mixing-components}
Let a regular graph have components with vertex proportions
$p_a\ge\omega>0$.
Suppose each component satisfies the preceding inequality
with parameters $\lambda,\beta_0$.
Then the whole graph satisfies it with
$\lambda,\beta_0/\omega$.
Convex combinations of walks satisfying the same inequality
also satisfy it.
\end{lemma}

\begin{proof}
For $f\ge0$, let $m_a$ be its mean on component $a$.
Summing the component inequalities leaves a mean term
$\beta_0\sum_a p_am_a^2$.
Since $p_a\ge\omega$ and all $m_a\ge0$,
\[
  \sum_a p_am_a^2
    =\sum_a\frac{(p_am_a)^2}{p_a}
    \le\frac1\omega\sum_a(p_am_a)^2
    \le\frac1\omega\left(\sum_a p_am_a\right)^2.
\]
The last sum is $\mathbb E f$.
The convex-combination assertion follows by linearity
in the walk operator.
\end{proof}

\begin{lemma}\label[lemma]{lem:arithmetic-directional-mixing}
There is $\beta>0$, independent of $\mathfrak r$, such that
every directional walk $W_j$ on $\widehat X_R$ or $X_R$
satisfies
$\langle f,W_jf\rangle
    \le\lambda_j\mathbb E(f^2)+\beta(\mathbb E f)^2$ for $f \ge 0$ and 
$\lambda_j=\frac{2\sqrt{s_j}}{s_j+1}$.
\end{lemma}

\begin{proof}
Decompose a function on $Z_{j,\mathfrak r}$ into its mean,
its component in the $-1$ eigenspace, and the orthogonal
remainder.
By \Cref{lem:one-direction-ramanujan}, its quadratic form
is at most
$\lambda_j\mathbb E(f^2)+(\mathbb E f)^2$.

For a chosen $y$, the graph
$Y_{j,y,\mathfrak r}:=
    (\Gamma_{j,y}\cap\ker\operatorname{red}_{\mathfrak r})
       \backslash T_j$
covers $Z_{j,\mathfrak r}$ with at most $k_{j,y}$ sheets,
and it covers the component
$(\Lambda_y\cap\ker\operatorname{red}_{\mathfrak r})
 \backslash T_j$ of $\widehat X_R$.
All the groups act freely and preserve colors, so these
are graph covers.
By \Cref{lem:nonnegative-mixing-covers}, this component
satisfies the required inequality with parameters
$\lambda_j,k_{j,y}$.

By \Cref{lem:arithmetic-surjectivity}, its vertex proportion
is that of $\Lambda_y\backslash T_j$ in $X_0$.
The minimum $\omega_j$ of these finitely many positive
proportions is independent of $\mathfrak r$.
\Cref{lem:mixing-components} gives the inequality on the
whole direction-$j$ graph, with
$\beta_j=\max_y k_{j,y}/\omega_j$.
Take $\beta=\max_j\beta_j$.
Finally, the direction-$j$ graph on $\widehat X_R$ covers
the one on $X_R$, so the quotient assertion of
\Cref{lem:nonnegative-mixing-covers} gives the same bound
on $X_R$.
\end{proof}

Taking $\lambda=\max_j\lambda_j$ and using
\Cref{lem:mixing-components} proves the assertion for every
convex combination.
In particular, the walk choosing an incident edge uniformly
is $\sum_j(n_j/n_\Sigma)W_j$, where $n_\Sigma=\sum_jn_j$.

\subsection{Injectivity Radius and Finite Data}
\label{subsec:arithmetic-injectivity}

Distances on the vertices of $\mathcal T$ are measured in
its one-skeleton, so they add across the six factors.

\begin{lemma}\label[lemma]{lem:arithmetic-injectivity}
For every $L$, there is $r_0$ such that, for every even
$r\ge r_0$ and every degree-$r$ prime $\mathfrak r$,
no nonidentity element of $\Lambda(\mathfrak r)$ or
$\Lambda_U(\mathfrak r)$ moves a vertex by distance
at most $L$.
\end{lemma}

\begin{proof}
Fix representatives of the finitely many $\Lambda$-orbits
of vertices.
For each representative, only finitely many elements of
$\Lambda$ move it by distance at most $L$.
The ball is finite, and freeness on vertices allows at
most one element for each image vertex.
Let $E_L$ be the union of these finite sets with the
identity removed.

Each $\delta\in E_L$ has nonidentity projective reduction
outside a finite set of primes.
Indeed, in a faithful rational matrix representation some
entry of $\delta-I$ is a nonzero rational function.
Its zeros and poles form a finite set.
Exclude these primes for every $\delta\in E_L$.
Normality of $\Lambda(\mathfrak r)$ now transfers the
conclusion from the chosen representatives to every vertex.

For the second subgroup, choose a quaternion representative
$d_\delta\in D^\times$ and put
$J(\delta):=
    \frac{\operatorname{Trd}(d_\delta)^2}
         {\operatorname{Nrd}(d_\delta)}$.
This rational function is unchanged by scaling the
representative or conjugating $\delta$.
It is nonzero.
Otherwise the reduced characteristic equation would give
$d_\delta^2\in\bbF^\times$, and the nonidentity projective
class $\delta$ would have order two.
An involution of a tree fixes the midpoint of the path
from a vertex to its image.
If that midpoint is an edge midpoint, the involution
exchanges its endpoints and reverses the bipartition.
Since $\delta$ preserves types, it therefore fixes a
vertex in each factor.
This gives a fixed vertex in $\mathcal T$, contradicting
the free action of $\Lambda$.

Exclude the zeros and poles of $J(\delta)$ for all
$\delta\in E_L$.
If a nonidentity $\gamma\in\Lambda_U(\mathfrak r)$ moved
some vertex by at most $L$, conjugating it by an element
of $\Lambda$ would produce $\delta\in E_L$.
The reduction of $\delta$ would belong to a conjugate
of $U_R$.
Every matrix in $U_R$ has trace zero in characteristic
two, so $J(\delta)$ would reduce to zero at
$\mathfrak r$, a contradiction.
All the excluded sets are finite, and hence their prime
degrees are bounded.
\end{proof}

A closed vertex star in $\mathcal T$ has vertex diameter
at most $12$.
If two of its cells had the same image in a quotient,
a group element would send a vertex of the first cell
to a vertex of the second, moving it by at most $12$.
\Cref{lem:arithmetic-injectivity} therefore gives the
embedded stars required by
\Cref{thm:chamber-distance-testability}.

\begin{lemma}\label[lemma]{lem:arithmetic-finite-data}
Once the fixed arithmetic data are chosen, the cells,
incidences, group action, and stalk matrices of
$\widehat X_R$ and $X_R$ can be computed deterministically
in time polynomial in $R$.
\end{lemma}

\begin{proof}
Fix representatives of the cells of $X_0$ in $\mathcal T$,
together with their face incidences.
Each incidence is specified by an element of $\Lambda$
carrying the chosen lift of a face to the corresponding
face of a chosen cell.
There are finitely many such elements.
Their quaternion coordinates, and the matrices of the
stalks on $X_0$, are part of the fixed data.

Given $\mathfrak r$, form
$\bbF_R=\bbF_2[t]/(\mathfrak r)$.
A splitting
$\mathcal O_D/\mathfrak r\mathcal O_D\simeq M_2(\bbF_R)$
can be found by enumerating the images of the generators
and checking their defining relations and linear
independence.
There are $R^{O(1)}$ choices, and the splitting calculation
above guarantees success.
Reduce the fixed incidence elements and normalize their
matrix representatives to determinant one.

For the principal cover, the lifts of each base cell are
indexed by $g\in G_R$, and its face incidences are
obtained by multiplying $g$ by the corresponding
reduced incidence elements.
For the intermediate cover, use $U_Rg$ in place of $g$.
Enumerating the matrices of $G_R$, their products, and
these cell incidences takes polynomial time in $R$.
The stalks are the pullbacks of fixed stalks under the
bijections of chamber stars.
Thus their matrices are obtained by reindexing fixed
matrices.
The deck action is left multiplication on the group
coordinate.
\end{proof}

\subsection{Routing and the Choice of Levels}
\label{subsec:arithmetic-routing}

We now construct generating sets for the deck groups on which every
permutation has a short routing schedule. The construction uses a
fixed-degree Cayley graph, independently of the six directions of
the cubical complex.

\begin{theorem}[{Adapted from
\cite[Theorem~5.13 and equation~(22)]{Mor94}}]
\label[theorem]{thm:morgenstern-input}
Let $q$ be a power of two, and choose $\varepsilon\in\bbF_q$ such
that $Z^2+Z+\varepsilon$ is irreducible over $\bbF_q$.
For every sufficiently large even integer $m$, let
$x\in\bbF_{q^m}$ have degree $m$ over $\bbF_q$, and let
$i\in\bbF_{q^m}$ satisfy $i^2+i+\varepsilon=0$.
For $\gamma,\delta\in\bbF_q$ with
$\gamma^2+\gamma\delta+\varepsilon\delta^2=1$, put
\begin{equation}\label{eq:morgenstern-generators}
  t_{\gamma,\delta}:=
  (1+x)^{-1/2}
  \begin{pmatrix}
    1&\gamma+\delta i+\delta\\
    (\gamma+\delta i)x&1
  \end{pmatrix}.
\end{equation}
These matrices form a symmetric generating set $T$ of size $q+1$
for $\mathrm{SL}_2(\bbF_{q^m})$.
The graph with edges $\set{g,gt}$, for
$g\in\mathrm{SL}_2(\bbF_{q^m})$ and $t\in T$, is simple,
connected, and nonbipartite. Its adjacency eigenvalue $q+1$ is
simple, and all other eigenvalues have absolute value at most
$2\sqrt q$.
\end{theorem}

The matrix before normalization in
\eqref{eq:morgenstern-generators} has determinant $1+x$ and square
$(1+x)I_2$. Thus the normalized generators have determinant one
and are involutions. The square root is unique in
$\bbF_{q^m}$. The equation on $\gamma,\delta$ is the norm-one
equation for $\bbF_{q^2}/\bbF_q$, so it has $q+1$ solutions.

For a finite graph, a round of routing swaps the endpoints of
every edge in a chosen matching and fixes the remaining vertices.

\begin{theorem}[{Adapted from
\cite[Theorem~1.1 and Section~4]{Nen23}
and \cite[Theorem~9]{ABC96}}]
\label[theorem]{thm:nenadov-input}
There are absolute constants $d_0$ and $C$ with the following
property. Let $H$ be a simple $d$-regular graph on $n$ vertices,
where $d\ge d_0$, whose adjacency eigenvalue $d$ is simple and
whose other adjacency eigenvalues have absolute value at most
$\lambda$, where $0<\lambda<d/72$.
Every permutation of $V(H)$ can be realized in at most
$C\frac{\log_2 n}{\log_2(d/\lambda)}$
rounds of routing.
For an involution $\pi$, such a routing can be computed
deterministically in time polynomial in $n$ if a partition
$V(H)=V_0\sqcup V_1$ is given such that
$\pi(V_j)=V_j$ for $j\in\set{0,1}$ and every vertex has at least
$d/3$ neighbors in each part.
\end{theorem}

We include the construction of the partition to make the routing
algorithm deterministic.

\begin{theorem}[{Adapted from \cite[Theorem~5]{CGH13}}]
\label[theorem]{thm:deterministic-lll-input}
Let $A_1,\ldots,A_M$ be events with specified sets of determining
variables among $N$ independent unbiased bits.
Suppose each event shares variables with at most
$\Delta$ other events, where $\Delta\ge 1$, and
$\Pr(A_j)\le \{e(\Delta+1)\}^{-3/2}$ for $j \in [M]$.
If the conditional probability of each event can be computed
exactly in time polynomial in $N+M$ under every partial assignment
of the bits, then a deterministic algorithm finds an assignment
avoiding all the events in time polynomial in $N+M+\Delta$.
\end{theorem}

To obtain this form, take $x(A_j)=1/(\Delta+1)$ and
$\epsilon=1/2$ in the source theorem. Its required inequality
follows from
\[
  x(A_j)\prod_{A_k\sim A_j}(1-x(A_k))
  \ge \frac{1}{\Delta+1}
       \left(\frac{\Delta}{\Delta+1}\right)^\Delta
  \ge \frac{1}{e(\Delta+1)}.
\]
The parameters in its running-time bound are polynomial in
$N+M+\Delta$.

\begin{lemma}\label[lemma]{lem:balanced-routing-partition}
For every fixed integer $b\ge 1$, there is an integer $d_b$ with
the following property. Let $H$ be a simple $d$-regular graph on
$n$ vertices, with $d\ge d_b$, and let $\mathcal B$ be a partition
of $V(H)$ into sets of size at most $b$.
A deterministic algorithm, in time polynomial in $n$, finds a
partition $V(H)=V_0\sqcup V_1$ such that each member of
$\mathcal B$ is contained in one part and every vertex has at
least $d/3$ neighbors in each part.
\end{lemma}

\begin{proof}
Assign each $B\in\mathcal B$ independently to $V_0$ or $V_1$
with equal probability, and write $\xi_B\in\set{0,1}$ for its
choice. For a vertex $v$, put
$w_{v,B}:=|N_H(v)\cap B|$. The number $S_v$ of its neighbors in
$V_1$ is
\[
  S_v=\sum_{B\in\mathcal B}w_{v,B}\xi_B,
  \qquad
  \sum_{B\in\mathcal B}w_{v,B}=d,
  \qquad
  \sum_{B\in\mathcal B}w_{v,B}^2\le bd.
\]
Let $A_v$ be the event that $S_v<d/3$ or $S_v>2d/3$.
For every $\theta>0$, independence and
$\cosh z\le\exp(z^2/2)$ give
\[
  \mathbb E\exp\bigl(\theta(S_v-d/2)\bigr)
  =\prod_{B\in\mathcal B}
       \cosh\bigl(\theta w_{v,B}/2\bigr)
  \le\exp\left(\frac{\theta^2}{8}
                  \sum_{B\in\mathcal B}w_{v,B}^2\right)
  \le\exp(\theta^2bd/8).
\]
Applying Markov's inequality with $\theta=2/(3b)$ bounds either
tail by $\exp(-d/(18b))$. Therefore
$\Pr(A_v)\le 2\exp(-d/(18b))$.
The event $A_v$ depends on at most $d$ bits.
A bit $\xi_B$ can affect $A_w$ only when $w$ has a neighbor in
$B$, so it affects at most $bd$ events. Thus every $A_v$ shares
variables with at most $bd^2$ other events. Choose $d_b$ large
enough that, for every $d\ge d_b$,
\[
  2\exp(-d/(18b))
    \le\{e(bd^2+1)\}^{-3/2}.
\]

It remains to compute conditional probabilities. After some bits
have been fixed, their contribution to $S_v$ is known. For the
remaining bits on which $A_v$ depends, the coefficient of $Z^a$
in
\[
  \prod_{B:\,\xi_B\text{ is unfixed},\ w_{v,B}>0}
        (1+Z^{w_{v,B}})
\]
counts the assignments contributing $a$ to $S_v$.
This polynomial has degree at most $d$, and its coefficients can
be computed by successive multiplication in time polynomial in
$d$. Summing the coefficients corresponding to $A_v$ and
dividing by the number of assignments gives its conditional
probability exactly. The coefficients have at most $d+1$ bits.
\Cref{thm:deterministic-lll-input} now supplies an assignment
avoiding all $A_v$, which gives the required partition.
\end{proof}

\begin{proposition}\label[proposition]{prop:deterministic-spectral-routing}
There are absolute constants $d_*$ and $C_*$ such that the
following holds. Let $H$ be a simple $d$-regular graph on $n$
vertices, with $d\ge d_*$, whose adjacency eigenvalue $d$ is
simple and whose other adjacency eigenvalues have absolute value
at most $\lambda$, where $0<\lambda<d/72$.
Given any permutation of $V(H)$, a deterministic algorithm
constructs at most
$C_*\frac{\log_2 n}{\log_2(d/\lambda)}$
rounds of routing realizing it, in time polynomial in $n$.
\end{proposition}

\begin{proof}
Every permutation is the composition of two involutions, which
can be found from its cycles. On a cycle
$\set{v_j}_{j\in\mathbb Z/k\mathbb Z}$ on which
$\pi(v_j)=v_{j+1}$, define
$\alpha(v_j)=v_{-j}$ and $\beta(v_j)=v_{1-j}$.
Then $\alpha^2=\beta^2=\id$ and
$\pi=\beta\circ\alpha$.
Applying these definitions on all cycles gives the two
involutions on $V(H)$.
Choose $d_*\ge\max\set{d_0,d_2}$, where $d_0$ is from
\Cref{thm:nenadov-input} and $d_2$ is from
\Cref{lem:balanced-routing-partition}.
For each involution, apply that lemma to the partition into its
orbits, whose sizes are at most two. The resulting partition
satisfies the algorithmic hypotheses of
\Cref{thm:nenadov-input}. Compute the two routing schemes and
concatenate them. Their total length has the asserted bound
with $C_*=2C$.
\end{proof}

\begin{lemma}\label[lemma]{lem:arithmetic-routing}
There is an integer $k_0\ge 1$ such that, for every sufficiently
large $r$ divisible by $2k_0$, the group
$G_R=\mathrm{SL}_2(\bbF_R)$, where $R=2^r$, has a symmetric
generating set $T_R\subseteq G_R\setminus\set{1}$ of size
$2^{k_0}+1$ on which every permutation of $G_R$ has a
$T_R$-schedule of length $O(\log|G_R|)$ in the sense of
\Cref{def:routing-schedule}.
The generating set and a schedule for any given permutation
can be constructed deterministically in time polynomial in $R$.
\end{lemma}

\begin{proof}
Choose $k_0$ so that $q:=2^{k_0}$ satisfies
$q+1\ge d_*$ and $2\sqrt q<(q+1)/72$,
where $d_*$ is from
\Cref{prop:deterministic-spectral-routing}.
Write $r=k_0m$, so $m$ is even.
\Cref{thm:morgenstern-input} gives a generating set $T_R$ of
size $q+1$ whose right Cayley graph on $G_R$ satisfies the
hypotheses of \Cref{prop:deterministic-spectral-routing}, with
$d=q+1$ and $\lambda=2\sqrt q$.
The denominator $\log_2((q+1)/(2\sqrt q))$ is a positive
constant, so that proposition gives $O(\log|G_R|)$ rounds.

The schedules of \Cref{def:routing-schedule} use left
multiplication. The bijection $\iota(g)=g^{-1}$ takes a right
Cayley edge to
$\iota\bigl(\set{g,gt}\bigr)
    =\set{g^{-1},t^{-1}g^{-1}}$,
which is a left Cayley edge because $T_R$ is symmetric.
For a desired permutation $\pi$, route
$\iota\circ\pi\circ\iota^{-1}$ on the right Cayley graph and
apply $\iota$ to each matching. This gives a $T_R$-schedule
realizing $\pi$.

For the construction, the subfield $\bbF_q\subseteq\bbF_R$
is the set of roots of $Z^q-Z$ and can be enumerated by
searching $\bbF_R$.
Choose $\varepsilon\in\bbF_q$ for which
$Z^2+Z+\varepsilon$ has no root in $\bbF_q$, and find a root
$i\in\bbF_R$ by search. Such a root exists because $m$ is
even. Find $x\in\bbF_R$ of degree $m$ over $\bbF_q$ by
enumeration, testing the length of its orbit under
$z\mapsto z^q$.
These searches and the evaluation of
\eqref{eq:morgenstern-generators} take time polynomial in $R$.
The group has $R(R^2-1)$ elements, so its Cayley graph and the
deterministic routing algorithm can also be constructed and
run in time polynomial in $R$.
\end{proof}

We henceforth restrict to sufficiently large exponents
$r\in 2k_0\mathbb N$. For consecutive such exponents,
\[
  \frac{|G_{2^{r+2k_0}}|}{|G_{2^r}|}
    =2^{2k_0}\frac{2^{2r+4k_0}-1}{2^{2r}-1}
    <2^{6k_0+1}.
\]
Thus the restriction needed for routing preserves a bounded
ratio between consecutive group orders.

\begin{corollary}\label[corollary]{cor:arithmetic-chamber-codes}
Put $b=|X_0(6)|$, $v_*=|X_0(0)|$,
$N_*=\prod_i n_i$, and $n_\Sigma=\sum_i n_i$.
For every sufficiently large level considered above, the
codes
$C_\ell(\widehat X_R),\ C_{2\ell}(\widehat X_R)
     \le\bbK^{[b]\times G_R}$
are $G_R$-invariant and satisfy
$C_\ell(\widehat X_R)*C_\ell(\widehat X_R)
 \le C_{2\ell}(\widehat X_R)$.
Both have relative distance at least a fixed $\delta>0$.
Each has a translated tester with $v_*N_*$ checks of width
at most $N_*$, perfect completeness, and rejection probability
at least $\sigma\dist(y,C_a(\widehat X_R))$ for
$y\notin C_a(\widehat X_R)$, where $\sigma>0$ is fixed.
The check lists are computable in time polynomial in $R$.
\end{corollary}

\begin{proof}
The group coordinates and product inclusion follow from
\Cref{thm:multiplicative-chamber-codes} and the charts
constructed above.
For $a\in\set{\ell,2\ell}$,
\[
  ae_i\le 2\ell e_i
    =\frac{2^h-4}{3(2^h+1)}(s_i-1)
    <\frac{s_i-1}{3}.
\]
By \Cref{lem:prs-properties}, the factor distances are
greater than $2/3$, so we may take $\theta=1/2$.
Let $\rho=\max_i n_i/\min_i n_i$.
The parameters of
\Cref{thm:chamber-distance-testability} obey
\[
  \rho\le2^{10h},\qquad
  \alpha=\frac{\min_i n_i}{96\rho n_\Sigma}
      \ge\frac1{576\rho^2}
      \ge\frac{2^{-20h}}{576},
  \qquad
  \lambda\le2^{1-65h}<\frac{\alpha}{2}.
\]
The last inequality holds for our choice of $h$.
By \Cref{lem:arithmetic-directional-mixing,lem:arithmetic-injectivity},
the hypotheses of
\Cref{thm:chamber-distance-testability} hold on both covers.
In particular, put
\[
  c=\frac{\alpha-\lambda}{\beta},\qquad
  c_{\mathrm{rep}}=\max\set{1,2/c},\qquad
  \delta=\frac{2^{-6}(1/2-\lambda)}{\beta},\qquad
  \sigma=\frac1{N_*(1+c_{\mathrm{rep}}n_\Sigma)}.
\]
These are positive constants.
The distance bound and
\Cref{cor:chamber-translated-tester} give the claims.
The construction of the lists is linear algebra on the
data supplied by \Cref{lem:arithmetic-finite-data}.
\end{proof}

\begin{proof}[Proof of~\Cref{thm:arithmetic-covers}]
\Cref{prop:quaternion-lattice} constructs $\Lambda$ and
the fixed finite quotient.
The charts following \Cref{thm:arithmetic-chart-input}
give the multiplicative family.
\Cref{lem:arithmetic-surjectivity} gives the full deck
group and the two fibre descriptions.
\Cref{lem:arithmetic-directional-mixing,lem:mixing-components}
give the asserted mixing inequality.
\Cref{lem:arithmetic-injectivity} gives the displacement
and embedded-star assertions.
The algorithms are supplied by
\Cref{lem:arithmetic-finite-data,lem:arithmetic-routing}.
Finally, the parameter calculation in the preceding proof
applies on both $\widehat X_R$ and $X_R$.
\Cref{thm:chamber-distance-testability} therefore gives
positive relative distance and local-test soundness
independent of $R$ for the two specified grades on
both covers.
\end{proof}

\section{Large Restriction Spaces from Determinant Sections}
\label{sec:determinant-restrictions}

We construct many independent functions on a chamber fibre of the
coset cover $X_R$. The functions arise from determinants of a matrix
whose entries are Frobenius powers of the two fibre coordinates.
The degree of a determinant section is independent of the size of
its minor.

Retain the covers and codes of \Cref{sec:arithmetic-covers}.
For a chamber $q$ of $X_0$, choose a lift $\widehat q$ to
$\widehat X_R$. Its fibre in $X_R$ is indexed by
$U_R\backslash G_R$, which we identify with
$\Omega:=\bbF_R^2\setminus\set{0}$ by
$U_Rg\mapsto g^{-1}e_1$.
Write $z|_q$ for restriction to this fibre and
$\res_q(\widehat z)(g):=\widehat z(g\widehat q)$ on the
principal cover. Thus pullback satisfies
\begin{equation}\label{eq:coset-fibre-pullback}
  \res_q(\widehat z)(g)=(z|_q)(g^{-1}e_1).
\end{equation}
For an extension $\bbL$ of $\bbK$, put
$C_{a,\bbL}(Y):=C_a(Y)\otimes_{\bbK}\bbL$.

\begin{theorem}\label[theorem]{thm:large-fibre-restriction}
There are a chamber $q$ of $X_0$ and a codeword
$A\in C_1(X_0)$ with $A(q)\ne0$ such that, for every
sufficiently large even $r$, there is a monic irreducible
$\mathfrak r\in\bbF_2[t]$ of degree $r$ for which the following
holds. Put $R=2^r$ and let $\bbL=\bbK\bbF_R$ be the compositum.
The restriction space
\[
  W:=\set*{z|_q:z\in C_{140,\bbL}(X_R)}\le \bbL^\Omega
\]
contains the constants, separates the points of $\Omega$, and
satisfies $\dim_{\bbL} W\ge2^{-17}R^2$.
The pullback of $A$ to $X_R$ has constant restriction
$A(q)\Ind_\Omega$.
\end{theorem}

\subsection{The Determinant Sections}
\label{subsec:determinant-sections}

Fix $\omega\in\bbF_4\setminus\bbF_2$ and put
$S_0=\left(\begin{smallmatrix}1&\omega\\1&\omega^2\end{smallmatrix}\right)$,
so $\det S_0=1$.
Take $q$ to be the chamber with labels
$\set{\infty,\ldots,\infty}$ at the type-zero origin.
For $\zeta\in\bbF_R$ of degree $r$ over $\bbF_2$, take
$\mathfrak r$ to be its minimal polynomial and identify the
residue of $t$ with $\zeta^4$.
For $0\le j<r$, put
$\tau_j:=\zeta^{2^{j+1}}$ and
$\xi_j:=(x^{2^j},y^{2^j})^{\mathsf T}$.
The nodes $\tau_j$ are nonzero and distinct.

We also use the frame and covariance formulas in
\cite[equations~(3.129), (3.131)--(3.138)]{LLHL26}.

\begin{theorem}[{Adapted from
\cite[Lemma~3.12 and Proposition~3.13]{LLHL26}}]
\label[theorem]{thm:determinant-section-input}
There are $A\in C_1(X_0)$ and $K_0\in C_3(X_0)$ with
$A(q)K_0(q)\ne0$, a polynomial $P_{\mathrm{ref}}\in\bbK[z]$
with $P_{\mathrm{ref}}(0)\ne0$, and a polynomial matrix
$B(t)\in\bbK[t]^{2\times2}$ with
$\det B(z^2)=P_{\mathrm{ref}}(z)^2$.
Put $D(z):=\operatorname{diag}(1,z^{-1})$ and
\[
  T_0(z):=
  \frac{D(z)B(z^2)D(z)^{-1}S_0}{P_{\mathrm{ref}}(z)}.
\]
These data are independent of $r$.
If $P_{\mathrm{ref}}(\tau_j)\ne0$ for all $j$, define
$(U_j,V_j)^{\mathsf T}:=T_0(\tau_j)\xi_j$ and the symmetric
matrix $K$ by
\begin{equation}\label{eq:determinant-matrix}
  K_{jj}:=U_jV_j,\qquad
  K_{ij}:=c_{ij}(U_iV_j+V_iU_j),\qquad
  c_{ij}:=\frac{\sqrt{\tau_i\tau_j}}{\tau_i+\tau_j}
  \quad(i\ne j).
\end{equation}
There are codewords $f_j\in C_{1,\bbL}(X_R)$,
$t_j\in C_{3,\bbL}(X_R)$, and
$D_{I,J}\in C_{8,\bbL}(X_R)$ for all equal-size subsets
$I,J\subseteq\set{0,\ldots,r-1}$, with restrictions
\[
  f_j|_q=A(q)U_j,\qquad
  t_j|_q=K_0(q)V_j,\qquad
  D_{I,J}|_q=(A(q)K_0(q))^2\det K[I,J].
\]
Here the forms on the right are evaluated on $\Omega$, and
the empty minor is one.
\end{theorem}

We explain the specialization to $q$.
The Moore determinant defining $A$ in
\cite[equations~(3.45), (3.49)]{LLHL26} becomes, at the
infinity labels, the Vandermonde determinant of the distinct
fixed square-root nodes. Its chart scalars are nonzero, so
$A(q)\ne0$.
The identity in \cite[equation~(3.171)]{LLHL26} then gives an
auxiliary section of grade three that is nonzero at $q$, which
we call $K_0$.
Both sections descend to $X_0$.
The normalization function $\chi$ of
\cite[Lemma~3.6]{LLHL26} is $\Lambda$-invariant, since
$\Lambda\subseteq\ker(\operatorname{lc}_\infty)$.
The empty auxiliary primitive $A$ is invariant by
\cite[equation~(3.129)]{LLHL26}.
For $K_0$, work temporarily on a cover recording the fixed
auxiliary level.
The squared row covariant and the adjacent Frobenius row
covariant in its defining contraction transform by the same
determinant-one Frobenius twist
\cite[equations~(3.129), (3.133)]{LLHL26}.
Their alternating pairing is invariant, so $K_0$ descends as well.
Consequently, both sections have constant restrictions on
the fibres over $q$.
The covariance formula makes the input at the lift
$U_Rg$ equal to $(g^{-1}e_1)^{[2^j]}$, giving the displayed
restrictions. The polynomial matrix $B$ absorbs the fixed
matrices in the source frame.

\begin{lemma}\label[lemma]{lem:determinant-grade-budget}
The space $W$ contains the constants and separates $\Omega$.
It also contains the restriction of every product of at most
seven minors of $K$ and at most twenty-eight coordinates
from $\set{U_j,V_j}_{0\le j<r}$.
\end{lemma}

\begin{proof}
The corresponding product of sections has grade at most
$7\cdot8+28\cdot3=140$.
Multiply it by a power of $A$ to reach grade $140$ and
divide by its nonzero constant factor on the fibre.
Multiplicativity gives the claim.
The empty product gives one, while $U_0,V_0$ determine
$(x,y)$ because $\det T_0(\tau_0)=1$.
\end{proof}

\subsection{A Frame Regular at Zero}
\label{subsec:regular-determinant-frame}

Put $L_0:=P_{\mathrm{ref}}(0)B(0)^{-1}
  =(\lambda_{ab})_{a,b\in[2]}$,
$H(z):=D(z)L_0D(z)^{-1}$, and $T(z):=H(z)T_0(z)$.
The determinant identity gives $\det L_0=1$.
Since $L_0B(z^2)=P_{\mathrm{ref}}(0)I_2+O(z^2)$,
conjugating by $D(z)$ gives
$D(z)L_0B(z^2)D(z)^{-1}
  =P_{\mathrm{ref}}(0)I_2+O(z)$.
Thus $T$ is regular at zero, $T(0)=S_0$, and $\det T=1$.
Its degrees are bounded independently of $r$.

\begin{lemma}\label[lemma]{lem:minor-rank-perturbation}
Let $M,N$ be matrices of the same size over a field, with
$\rank N\le k$.
Every minor of order $p$ of $M+N$ is a sum, with signs, of
products of a minor of order $p-j$ of $M$ and a minor of
order $j$ of $N$, where $0\le j\le\min\set{k,p}$.
\end{lemma}

\begin{proof}
Expand the determinant by multilinearity in its rows.
In each term, expand along the rows selected from $N$.
This gives the asserted products, and the terms with $j>k$
vanish.
\end{proof}

\begin{lemma}\label[lemma]{lem:normalized-determinant-sections}
Let $\widetilde K$ be defined by
\eqref{eq:determinant-matrix} using the spinors
$T(\tau_j)\xi_j$.
Every product of at most seven minors of $\widetilde K$,
restricted to $\Omega$, belongs to $W$.
\end{lemma}

\begin{proof}
For the original spinors, set
$a_j:=U_j/\sqrt{\tau_j}$ and $b_j:=\sqrt{\tau_j}V_j$.
Then $H(\tau_j)$ acts on $(a_j,b_j)^{\mathsf T}$ by $L_0$,
and
\[
  K_{jj}=a_jb_j,\qquad
  K_{ij}=\frac{\tau_i a_ib_j+\tau_j a_jb_i}{\tau_i+\tau_j}.
\]
Substitution, using
$\lambda_{11}\lambda_{22}+\lambda_{12}\lambda_{21}=1$,
gives $\widetilde K=K+XCX^{\mathsf T}$, where
\[
  X_{j,*}=(a_j,b_j),\qquad
  C=
  \begin{pmatrix}
    \lambda_{11}\lambda_{21}&\lambda_{12}\lambda_{21}\\
    \lambda_{12}\lambda_{21}&\lambda_{12}\lambda_{22}
  \end{pmatrix}.
\]
This identity also holds on the diagonal.
The correction has rank at most two.
Its minors of order at most two are combinations over $\bbL$
of products of at most four original spinor coordinates.
By \Cref{lem:minor-rank-perturbation}, a product of seven
minors of $\widetilde K$ is a combination of products of at
most seven minors of $K$ and twenty-eight coordinates.
Apply \Cref{lem:determinant-grade-budget}.
\end{proof}

Henceforth $K$ and $(U_j,V_j)^{\mathsf T}$ refer to the
normalized frame $T$.

\subsection{Independent Forms on One Frobenius Block}
\label{subsec:one-frobenius-block}

The rank argument in the remainder of this section adapts
\cite[\S3.3.1, equations~(3.191)--(3.208)]{LLHL26}
to the fibre of the fixed chamber $q$.
We use the regular frame above and a direct valuation
calculation.

For now, $\zeta$ is an indeterminate and the coefficients
lie in $\bbK(\zeta)$.
Use $\tau_j=\zeta^{2^{j+1}}$ for all integers $j\ge0$.
Write $\operatorname{ord}_\zeta$ for the order of vanishing
at zero.
For a finite index set $I$, put
$\langle I\rangle:=\sum_{i\in I}2^i$.
The minor $\det K[I,J]$ is homogeneous of degree
$\langle I\rangle+\langle J\rangle$.
For $n\ge0$, let $I_n$ be its binary support and put
$P_n:=\det K[I_n,I_n]$ and
$Q_n:=\det K[I_n,I_n+1]$.
Their degrees are $2n$ and $3n$.
Also put $N(x,y):=x^2+xy+y^2=(x+\omega y)(x+\omega^2y)$.

\begin{lemma}\label[lemma]{lem:minor-reduction}
The coefficients of $P_n$ and $\zeta^{-n}Q_n$ are regular
at $\zeta=0$.
Modulo $\zeta$, the former reduces to $N(x,y)^n$ and the
latter takes the value one at $(x,y)=(\omega,1)$.
\end{lemma}

\begin{proof}
The spinor $T(\tau_j)\xi_j$ reduces to $S_0\xi_j$, so
$U_jV_j$ reduces to $N(x,y)^{2^j}$.
For $i\ne j$,
\[
  c_{ij}=
  \frac{\zeta^{2^i+2^j}}{(\zeta^{2^i}+\zeta^{2^j})^2},
  \qquad
  \operatorname{ord}_\zeta(c_{ij})=|2^i-2^j|,
\]
with leading coefficient one.
The off-diagonal entries vanish modulo $\zeta$, proving
the assertion for $P_n$.

For $Q_n$, multiply row $i$ by $\zeta^{2^i}$ and column
$j$ by $\zeta^{-2^j}$.
The resulting determinant is $\zeta^{-n}Q_n$.
Every off-diagonal scalar has order
$|2^i-2^j|+2^i-2^j\ge0$, and the diagonal is unchanged,
so all entries are regular.
At $(\omega,1)$, the reduced spinor is $(0,1)^{\mathsf T}$
for even $i$ and $(1,0)^{\mathsf T}$ for odd $i$.
Order the rows by $i\in I_n$ and the columns by
$j=k+1$, $k\in I_n$.
If $k<i$, then $j\le i$.
For $j<i$ the scalar has positive order, while for $j=i$
the entry reduces to $N(\omega,1)^{2^i}=0$.
For $k=i$, the scalar has leading coefficient one and the
two spinors are the distinct standard basis vectors.
The reduced matrix is therefore upper triangular with
diagonal entries one.
\end{proof}

We use the following form of the leading-term argument.

\begin{lemma}\label[lemma]{lem:valuation-independence}
Let $\bbF$ be a field and $p_1,\ldots,p_m\in \bbF(z)[x]$.
Suppose there are integers $a_i$ such that $z^{a_i}p_i$
has coefficients regular at zero, with reductions
$\overline p_i\ne0$.
If the orders of vanishing of $\overline p_i$ at some
$u\in \bbF$ are pairwise distinct, then the $p_i$ are
linearly independent over $\bbF(z)$.
\end{lemma}

\begin{proof}
The reductions are independent over $\bbF$, because the
smallest order in a nonzero combination cannot cancel.
A relation over $\bbF(z)$ between the normalized polynomials
can be scaled so that its coefficients are regular and at
least one is nonzero modulo $z$.
Reduction would give a nonzero relation between the
$\overline p_i$, a contradiction.
\end{proof}

Fix $m_0\ge1$, put $M:=2^{m_0}$, and for $0\le t<M$ set
$E_t:=P_{6M-6-6t}Q_{4t+2}$ and $d_{\mathrm{blk}}:=12M-6$.
These forms have degree $d_{\mathrm{blk}}$ and use only
indices $0,\ldots,m_0+2$.

\begin{lemma}\label[lemma]{lem:one-block-independence}
The forms $\set{E_t}_{0\le t<M}$ are linearly independent
over $\bbK(\zeta)$.
\end{lemma}

\begin{proof}
By \Cref{lem:minor-reduction}, the reduction of
$\zeta^{-(4t+2)}E_t(x,1)$ has order exactly
$6M-6-6t$ at $x=\omega$.
These orders are distinct, since $\omega$ is a simple
root of $N(x,1)$.
Apply \Cref{lem:valuation-independence}.
\end{proof}

\subsection{Specialization and Three Blocks}
\label{subsec:determinant-specialization}

Take $m_0:=\lfloor(r-13)/3\rfloor$ and $M:=2^{m_0}$, so
$R/2^{15}\le M^3\le R/2^{13}$.
Choose $d_T\in\bbK[z]\setminus\set{0}$ such that $d_TT$
has polynomial entries.
The degrees of $d_T$ and these entries are fixed.

\begin{lemma}\label[lemma]{lem:determinant-specialization-bound}
There is a nonzero polynomial $H_T\in\bbK[\zeta]$ of degree
$O(rM^2)$ such that, whenever $H_T(\zeta)\ne0$, the forms
$\set{E_t}_{0\le t<M}$ are defined and independent after
specialization.
The same bound holds for every coefficient Frobenius twist
of $T$.
\end{lemma}

\begin{proof}
Use the common denominator
\[
  D_T(\zeta):=
  \prod_{j=0}^{m_0+2}d_T(\tau_j)^4
  \prod_{0\le i<j\le m_0+2}(\tau_i+\tau_j)^8.
\]
In a determinant term, each index occurs at most once as a
row and once as a column.
Thus a term of $E_t$ uses at most four spinor coordinates
with any index and at most four factors $c_{ij}$ for any
unordered pair.
Their numerators are polynomial in $\zeta$, including
$\sqrt{\tau_i\tau_j}=\zeta^{2^i+2^j}$.
Consequently $D_TE_t(x,1)$ has polynomial coefficients.
The sums $\sum_{j\le m_0+2}2^j=O(M)$ and
$\sum_{i<j\le m_0+2}2^j=O(rM)$ show that $D_T$ and all
these coefficients have degree $O(rM)$.

By \Cref{lem:one-block-independence}, the coefficient
matrix of the $E_t(x,1)$ has a nonzero $M\times M$ minor.
Multiply each row by $D_T$ and take that minor.
Its determinant is a nonzero polynomial of degree
$O(rM^2)$.
Its product with $D_T$ gives $H_T$.
Frobenius twists preserve degrees, and they send $S_0$ to
itself or to the matrix with its rows exchanged.
Exchanging $U_j,V_j$ for every $j$ leaves $K$ unchanged,
so the same independence argument applies to every twist.
\end{proof}

Let $\operatorname{Fr}$ be squaring on $\bbK$.
For $s\in\set{0,m_0+4,2m_0+8}$, write $T^{(-s)}$ for
the frame obtained by applying $\operatorname{Fr}^{-s}$
to its coefficients.
Let $E_t^{(s)}$ be the product defining $E_t$ with every
row and column index increased by $s$.
Since $\tau_{j+s}=\tau_j^{2^s}$, we have
\begin{equation}\label{eq:twisted-blocks}
  E_t^{(s)}
   =\bigl(E_t\text{ formed with }T^{(-s)}\bigr)^{2^s}.
\end{equation}
All indices used are at most $3m_0+10\le r-3$.
The coefficient twist in \eqref{eq:twisted-blocks} accounts
for the coefficients of $T$ that are not fixed by
$\operatorname{Fr}^s$.

\begin{lemma}\label[lemma]{lem:separated-digit-products}
Let $\bbF$ be a field, $B\ge2$ an integer, and
$V_j\le \bbF[x^{B^j},y^{B^j}]$ a space of homogeneous
polynomials of degree $a_jB^j$, where $0\le a_j<B$.
Multiplication from $\bigotimes_{j=0}^{t-1}V_j$ to
$\bbF[x,y]$ is injective.
\end{lemma}

\begin{proof}
For the ambient monomial spaces, a product has $x$-exponent
$\sum_j\alpha_jB^j$, with $0\le\alpha_j\le a_j<B$.
Its base-$B$ digits determine every $\alpha_j$.
Distinct basis tensors therefore give distinct monomials.
Restricting this injective map to the stated tensor product
preserves injectivity.
\end{proof}

\begin{lemma}\label[lemma]{lem:good-determinant-level}
For every sufficiently large even $r$, there is
$\zeta\in\bbF_R$ of degree $r$ such that
$P_{\mathrm{ref}}(\tau_j)\ne0$ for all $j<r$ and the
$M^3$ forms
\[
  E_{t_0}^{(0)}E_{t_1}^{(m_0+4)}E_{t_2}^{(2m_0+8)}
\]
are linearly independent over $\bbL$, for $0\le t_0,t_1,t_2<M$.
Their common degree $d_0$ is less than $3R/8$.
\end{lemma}

\begin{proof}
For each of the three twisted frames, exclude the roots in
$\bbF_R$ of the polynomial supplied by
\Cref{lem:determinant-specialization-bound}.
This excludes $O(rM^2)=o(R)$ elements, since
$M^2\le R^{2/3}$.
The conditions $P_{\mathrm{ref}}(\zeta^{2^{j+1}})\ne0$
exclude $O(r)$ further elements, since each Frobenius map
permutes $\bbF_R$.
The union of its proper subfields has size less than
$\sum_{d=1}^{r/2}2^d<2\sqrt R$.
Some degree-$r$ element therefore avoids all excluded sets.
Its nodes are nonzero and distinct.

By \eqref{eq:twisted-blocks} and perfection of $\bbL$, each
shifted block spans a space of dimension $M$.
Put $B:=16M$, so the three blocks lie in
$\bbL[x^{B^j},y^{B^j}]$ and have degrees
$d_{\mathrm{blk}}B^j$ for $j\in\set{0,1,2}$.
Since $d_{\mathrm{blk}}<B$,
\Cref{lem:separated-digit-products} gives independence of
their products.
Finally,
\[
  d_0=(12M-6)(1+16M+256M^2)
       <3072M^3\le3R/8.
\]
\end{proof}

\subsection{Principal Minors and the Dimension Bound}
\label{subsec:principal-minors}

The remaining rank gain comes from principal minors of
different degrees.
We use the argument of
\cite[equations~(3.206)--(3.208)]{LLHL26};
the same proof applies to arbitrary index sets and
invertible frames.

\begin{lemma}\label[lemma]{lem:principal-minor-nonvanishing}
Let $\bbF$ be a field of characteristic two, let
$J\subseteq\bbZ_{\ge0}$ be finite, and choose distinct
nonzero squares $\tau_j\in \bbF$ and matrices $T_j\in\operatorname{GL}_2(\bbF)$.
Put $(U_j,V_j)^{\mathsf T}:=T_j(x^{2^j},y^{2^j})^{\mathsf T}$
and form $K[J,J]$ by \eqref{eq:determinant-matrix}.
Its determinant is a nonzero homogeneous polynomial of
degree $2\sum_{j\in J}2^j$.
\end{lemma}

\begin{proof}
The empty case is immediate.
Put $\phi(z):=\prod_{j\in J}(z+\tau_j)$.
On polynomials of degree less than $|J|$, the operator
$z\,d/dz$ is idempotent in characteristic two.
In the basis of evaluations at the $\tau_j$, Lagrange
interpolation gives its off-diagonal entry
\[
  \frac{\tau_i\phi'(\tau_i)}
       {(\tau_i+\tau_j)\phi'(\tau_j)}.
\]
Conjugate by the diagonal matrix with entries
$\sqrt{\tau_j}\phi'(\tau_j)$.
The resulting matrix $\Pi$ is idempotent with off-diagonal
entries $c_{ij}$, and is therefore symmetric.
Put $\Pi':=I+\Pi$, and let $\mathcal U,\mathcal V$ be the
diagonal matrices of $U_j,V_j$.
Then
\[
  K[J,J]
   =(\Pi'\mathcal U+\Pi\mathcal V)^{\mathsf T}
      (\Pi\mathcal U+\Pi'\mathcal V).
\]
Indeed, $\Pi\Pi'=0$, so the product is
$\mathcal U\Pi'\mathcal V+\mathcal V\Pi\mathcal U$.
Its off-diagonal entries are those of $K$, and its
diagonal entries are $U_jV_j$.

Regard the $U_j,V_j$ first as independent variables.
Each factor determinant is homogeneous of degree one in
each pair, and is one when every pair equals $(1,1)$.
Both are nonzero, and remain so after the invertible
substitutions $(U_j,V_j)^{\mathsf T}=T_j(X_j,Y_j)^{\mathsf T}$.
The substitution $(X_j,Y_j)=(x^{2^j},y^{2^j})$ is injective
on this space: its monomials have $x$-exponents
$\sum_{j\in J'}2^j$ for distinct subsets $J'\subseteq J$.
Thus both factors remain nonzero in $\bbF[x,y]$, as does
their product. Its degree is $2\sum_{j\in J}2^j$.
\end{proof}

\begin{proof}[Proof of~\Cref{thm:large-fibre-restriction}]
Choose $\zeta$ as in \Cref{lem:good-determinant-level} and
let $\mathfrak r$ be its minimal polynomial.
The determinant sections are defined at every node.
Their grade-one section $A$ and
\Cref{lem:determinant-grade-budget} give the remaining
claims except for the dimension bound.

Let $W_0\le \bbL[x,y]$ be the span of the three-block products.
It has dimension $M^3$ and consists of forms of degree
$d_0<3R/8$.
For $J\subseteq\set{0,\ldots,r-3}$, put
$P_J:=\det K[J,J]$.
By \Cref{lem:principal-minor-nonvanishing}, multiplication
by $P_J$ is injective and its image $P_JW_0$ has degree
$d_0+2\langle J\rangle$.
These degrees are distinct for distinct $J$.
Consequently the sum
\[
  \mathcal V:=
   \bigoplus_{J\subseteq\set{0,\ldots,r-3}}P_JW_0
  \quad\text{has dimension}\quad
  \frac R4M^3\ge2^{-17}R^2.
\]
Every polynomial in $\mathcal V$ has total degree less than
$3R/8+R/2<R$ and vanishes at the origin.
Evaluation on $\Omega$ is injective on this space.
Indeed, vanishing on $\Omega$ implies vanishing on
$\bbF_R^2$. For each fixed $y\in\bbF_R$, a polynomial in
$x$ of degree less than $R$ with $R$ roots is zero.
Its coefficient polynomials in $y$, also of degree less
than $R$, then vanish at all $R$ values and are zero.

Each generator of $\mathcal V$ is a product of seven
minors of the normalized matrix $K$.
Its restriction belongs to $W$ by
\Cref{lem:normalized-determinant-sections}.
The injectivity just proved gives
$\dim_{\bbL}W\ge\dim_{\bbL}\mathcal V\ge2^{-17}R^2$.
\end{proof}

\section{Interpolation and Equivariant Encoding}
\label{sec:fibre-interpolation}

The restriction space constructed in \Cref{sec:determinant-restrictions}
has dimension proportional to the size of the coset fibre.
We use products to obtain many point indicators on that fibre.
Two translated indicators then isolate a group element in the
principal fibre, giving an equivariant encoder.

Keep the arithmetic data and the chamber $q\in X_0(6)$ of
\Cref{thm:large-fibre-restriction} fixed.
Write $b=|X_0(6)|$, and choose the group coordinates on
$\widehat X_R(6)$ with the lifts of $q$ as the first fibre.
Thus $\res_q:\bbK^{\widehat X_R(6)}\to\bbK^{G_R}$ is the
map $\res$ of \Cref{subsec:group-actions}.

\begin{theorem}\label[theorem]{thm:fibre-interpolation}
For every sufficiently large even $r$, there is a monic
irreducible $\mathfrak r\in\bbF_2[t]$ of degree $r$ such that,
with $R=2^r$, there are codewords
\[
  \eta_\ell\in C_\ell(\widehat X_R),\qquad
  \eta_{2\ell}:=\eta_\ell\odot\eta_\ell
     \in C_{2\ell}(\widehat X_R),\qquad
  \res_q(\eta_\ell)=\res_q(\eta_{2\ell})=\delta_{1_{G_R}}.
\]
For $a\in\set{\ell,2\ell}$, the map
$\Enc_a:\bbK^{G_R}\to C_a(\widehat X_R)$ given by
\[
  \Enc_a(f):=\sum_{g\in G_R}f(g)(g\cdot\eta_a)
\]
is $G_R$-equivariant and satisfies $\res_q\circ\Enc_a=\id$.
Both codes have dimension at least $|G_R|$ and rate at least
$1/b$.
For a selected prime $\mathfrak r$, the codewords and the
matrices of these encoders can be computed in time polynomial
in $R$ from the cover and stalk matrices.
\end{theorem}

\subsection{Products and Linear Descent}
\label{subsec:indicator-products}

For a field $\bbL$, a finite set $\Omega$, and a subspace
$W\le\bbL^\Omega$, for $s \ge 1$ let
\[
  W^{\langle s\rangle}:=
    \linspan_\bbL\set*{f_1\odot\cdots\odot f_s:
      f_1,\ldots,f_s\in W}. 
\]
We say that $W$ separates points if, for any distinct
$u,v\in\Omega$, some $f\in W$ has $f(u)\ne f(v)$.

\begin{theorem}[{Adapted from \cite[Lemma~3.16]{LLHL26}}]
\label[theorem]{thm:bounded-product-indicators}
Let $\bbL$ be a field, let $\Omega$ be a finite nonempty set,
and let $W\le\bbL^\Omega$ contain $\Ind_\Omega$ and separate
points. For every integer $s\ge2$, the space
$W^{\langle s\rangle}$ contains the indicators of at least
\[
  \dim_\bbL W-\frac{2(|\Omega|-\dim_\bbL W)}{s}
\]
points of $\Omega$.
\end{theorem}

We will obtain the indicators over an extension of $\bbK$.
The following observation allows their preimages to be chosen
over $\bbK$.

\begin{lemma}\label[lemma]{lem:linear-descent}
Let $\bbL/\bbK$ be a field extension, let
$H\in\bbK^{m\times n}$, and let $b\in\bbK^m$.
If $Hx=b$ has a solution in $\bbL^n$, it has a solution in
$\bbK^n$.
\end{lemma}

\begin{proof}
Row-reduce the augmented matrix $[H\mid b]$ over $\bbK$.
An inconsistent row would remain inconsistent over $\bbL$, so
none occurs. Setting the free variables to zero and solving
for the pivot variables gives a solution over $\bbK$.
\end{proof}

Put $s_*=2^{20}$ and $d_{\mathrm{ind}}=140s_*$.
The choice of $\ell$ gives $2d_{\mathrm{ind}}\le\ell$.
For a prime supplied by \Cref{thm:large-fibre-restriction},
identify the fibre of $X_R$ over $q$ with
$\Omega=\bbF_R^2\setminus\set{0}$, and let
\[
  S:=\set*{u\in\Omega:
      \text{there is }c_u\in C_{d_{\mathrm{ind}}}(X_R)
      \text{ with }c_u|_q=\delta_u}.
\]

\begin{lemma}\label[lemma]{lem:dense-fibre-indicators}
The set $S$ has size at least $2^{-18}R^2$.
\end{lemma}

\begin{proof}
Let $\bbL$ and $W\le\bbL^\Omega$ be as in
\Cref{thm:large-fibre-restriction}.
The space $W$ contains the constants, separates points, and
has dimension at least $2^{-17}R^2$.
By \Cref{thm:bounded-product-indicators},
$W^{\langle s_*\rangle}$ contains at least
\[
  2^{-17}R^2-\frac{2R^2}{2^{20}}
    \ge 2^{-18}R^2
\]
point indicators.
Every product of $s_*$ elements of $W$ lifts to a product
of $s_*$ codewords in $C_{140,\bbL}(X_R)$.
Multiplicativity places that product in
$C_{d_{\mathrm{ind}},\bbL}(X_R)$.
Taking linear combinations shows that each of the indicated
point indicators has a preimage in this code.

Stalk membership and the condition $c_u|_q=\delta_u$ are
linear equations whose coefficients and right-hand sides
belong to $\bbK$.
By \Cref{lem:linear-descent}, a preimage exists over
$\bbK$ whenever one exists over $\bbL$.
Thus all these points belong to $S$.
\end{proof}

\subsection{Separating Group Elements}
\label{subsec:separating-group-elements}

The passage from a coset indicator to a group indicator uses
tuples with trivial pointwise stabilizer.
The following statement applies to any finite group action.

\begin{lemma}\label[lemma]{lem:stabilizer-free-indicators}
Let a finite group $G$ act on a finite set $\Omega$, and let
$\set{D_a\le\bbK^{[b]\times G}}_{a=0}^A$ be $G$-invariant
linear codes over a field $\bbK$, with
$D_a*D_{a'}\le D_{a+a'}$ whenever $a+a'\le A$.
Let $\res$ restrict to the first fibre.
Let $t\ge1$. Suppose $a_1,\ldots,a_t\in\Omega$ have trivial pointwise
stabilizer, and let $u_j=g_0^{-1}a_j$ for some $g_0\in G$.
If $F_j\in D_{d_j}$, with integers $d_j\ge0$ and
$d:=\sum_{j=1}^t d_j\le A$, satisfy
$\res(F_j)(g)=\Ind_{\set{u_j}}(g^{-1}a_j)$ for $g \in G$ and $j \in [t]$,
then $F:=F_1\odot\cdots\odot F_t$ belongs to $D_d$ and
\[
  \res(F)=\delta_{g_0},\qquad
  \res(g_0^{-1}\cdot F)=\delta_{1_G}.
\]
\end{lemma}

\begin{proof}
Multiplicativity gives $F\in D_d$.
Its restriction is one precisely when
$g^{-1}a_j=g_0^{-1}a_j$ for every $j$.
Then $gg_0^{-1}$ fixes every $a_j$, so $g=g_0$.
At every other group element at least one factor vanishes.
Finally, invariance gives $g_0^{-1}\cdot F\in D_d$, and
the left action sends the indicator at $g_0$ to that at $1_G$.
\end{proof}

For $\operatorname{SL}_2(\bbF_R)$ acting on nonzero vectors,
the two standard basis vectors have trivial pointwise stabilizer.
The next lemma finds a pair in the required orbit inside any
sufficiently large set. It holds in every characteristic.

\begin{lemma}\label[lemma]{lem:determinant-incidences}
Let $\bbF_R$ be a finite field, let $a\in\bbF_R^\times$,
and put $\Omega=\bbF_R^2\setminus\set{0}$.
For $S,T\subseteq\Omega$, let
$E_a(S,T):=|\set{(u,v)\in S\times T:\det(u,v)=a}|$.
Then
\[
  \left|E_a(S,T)-\frac{R|S||T|}{R^2-1}\right|
    \le\sqrt{R|S||T|}.
\]
In particular, if $|S|>R^{3/2}$, there are $u,v\in S$
with $\det(u,v)=a$.
\end{lemma}

\begin{proof}
Let $N$ be the real matrix indexed by $\Omega$ with
$N_{uv}=1$ when $\det(u,v)=a$ and $N_{uv}=0$ otherwise.
Each row and each column has $R$ ones.
Two rows indexed by nonproportional vectors have exactly one
common one, because their two determinant equations form an
invertible linear system in $v$.
Distinct proportional vectors have no common one, since
$a\ne0$.
Consequently,
\[
  NN^{\mathsf T}=RI+J-B,\qquad
  B:=\bigoplus_{L\in\bbP^1(\bbF_R)}J_{R-1},
\]
where $J$ is the all-ones matrix and each block of $B$ is
indexed by the nonzero vectors on a line $L$.

The space of real functions on $\Omega$ is the orthogonal
sum of the constants, the functions constant on each line
and summing to zero, and the functions summing to zero on
each line.
On these three spaces, $NN^{\mathsf T}$ acts by $R^2$, $1$,
and $R$, respectively.
Thus $\|N^{\mathsf T}f\|_2\le\sqrt R\,\|f\|_2$ for
$f$ orthogonal to the constants, using the counting inner
product and norm.

Write $\Ind_S=p\Ind_\Omega+f$ and
$\Ind_T=q\Ind_\Omega+g$, where
$p=|S|/(R^2-1)$, $q=|T|/(R^2-1)$, and $f,g$ sum to zero.
The row and column sums eliminate the cross terms, giving
\[
  E_a(S,T)-\frac{R|S||T|}{R^2-1}
    =\langle N^{\mathsf T}f,g\rangle.
\]
Cauchy--Schwarz bounds its absolute value by
$\sqrt R\,\|f\|_2\|g\|_2\le\sqrt{R|S||T|}$.
Taking $T=S$ yields
\[
  E_a(S,S)\ge\frac{R|S|^2}{R^2-1}-\sqrt R\,|S|>0
\]
when $|S|>R^{3/2}$, proving the last assertion.
\end{proof}

\subsection{Interpolation on the Principal Fibre}
\label{subsec:principal-fibre-interpolation}

\begin{proof}[Proof of~\Cref{thm:fibre-interpolation}]
Choose the prime at degree $r$ supplied by
\Cref{thm:large-fibre-restriction}, with $R>2^{36}$.
By \Cref{lem:dense-fibre-indicators},
$|S|\ge2^{-18}R^2>R^{3/2}$.
Apply \Cref{lem:determinant-incidences} with $a=1$ to obtain
$u_1,u_2\in S$ with $\det(u_1,u_2)=1$.
Choose $c_{u_1},c_{u_2}\in C_{d_{\mathrm{ind}}}(X_R)$
with the prescribed coset-fibre indicators, and let
$\widehat c_1,\widehat c_2$ be their pullbacks to
$\widehat X_R$.
By \eqref{eq:coset-fibre-pullback},
$\res_q(\widehat c_j)(g)
    =\Ind_{\set{u_j}}(g^{-1}e_1)$ for $j\in\set{1,2}$.
Put $e_2=(0,1)^{\mathsf T}$ and
$s_0=\left(\begin{smallmatrix}0&1\\1&0\end{smallmatrix}\right)$.
Since the characteristic is two, $s_0\in G_R$ and
$s_0e_1=e_2$.
Using the left action from \Cref{subsec:group-actions},
\[
  \res_q(s_0\cdot\widehat c_2)(g)
    =\res_q(\widehat c_2)(s_0^{-1}g)
    =\Ind_{\set{u_2}}(g^{-1}e_2).
\]
There is $g_0\in G_R$ with inverse matrix having columns
$u_1,u_2$.
By \Cref{lem:stabilizer-free-indicators},
\[
  F:=g_0^{-1}\cdot
       \bigl(\widehat c_1\odot(s_0\cdot\widehat c_2)\bigr)
    \in C_{2d_{\mathrm{ind}}}(\widehat X_R),
  \qquad \res_q(F)=\delta_{1_{G_R}}.
\]

Let $\widehat A$ be the pullback of the grade-one section
from \Cref{thm:large-fibre-restriction}.
Its restriction to the principal fibre is the nonzero
constant $A(q)\in\bbK$.
The parameter choice \eqref{eq:arithmetic-parameters} gives
$2d_{\mathrm{ind}}\le\ell$.
Define the pointwise products
\[
  \eta_\ell:=
    \left(\frac{\widehat A}{A(q)}\right)^{\ell-2d_{\mathrm{ind}}}
       \odot F,
  \qquad
  \eta_{2\ell}:=\eta_\ell\odot\eta_\ell.
\]
Multiplicativity puts these words in the asserted grades.
The padding factor is one on the fibre, so
$\res_q(\eta_\ell)=\delta_{1_{G_R}}$.
Squaring preserves this indicator, giving the same restriction
for $\eta_{2\ell}$.
All grades used are at most $2\ell<2^h+1$.
The encoder formulas and the rate bounds now follow from
\Cref{lem:equivariant-encoding}.

For the computational assertion, use the stalk matrices to
write the linear system defining $C_\ell(\widehat X_R)$,
and add the equations prescribing its first-fibre values
to be $\delta_{1_{G_R}}$.
The construction above proves that this system is consistent.
It has polynomially many rows and columns in $R$, so row
reduction over the fixed field $\bbK$ finds $\eta_\ell$ in
polynomial time.
Squaring gives $\eta_{2\ell}$, and translating the two words
gives the columns of the two encoders.
\end{proof}

\section{Explicit Equivariant Multiplication Codes and PCPs}
\label{sec:explicit-codes-pcps}

We combine the chamber codes with fibre interpolation to obtain
equivariant multiplication pairs. The construction of a member
reduces to finite-field computations and linear algebra.

\begin{theorem}\label[theorem]{thm:explicit-multiplication-codes}
There are a finite field $\bbK$ of characteristic two, integers
$\ell,b,w,h_{\mathrm{chk}},k_0\ge1$, and constants
$\Delta,\sigma\in(0,1]$ with the following property.
For every sufficiently large $r\in2k_0\mathbb N$, put
$R=2^r$ and $G_R=\operatorname{SL}_2(\bbF_R)$.
There are linear codes
$C_\ell,C_{2\ell}\le\bbK^{[b]\times G_R}$ such that:
\begin{itemize}
  \item[\textnormal{(1)}]
  Both codes are $G_R$-invariant and have relative distance
  at least $\Delta$, and $C_\ell*C_\ell\le C_{2\ell}$.
  \item[\textnormal{(2)}]
  There are codewords
  $\eta_\ell\in C_\ell$ and
  $\eta_{2\ell}=\eta_\ell\odot\eta_\ell\in C_{2\ell}$
  with $\res(\eta_\ell)=\res(\eta_{2\ell})=\delta_{1_{G_R}}$.
  For $a\in\set{\ell,2\ell}$, the map
  \[
    \Enc_a(f):=\sum_{g\in G_R}f(g)(g\cdot\eta_a)
  \]
  is a $G_R$-equivariant linear map from $\bbK^{G_R}$ to
  $C_a$ satisfying $\res\circ\Enc_a=\id$.
  Both codes have rate at least $1/b$.
  \item[\textnormal{(3)}]
    For each $a\in\set{\ell,2\ell}$, the code $C_a$ has a list of $h_{\mathrm{chk}}$ linear checks
  of width at most $w$ whose translated tester accepts every
  $y\in C_a$ with probability one and rejects every
  $y\notin C_a$ with probability at least
  $\sigma\dist(y,C_a)$.
\end{itemize}
The multiplication table of $G_R$, the check lists, and the
codewords $\eta_\ell,\eta_{2\ell}$ can be constructed
deterministically in time polynomial in $R$.
The two encoders can be evaluated within the same time bound.
In particular, these data give equivariant multiplication pairs
with common parameters $(b,\Delta,\sigma,w,h_{\mathrm{chk}})$
in the sense of \Cref{def:multiplication-pair}.
\end{theorem}

\subsection{Constructing a Level}
\label{subsec:explicit-code-construction}

Fix the arithmetic data of \Cref{sec:arithmetic-covers}, with
the fixed places and chamber $q$ of
\Cref{thm:fibre-interpolation}.
Choose the group coordinates on $\widehat X_R(6)$ so that
the lifts of $q$ form the data fibre.
Put
\[
  b:=|X_0(6)|,\qquad
  w:=N_*:=\prod_{i=1}^6n_i,\qquad
  h_{\mathrm{chk}}:=|X_0(0)|N_*.
\]
These numbers, the field $\bbK$, and the cell and stalk data
on $X_0$ are fixed throughout the construction.

\begin{lemma}\label[lemma]{lem:explicit-arithmetic-level}
For every sufficiently large $r\in2k_0\mathbb N$, a
deterministic algorithm computes a monic irreducible
$\mathfrak r\in\bbF_2[t]$ of degree $r$, the multiplication
table of $G_R$, the chamber coordinates
$\widehat X_R(6)=[b]\times G_R$, the two check lists of
\Cref{cor:arithmetic-chamber-codes}, and a word
$\eta_\ell\in C_\ell(\widehat X_R)$ satisfying
$\res(\eta_\ell)=\delta_{1_{G_R}}$.
Its running time is polynomial in $R=2^r$.
\end{lemma}

\begin{proof}
Enumerate the $2^r$ monic polynomials of degree $r$ over
$\bbF_2$ and test irreducibility by trial division by nonconstant
monic polynomials of degree at most $r/2$.
This takes time polynomial in $R$.
For each irreducible candidate $\mathfrak r$, form
$\bbF_R=\bbF_2[t]/(\mathfrak r)$ and apply
\Cref{lem:arithmetic-finite-data} to construct the group,
the principal cover, and its stalk matrices.
\Cref{cor:arithmetic-chamber-codes} then constructs the two
check lists in time polynomial in $R$.

Write the grade-$\ell$ list as
$\set{\chi_{\ell,\rho}}_{\rho\in[h_{\mathrm{chk}}]}$.
Solve the following linear equations over $\bbK$ for
$z\in\bbK^{[b]\times G_R}$:
\[
  \chi_{\ell,\rho}(s\cdot z)=0
    \quad(\rho\in[h_{\mathrm{chk}}],\ s\in G_R),
  \qquad
  z(1,g)=\delta_{1_{G_R}}(g)
    \quad(g\in G_R).
\]
Perfect completeness and positive soundness imply that the
first set of equations is equivalent to
$z\in C_\ell(\widehat X_R)$.
There are $(h_{\mathrm{chk}}+1)|G_R|$ equations in
$b|G_R|$ unknowns, where $|G_R|=R(R^2-1)$.
Gaussian elimination therefore takes time polynomial in $R$.
If the system is consistent, output the candidate and a
solution $\eta_\ell$.
Otherwise continue to the next candidate.
By \Cref{thm:fibre-interpolation}, at least one candidate
succeeds for every sufficiently large eligible $r$.
There are at most $R$ candidates, so the complete search
takes time polynomial in $R$.
\end{proof}

\begin{proof}[Proof of~\Cref{thm:explicit-multiplication-codes}]
Use the prime and word supplied by
\Cref{lem:explicit-arithmetic-level}, and set
$C_a=C_a(\widehat X_R)$ for $a\in\set{\ell,2\ell}$.
\Cref{cor:arithmetic-chamber-codes} gives invariance,
multiplication, the common positive distance bound, and
the translated testers with the stated list sizes and widths.
Take $\Delta$ and $\sigma$ to be their common bounds,
decreased to at most one if necessary.

Multiplication gives
$\eta_{2\ell}:=\eta_\ell\odot\eta_\ell\in C_{2\ell}$,
and restriction gives
$\res(\eta_{2\ell})=\delta_{1_{G_R}}$.
For both grades, \Cref{lem:equivariant-encoding} gives the
asserted encoder and rate bound.
In particular,
\[
  \res(\Enc_a(f))
    =\sum_{g\in G_R}f(g)(g\cdot\delta_{1_{G_R}})=f,
  \qquad
  \frac{\dim_\bbK C_a}{b|G_R|}\ge\frac1b.
\]
The construction lemma supplies the required data.
Squaring $\eta_\ell$ computes $\eta_{2\ell}$, and each
encoder is evaluated by summing $|G_R|$ translates of a
word of length $b|G_R|$.
These operations take time polynomial in $R$.
Indexing the family by the binary representations of the
eligible exponents $r$ gives the algorithmic family of
\Cref{def:local-testability}.
\end{proof}

\subsection{Admissibility and the PCP}
\label{subsec:arithmetic-admissibility}

\begin{lemma}\label[lemma]{lem:admissible-arithmetic-family}
The pairs of \Cref{thm:explicit-multiplication-codes}, for
all sufficiently large $r\in2k_0\mathbb N$, form an
admissible family in the sense of
\Cref{def:admissible-family}.
\end{lemma}

\begin{proof}
Let $r_1\in2k_0\mathbb N$ be large enough for
\Cref{thm:explicit-multiplication-codes,lem:arithmetic-routing}
to apply at every eligible exponent $r\ge r_1$.
All pairs have the same field and parameters, with
$C_1=C_\ell$, $C_2=C_{2\ell}$, and $\eta=\eta_\ell$.

For \textnormal{(A1)}, given $u\ge1$, choose the least
$r\in2k_0\mathbb N$ with $r\ge r_1$ and $|G_{2^r}|\ge u$.
The bound on consecutive group orders at the end of
\Cref{subsec:arithmetic-routing} gives
\[
  u\le |G_{2^r}|
    \le \max\set{|G_{2^{r_1}}|,2^{6k_0+1}}\,u.
\]
Indeed, if $r>r_1$, the preceding eligible group has order
less than $u$, and if $r=r_1$ the bound follows from $u\ge1$.
\Cref{lem:explicit-arithmetic-level} constructs the
multiplication table, word, and check lists, while
\Cref{lem:arithmetic-routing} constructs a symmetric
generating set $T_R$ of size $2^{k_0}+1$.
Since $R\le|G_R|=O(u)$, the total running time is
polynomial in $u$.

For \textnormal{(A2)}, \Cref{lem:arithmetic-routing}
constructs a $T_R$-schedule for every permutation of $G_R$
in time polynomial in $|G_R|$, with length
$O(\log|G_R|)$.
Choose the constant $c$ in \Cref{def:admissible-family}
large enough to bound the group-size ratio, the generating
set size, and the coefficient of
$\lceil\log_2|G_R|\rceil$ in the schedule length.
Both conditions then hold with this $c$.
\end{proof}

\begin{theorem}\label[theorem]{thm:explicit-pcp}
\textnormal{Circuit-SAT} has a polynomial-time PCP verifier
with perfect completeness and soundness error at most $1/2$.
For a circuit $C$ of size $n$, the proof has length
$O(n\log n)$ bits, and the verifier uses $O(\log n)$ random
bits and makes $O(1)$ nonadaptive bit queries.
An accepting proof can be constructed deterministically
from $\cenc(C)$ and a satisfying assignment in time
polynomial in $n$.
\end{theorem}

\begin{proof}
Apply \Cref{thm:code-to-pcp} to the admissible family of
\Cref{lem:admissible-arithmetic-family}.
\end{proof}

\section*{AI Disclosure}

The overall direction of starting from our previous qLTCs (which combine locally testability with local Reed-Solomon codes) towards a more efficient classical PCPs came from the authors. The results in this paper came from several rounds of interactions with ChatGPT Astra 6 Pro and Astra 6.2 Ultra. For editorial assistance, ChatGTP Astra, 6 and 6.2, and Claude Fable 5.2 were used. The authors verified the results and take responsibility for them.

\bibliographystyle{alpha}
\bibliography{refs}

\appendix
\section{Proofs of the preliminary lemmas}
\label{app:preliminaries}

\begin{proof}[Proof of~\Cref{lem:information-sets}]
A codeword in the kernel of restriction $C\to\bbK^S$ is supported on
$I\setminus S$. Its relative weight is therefore strictly less than
$\Delta(C)$, so it is zero. Thus restriction is injective.
The coordinate functionals indexed by $S$ consequently span the dual
space $C^*$. Choose a basis of $C^*$ among these functionals, and let
$A\subseteq S$ be the corresponding coordinate set. Then
$|A|=\dim_{\bbK}C$, and restriction $C\to\bbK^A$ is an isomorphism.
This also covers $C=\set{0}$, for which the chosen basis and $A$ are empty.
\end{proof}

\begin{proof}[Proof of~\Cref{lem:tensor-codes}]
Choose a basis of each $C_i$ and extend it to a basis of
$\bbK^{I_i}$. Their products form a basis of
$\bbK^{\prod_i I_i}$. For a fixed $i$, expanding in the coordinate
bases of the other factors shows that the $i$-line conditions are
equivalent to membership in
\[
  \bbK^{I_1}\otimes\cdots\otimes\bbK^{I_{i-1}}
  \otimes C_i\otimes
  \bbK^{I_{i+1}}\otimes\cdots\otimes\bbK^{I_d}.
\]
In the chosen product basis, this condition eliminates exactly the
coefficients using a basis vector outside $C_i$ in position $i$.
Imposing the conditions for every $i$ leaves precisely the product
basis of $\bigotimes_i C_i$. This proves the line criterion and the
dimension formula.

For the distance bound, first consider a nonzero word
$z\in C\otimes D\subseteq\bbK^{I\times J}$.
A nonzero slice $z(i,\cdot)\in D$ has at least
$\Delta(D)|J|$ nonzero entries. Hence at least that many slices
$z(\cdot,j)\in C$ are nonzero, and each has at least
$\Delta(C)|I|$ nonzero entries. Thus
\[
  \weight(z)\ge\Delta(C)\Delta(D).
\]
Iteration gives the bound for any number of nonzero factors. A tensor
of minimum-weight codewords attains it, since the support of a simple
tensor is the product of the supports of its factors. If a factor is
zero, then the tensor code is zero; its relative distance is $1$ by
convention, while every factor has relative distance at most $1$.
The asserted inequality therefore holds in this case as well.
\end{proof}

\begin{proof}[Proof of~\Cref{lem:parity-check-test}]
Construct $H_i$ and choose one of its $m_i$ rows uniformly. Query the
coordinates on which that row is supported and accept if the resulting
linear check is zero. This uses at most $q$ queries, chosen before any
answers are read. Every codeword passes because every row is a parity
check for $C_i$. For an arbitrary word $x$, the ideal rejection
probability is
\[
  \frac{|\supp(H_i x)|}{m_i}
  \ge\sigma\,\dist(x,C_i).
\]

For an implementation using a bounded number of unbiased random bits,
put $k=\lceil\log_2 m_i\rceil$ and use $k$ bits to choose
$U\in\set{0,\ldots,2^k-1}$ uniformly. If $U\ge m_i$, accept without
querying the word. Otherwise, check row $U+1$ as above. Each violated
row is selected with probability $2^{-k}$, so the rejection probability
is
\[
  \frac{|\supp(H_i x)|}{2^k}
  =\frac{m_i}{2^k}\,
    \frac{|\supp(H_i x)|}{m_i}
  \ge\frac{\sigma}{2}\,\dist(x,C_i).
\]
When $m_i$ is a power of two, $2^k=m_i$ and the soundness parameter is
$\sigma$. The case $m_i=1$ uses no random bits. Perfect completeness
and nonadaptivity hold in every case. The uniform construction of
$H_i$ and the subsequent row selection and evaluation give the claimed
polynomial-time algorithm.
\end{proof}

\begin{proof}[Proof of~\Cref{lem:equivariant-encoding}]
Surjectivity of $\res|_C$ gives a word $\eta\in C$ restricting to
$\delta_{1_G}$. Conversely, suppose such a word exists. Invariance
implies $g\cdot\eta\in C$ for every $g\in G$, and equivariance of
restriction gives
\[
  \res(g\cdot\eta)=g\cdot\delta_{1_G}=\delta_g.
\]
Consequently the linear map
\[
  \Enc(f)=\sum_{g\in G}f(g)(g\cdot\eta)=f\cdot\eta
\]
takes values in $C$ and satisfies $\res\circ\Enc=\id_{\bbK^G}$.
For $s\in G$, changing variables $g=sh$ gives
\[
  \Enc(s\cdot f)
  =\sum_{g\in G}f(s^{-1}g)(g\cdot\eta)
  =s\cdot\sum_{h\in G}f(h)(h\cdot\eta)
  =s\cdot\Enc(f).
\]
Thus this right inverse is equivariant. Any right inverse implies
surjectivity and is injective. This proves the equivalence and, whenever
the equivalent conditions hold, the bound $\dim_{\bbK}C\ge|G|$.
\end{proof}

\begin{proof}[Proof of~\Cref{lem:translated-checks}]
Condition on a check $\chi_\rho$. A term of that check supported at
$(j,g)$ queries $(j,s^{-1}g)$ after translation. Since
$s\mapsto s^{-1}g$ is a bijection of $G$, this coordinate is uniform
on $\set{j}\times G$. Its probability of lying in $B$ is
\[
  \frac{|B\cap(\set{j}\times G)|}{|G|}
  \le\frac{|B|}{|G|}.
\]
A union bound over the at most $w$ supported terms gives
$w|B|/|G|$ as an upper bound on the probability of at least one query
in $B$. Averaging over the independently chosen check preserves this
bound.

For words $x,y$ at relative distance $\delta$, let
$B=\set{i\in I:x(i)\ne y(i)}$, so $|B|=b|G|\delta$.
Run the two tests with the same check and translation. Their outcomes
agree unless some queried coordinate lies in $B$. Hence the absolute
difference of their rejection probabilities is at most
\[
  \frac{w|B|}{|G|}=wb\delta. \qedhere
\]
\end{proof}

\end{document}